\documentclass[lettersize,journal]{IEEEtran}

\usepackage[hyphens]{url}
\usepackage[pdfborder={0 0 0}, citecolor=blue, linkcolor=blue, urlcolor=black, colorlinks=true]{hyperref}
\usepackage[hyphenbreaks]{breakurl}

\usepackage{graphicx}
\usepackage{amsmath}
\usepackage{multirow}
\usepackage{graphicx}
\usepackage{caption}
\usepackage{subfigure}
\usepackage{xcolor}
\usepackage{tcolorbox}
\usepackage{colortbl}
\usepackage{textcomp}
\usepackage{amssymb}
\usepackage{microtype}
\usepackage{tablefootnote}
\usepackage{listings}
\usepackage{float}
\usepackage{comment}
\usepackage{fancyvrb}
\usepackage{latexsym}
\usepackage{pifont}
\usepackage{tabu}
\usepackage{booktabs}
\usepackage{pifont}
\usepackage{makecell}
\usepackage{xspace}
\usepackage{threeparttable}
\usepackage{hyperref}
\usepackage{pdflscape}
\usepackage{enumitem}

\definecolor{LightGray}{gray}{0.92}

\newcommand{\ivn}{\scalebox{0.92}{{IVN}}}
\newcommand{\ecu}{\scalebox{0.92}{{ECU}}}
\newcommand{\can}{\scalebox{0.92}{{CAN}}}

\newcommand{\obd}{\scalebox{0.92}{OBD-II}}

\newcommand{\vids}{\scalebox{0.92}{{VIDS}}}

\newcommand{\ishighlight}{\iftrue}		
\newcommand{\deltext}[1]{{}}
\newcommand{\newtext}[1]{{#1}}
\newcommand{\revisedtext}[1]{{#1}}
\newcommand{\newadd}[1]{{#1}}
\newcommand{\revised}[1]{{#1}}
\newcommand{\rerevised}[1]{{#1}}
\newcommand{\rnewadd}[1]{{#1}}

\newcommand{\bibauthor}[1]{{{#1} et al.}}

\renewcommand{\smallskip}{\vspace{0.5mm}}

\begin{document}

\title{Vehicular Intrusion Detection System for Controller Area Network: A Comprehensive Survey and Evaluation}

\author{
    \IEEEauthorblockN{
        Yangyang Liu,
        Lei Xue,
        Sishan Wang,
        Xiapu Luo,
        Kaifa Zhao,
        Pengfei Jing,\\
        Xiaobo Ma,
        Yajuan Tang,
        Haiying Zhou
    }
    \IEEEcompsocitemizethanks{
        \IEEEcompsocthanksitem Yangyang Liu, Xiapu Luo, Kaifa Zhao, and Pengfei Jing are with Department of Computing, The Hong Kong Polytechnic University;
        \IEEEcompsocthanksitem Lei Xue is with School of Cyber Science and Technology, Sun Yat-Sen University;
        \IEEEcompsocthanksitem Sishan Wang and Haiying Zhou are with Hubei University of Automotive Technology;
        \IEEEcompsocthanksitem Xiaobo Ma is with Department of Computer Science and Technology, Xi'an Jiaotong University;
        \IEEEcompsocthanksitem Yajuan Tang is with College of Engineering, Shantou University.
    }
}

\maketitle

\begin{abstract} 
\label{sec:abstract}
The progress of automotive technologies has made cybersecurity a crucial focus, leading to various cyber attacks.
These attacks primarily target the Controller Area Network (CAN) and specialized Electronic Control Units (ECUs).
In order to mitigate these attacks and bolster the security of vehicular systems, numerous defense solutions have been proposed.These solutions aim to detect diverse forms of vehicular attacks. However, the practical implementation of these solutions still presents certain limitations and challenges.
In light of these circumstances, this paper undertakes a thorough examination of existing vehicular attacks and defense strategies employed against the CAN and ECUs. The objective is to provide valuable insights and inform the future design of Vehicular Intrusion Detection Systems (VIDS).
The findings of our investigation reveal that the examined VIDS primarily concentrate on particular categories of attacks, neglecting the broader spectrum of potential threats. 
Moreover, we provide a comprehensive overview of the significant challenges encountered in implementing a robust and feasible VIDS. Additionally, we put forth several defense recommendations based on our study findings, aiming to inform and guide the future design of VIDS in the context of vehicular security.

\end{abstract}

\begin{IEEEkeywords}
In-vehicle network, Intrusion Detection System, {\can} bus. 
\end{IEEEkeywords}

\section{INTRODUCTION}
The Global Automotive Cybersecurity market size is projected to reach USD 3574.5 million by 2028, from USD 571 million in 2021, at a CAGR of 29.6\% during 2022-2028~\cite{autosecuritytrend}.
Modern vehicles consist of 70 to 100 {\ecu}s that interface with the {\can}. These units work together to execute various vehicle functions, encompassing powertrain, chassis, and body systems~\cite{nilsson2008vehicle, tuohy2014intra}.
Traditional designs do not fully consider security issues, such as fake messages, making the vehicle vulnerable to cyberspace~\cite{al2019intrusion}, and many cyberattack surfaces are exposed~\cite{biham2008steal,hoppe2011security,miller2013adventures,miller2015remote}.
For example, {\can}, the current de facto standard for in-vehicle network ({\ivn}), is designed with multiple safety considerations but limited security considerations, such as the nature of the broadcast, the lack of network segmentation, the lack of authentication and data encryption, and the vulnerable arbitration mechanism, which lead to various security issues.

\IEEEpubidadjcol
\revisedtext{Recently, various attacks are launched against the real vehicles~\cite{greenberg2015hackers, kulandaivel2021cannon} exploiting the vehicular vulnerabilities.
In a notable case, researchers Don Bailey and Mathew Solnik from iSec gained unauthorized access to a vehicle and remotely started its engine. They exploited vulnerabilities in the protocols used for remote vehicle control, as documented in their research~\cite{remotecontrolapp}.
Similarly, researchers Miller and Valasek demonstrated their ability to remotely manipulate a Jeep Cherokee while it was traveling at a speed of 70 mph. Exploiting vulnerabilities in the vehicle's entertainment system, they gained control over critical functions such as steering and brake activation, as documented in their work~\cite{greenberg2015hackers}.}
Furthermore, researchers from Trend Micro showcased a potential attack vector in the realm of vehicular security. This demonstration involved the exploitation of inherent vulnerabilities in the error handling mechanisms of CAN protocols~\cite{trendmicroattacks}.
In addition, the Proof-of-Concept (PoC) attacks were conducted to assess the vulnerability of vehicles through the compromise of the Telematics Control Unit (TCU)~\cite{foster2015fast}.
In this comprehensive survey, our first focus entails an in-depth examination of the specific attacks that are targeted by IDSs as well as the CAN vulnerabilities that attackers exploit in their endeavors.

\revisedtext{
To address the vehicular security issues, especially the issues of {\can}, various defense mechanisms are proposed to improve the vehicle security leveraging four major types of techniques, including message encryption~\cite{farag2017cantrack,dariz2017trade}, {\ecu} authentication~\cite{carel2018design,palaniswamy2020efficient}, safety-related component isolation~\cite{macher2017automotive,hu2020cvshield}, and {\vids}~\cite{cho2016fingerprinting,cho2017viden}. 
Among these mechanisms, {\vids}s detect potential attacks by monitoring the {\ivn} traffic without modifying the existing {\ivn} architecture or incurring additional {\ivn} traffic, and so that they are more practical compared to other types of defense approaches~\cite{wen2020plug,jeong2020sensor,joo2020hold}.
Consequently, this survey focuses on the {\vids}s for {\can}.
}

\revisedtext{
More precisely, the {\vids}s usually detect anomalies leveraging the features and patterns of the characteristics of the {\ecu}s and in-vehicle traffic, such as the fingerprints of {\ecu}s, the signal features, the clock skews, and message payloads~\cite{cho2017viden, stabili2017detecting}.
It is worth noting that, although these {\vids} can improve the security of {\ivn}s, they are mainly proposed with the consideration of special issues of the {\ivn} and without comprehensive studies of the security limitations of the {\ivn}. 
Hence, when they are applied in practice, various challenges will be encountered, such as efficiency, feasibility, and stability.
So we need to collect these papers and compare them in detail to illustrate the limitations of existing methods. Researchers can also find and effectively detect methods based on these limitations.}

Although there are works that study the of-the-shelf {\vids}~\cite{avatefipour2018state,tomlinson2018towards,young2019survey,wu2019survey,karopoulos2022demystifying,rajapaksha2023ai}, to our best knowledge, they cannot provide a comprehensive and practical view of the {\vids} to shed light for future {\vids} design and implementation. 
First, many {\vids}s have been proposed defending the attacks against {\ivn}s, but the existing surveys just include a limited number of them, such as survey conducted by Tomlinson et al.~\cite{tomlinson2018towards}, studying only 17 {\vids}s and missing the state-of-art fingerprint-based {\vids}. 
Second, these papers do not offer a detailed description and classification of the attacks targeted by VIDSs.
Third, some surveys do not evaluate detection performance of these VIDSs, such as the survey conducted by Young et al.~\cite{young2019survey}, where no evaluation is performed.

Consequently, To address these gaps, we offer a comprehensive survey of existing VIDS, summarizing all CAN-related attacks and detection methods, and providing a detailed comparison. Following the evaluation, we discuss the challenges faced by these methods and outline future development trends. We hope that our survey will generate increased interest in the field of vehicle intrusion detection. We aim for other researchers to gain a comprehensive understanding of existing attacks and detection methods targeting the CAN through our survey. By reading our work, we hope they will discern the differences and limitations among various methods, enabling them to identify potential research directions.

In general, this survey has the following four major contributions.
\begin{itemize}
 \item [1)] We analyzed 34 research studies related to vehicle attacks and systematically classified them into 18 distinct attack types.
\item[2)]
	We examined 53 different {\vids}, carefully analyzing and comparing their threat models, defense scenarios, and defense mechanisms.
\item[3)]
	We reproduced {\vids} that can be compared using the same dataset and evaluated these detection methods using real-world vehicle data. 
\item[4)]
	In addition to the survey and evaluation results, we delve into a thorough examination of the constraints associated with the investigative defense approach. Subsequently, we explore forthcoming trends in vehicle advancements, elucidating their implications for the future of {\vids}.
\end{itemize}

The remainder of the paper is organized as follows.
Section \ref{sec:relate_work} introduces and compares some other surveys on intrusion detection of {\ivn}s.
Section \ref{sec:preliminaries} gives a brief overview of the {\ivn} composition and the vulnerabilities that make it vulnerable to attacks.
Section \ref{sec:threat_model} analyses the attack models against the collected {\vids}s, and Section \ref{sec:attack} details the specific attack scenarios.
Section \ref{sec:NIDS} details all the {\vids} we find, which are evaluated from different perspectives in Section \ref{sec:evaluation}.
We reproduce and evaluate some {\vids}s, and show the test results and the challenges encountered in the implementation in Section \ref{sec:experiment}.
Finally, Section \ref{sec:discussion} discusses the current issues and trends with existing {\vids}s. Simultaneously, for the sake of comprehensiveness, we introduce the intrusion detection of heavy-duty vehicle CAN and the intrusion detection system for the Internet of Vehicles (IoV).

\section{RELATED WORK}
\label{sec:relate_work}

\begin{table*}[]
\caption{Summary of previous survey works on the field of VIDS}
\centering
\begin{tabular}{cccccc}
\toprule
Survey   & Year & No of Works & Attack Description & Performance Comparison & Experiment \\
\toprule
Rajapaksha et al.\cite{rajapaksha2023ai}& 2023 & 40  & X                  & \checkmark                   & X          \\
Karopoulos et al.\cite{karopoulos2022demystifying} & 2022 & 40  & X                  & \checkmark                   & X          \\
Aliwa et al.\cite{aliwa2021cyberattacks}           & 2021 & 30  & \checkmark               & \checkmark                   & X          \\
Xie et al.\cite{xie2021cybersecurity}              & 2021 & 23  & X                  & \checkmark                   & X          \\
Hafeez et a.\cite{hafeez2020state}                & 2020 & 5   & X                  & \checkmark                   & X          \\
Wu et al.\cite{wu2019survey}                       & 2019 & 20  & \checkmark       & \checkmark                   & X          \\
Young et al.\cite{young2019survey}                 & 2019 & 15  & \checkmark      & \checkmark                   & X          \\
Al et al.\cite{al2019intrusion}                    & 2019 & 24  & \checkmark       & \checkmark                   & X          \\
Lokman et al.\cite{lokman2019intrusion}            & 2019 & 25  & \checkmark       & \checkmark                   & X          \\
Dupont et al.\cite{dupont2019survey}               & 2019 & 24  & \checkmark               & \checkmark                   & X          \\
Tomlinson et al.\cite{tomlinson2018towards}        & 2018 & 17  & X                  & \checkmark                   & X          \\
Avatefipour et al.\cite{avatefipour2018state}      & 2018 & 5   & X                  & X                  & X          \\
Liu et al.\cite{liu2017vehicle}                    & 2017 & 4   & \checkmark               & \checkmark                   & X   \\
\bottomrule
\end{tabular}
\label{tab:summary-of-survey}
\end{table*}


\revisedtext{Although there are several survey papers on existing {\vids}~\cite{liu2017vehicle,avatefipour2018state,tomlinson2018towards,dupont2019survey,lokman2019intrusion,young2019survey,wu2019survey,al2019intrusion,hafeez2020state,xie2021cybersecurity,xie2021cybersecurity,aliwa2021cyberattacks,karopoulos2022demystifying,rajapaksha2023ai}, they do not provide a comprehensive evaluation of existing {\vids}s. 
They have several limitations and we will describe them in detail.
For better illustration, we provide an overview of recent surveys on {\vids} and compare their contributions.
The results of the comparison are shown in Table~\ref{tab:summary-of-survey}.}

\revisedtext{First, it is necessary to include more state-of-the-art research works.
There are more than 50 {\vids}s and new {\vids}s are constantly appearing, but some surveys only contain a small portion.
For example, Liu et al. \cite{liu2017vehicle} merely describe 4 papers about {\vids} and the latest paper among them was published in 2016. 
\bibauthor{Avatefipour} \cite{avatefipour2018state} focus on introducing the {\can} bus and its vulnerabilities, and they only analyze 5 research works related to intrusion detection of the {\ivn}. Additionally, they do not give a further comparison and analysis of these papers.
Tomlinson et al. \cite{tomlinson2018towards} detail 17 research works which are published before 2018.
Young et al. \cite{young2019survey} introduce 15 {\vids} based on the detection feature which are published before 2018.
Rajapaksha et al. \cite{rajapaksha2023ai} primarily focused on introducing intrusion methods related to AI technology and did not comprehensively cover all IDS relevant to {\ivn}.}

\revisedtext{
Second, these papers do not offer a detailed description and classification of the attacks targeted by {\vids}s. While some survey works mention attacks, they either refer to previous research or provide a brief overview of common attack scenarios. They do not enumerate the most recently proposed significant attacks and lack a detailed classification of all attacks. For instance, Aliwa et al. \cite{aliwa2021cyberattacks} only list six common attack scenarios: CAN bus sniffing, CAN bus fuzzing attack, CAN bus frame falsifying attack, CAN bus injection attack, CAN bus DoS attack, and ECU impersonation. They do not include the latest attack scenarios, such as voltage corruption attacks \cite{bhatia2021evading}. Additionally, Young et al. \cite{young2019survey} only present three attack demonstrations to illustrate attacks on vehicles.
Similarly, Rajapaksha et al. \cite{rajapaksha2023ai} only choose to address 5 common attack scenarios.
Such a simple description is insufficient and detailed attack descriptions can help researchers understand the goals of defenses.}

Third, appropriate experiments can help researchers understand the advantages of different approaches.
All these surveys do not reproduce the {\vids}s they introduce, and they also do not evaluate the detection performance of these {\vids}s based on a large-scale dataset.
The evaluation under the same dataset can help researchers intuitively compare the pros and cons of different methods.

Based on the above limitations of these surveys, we take the further study at existing intrusion detection systems for {\ivn}.
First, we search for various paper repositories and relevant conferences/journals to find comprehensive research works.  We collect 53 specific {\vids}s and analyze them carefully.
Second, We analyze and summarise the threat models for all these {\vids}s, and we also provide a detailed description and classification of the attacks that these {\vids}s target.
Third, we classify and present these papers in detail based on the data features used by these {\vids}s (e.g., {\ecu} characteristics, semantic information, etc.).
Finally, we compare the effectiveness of these {\vids}s in various ways, including the features used, the detection technologies, the attacks included, the validation methods and the detection results. Furthermore, we also reproduce the {\vids}s that can be implemented in real cars and test their detection effectiveness based on the same dataset.

\section{PRELIMINARIES}
\label{sec:preliminaries}
\subsection{In-vehicle Network}
Due to the increase of electronic control system complexity and the number of in-vehicle {\ecu}s,
the in-vehicle wiring harnesses also increases, which introduces various new challenges to guarantee the reliability and security of the in-vehicle communications.
With the purpose of reducing {\ivn} wiring zones and achieving efficient data sharing and exchange, the automotive electronic network system, namely {\ivn}, was born mixed with a variety of network technologies~\cite{ramesh2004vehicle}.
In this section, we will present the preliminary knowledge of {\ivn} from two major perspectives, including both the commonly used {\ivn} technologies and the {\ecu}s connected to {\ivn}.

\subsubsection{Electronic Control Unit}
{\ecu}s play an indispensable role in controlling the vehicle.
Like an ordinary computer, an {\ecu} consists of a microprocessor (MCU), memory (ROM, RAM), input/output interface (I/O), analog-to-digital converter (A/D), and large-scale integrated circuits.
It adjusts and manipulates the running of vehicles with different sensors and controllers.
In modern automobiles, {\ecu}s are used in various modules, such as engine control module (ECM), Powertrain Control Module (PCM), Transmission Control Module (TCM), and so on.
Also, the {\ecu}s need to exchange information between each other during running.
For example, the {\ecu} that controls the dashboard display requires various vehicle states, such as vehicle speed.
Consequently, the {\ecu}s must possess a relatively stable and efficient network environment.

\begin{figure}[t]
	\centering
	\subfigure[The architecture of IVN with gateway~\cite{miller2014survey}.]
	{
		\includegraphics[width=.38\textwidth]{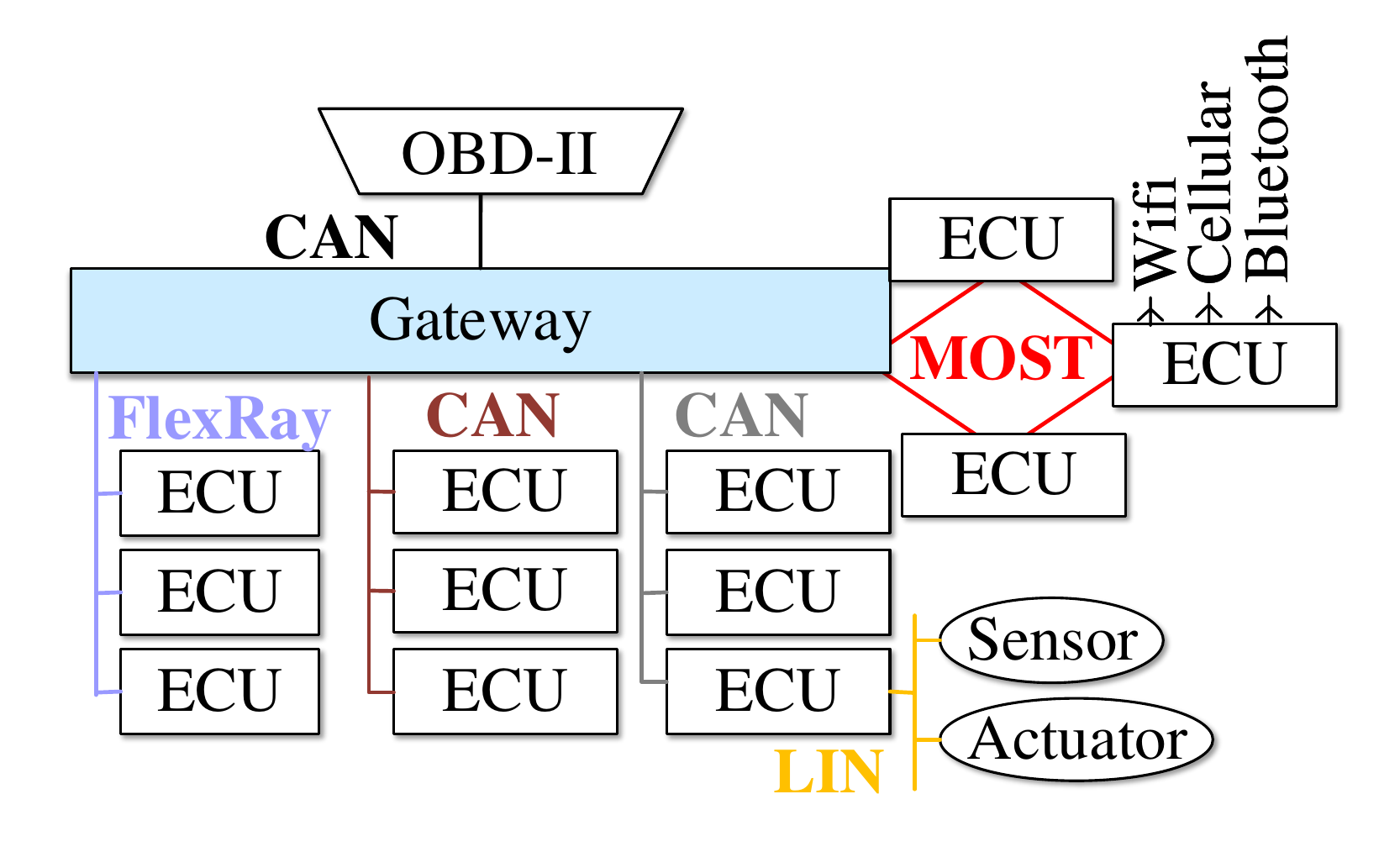}
		\label{fig:IVN_S_1}
	}
	\subfigure[An IVN architecture without gateways~\cite{miller2015remote}.]
	{
		\includegraphics[width=.38\textwidth]{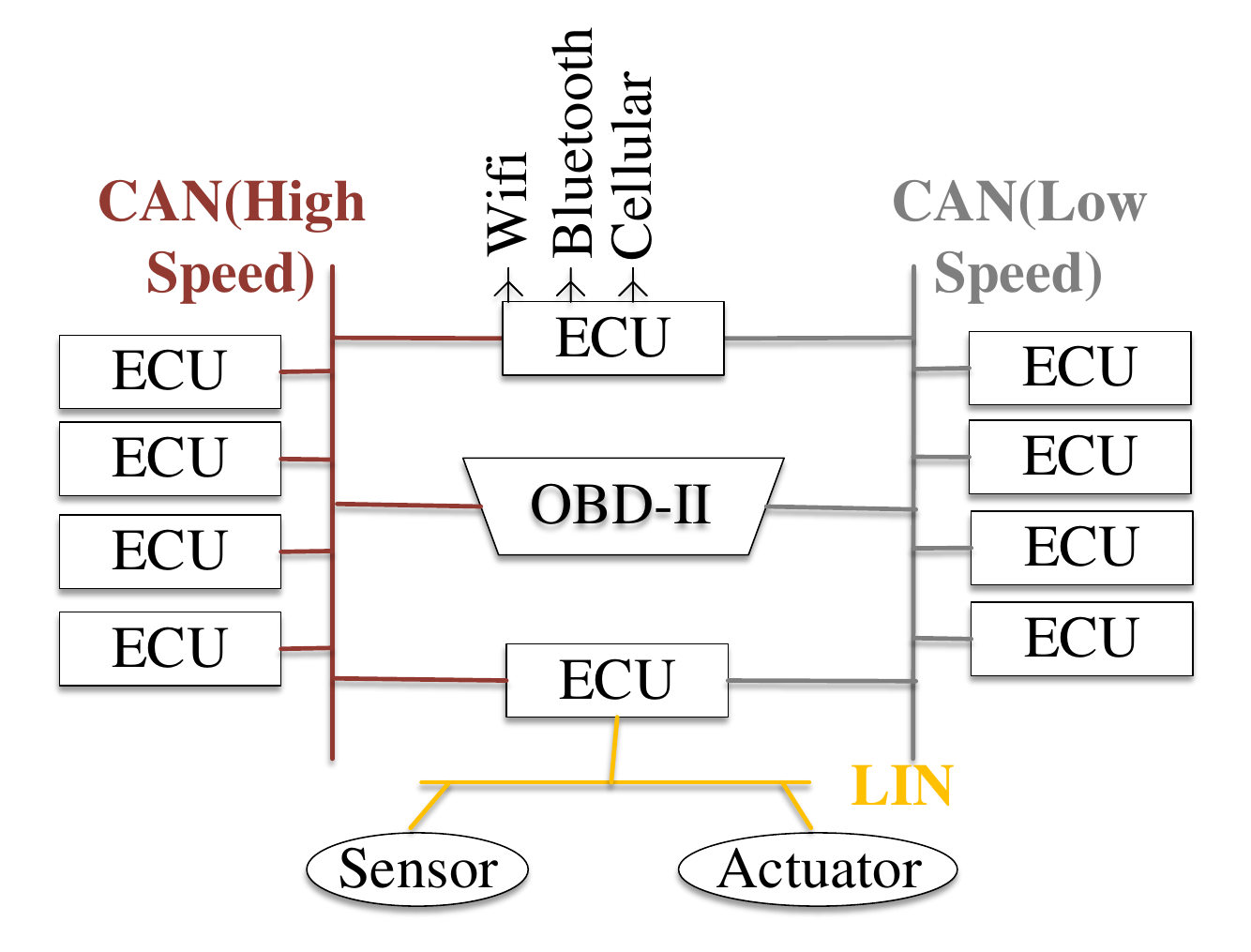}
		\label{fig:IVN_S_2}
	}
	\vspace{-1ex}
	\caption{Two common IVN architectures.}
	\vspace{-1ex}
	\label{fig:IVN_S}
\end{figure}

\revised{
\begin{table}[t]
	\centering
	\caption{Specifications of the widely used IVNs.}
	\begin{tabular}{c c c c}
		\toprule
		\makecell[c]{ Network \\Technologies} & Bitrate(Max) & Medium  & Standard \\
		\toprule
		\rowcolor{LightGray}
		LIN     & 19.2 Kbps & Single Wire                & Serial  \\
		{\can}  & 1 Mbps    & Twisted Pair               & CSMA/CR \\
		\rowcolor{LightGray}
            {\can}-FD & 8 Mbps    & Twisted Pair               & CSMA/CR \\
		FlexRay & 10 Mbps   & \makecell[c]{Twisted Pair \\or Optical Fibre} & TDMA    \\
            \rowcolor{LightGray}
		MOST    & 150 Mbps  & Optical Fibre              & TDMA    \\
            \makecell[c]{Automotive \\ Ethernet}  & 10 Gbps & Twisted Pair &\makecell[c]{Switched Full\\ Duplex} \\
		\bottomrule
	\end{tabular}
	\label{tab:IVN}
\end{table}
}

\subsubsection{IVN Technologies}
As shown in Figure~\ref{fig:IVN_S}, there are two common types of {\ivn} architectures.
In the first architecture (i.e., Figure~\ref{fig:IVN_S_1}), the in-vehicle control domains of {\ivn}s are connected to a central gateway, which provides an onboard diagnostic (\obd) port for diagnosing from outside of the vehicle~\cite{miller2014survey}.
In Figure~\ref{fig:IVN_S_2}, the {\obd} port directly connects to the {\ivn} without any gateway, and thus the external devices can easily monitor the in-vehicle communication data.

The in-vehicle {\ecu}s have different requirements on the speeds of the communication traffic.
For example, the body-related states (such as lights and door locks) can be transmitted in a low speed, and whereas the safety-related states (such as steering wheel angle and brake pedal pressure) need to be transmitted in a high speed.
Intuitively, the high communication speed require high cost and advanced techniques~\cite{cooperation2008most}.
Therefore, in order to reduce costs and meanwhile meet the various in-vehicle communication requirements between different ECUs, then vendors design the {\ivn}s with a mix of various network techniques.
Such as shown Figure~\ref{fig:IVN_S}, the most commonly used network technologies are {\can}~\cite{specification1991robert}, Local Interconnect Network (LIN)~\cite{package2003revision}, FlexRay~\cite{makowitz2006flexray}, and Media Oriented Systems Transport (MOST)~\cite{cooperation2008most}. 
Also, Table~\ref{tab:IVN} illustrates specifications of these {\ivn} techniques, including maximum transmission rate, transmission media, and the transmission standards.
For example, LIN is a low-speed serial communication protocol with the maximum transmission rate of 19.2 Kbps, and it uses single wire as the medium in the physical layer.

Among them, {\can} becomes a de facto {\ivn} protocol and it is proposed by Robert Bosch GmbH to define the layer-1 and layer-2 functionalities of the Open Systems Interconnection (OSI) network model in 1986~\cite{bosch1991can}.
{\can} is typically used to provide an efficient, stable, reliable, and economical communication method between {\ecu}s without a host computer, and it usually controls the core subsystem of vehicle, such as the engine power system, body control system, and electronic central electrical system.

LIN is a low-cost master-slave serial communication bus released in the late 1990s, and it is designed to serve as a cheap alternative to {\can} in {\ivn}s~\cite{ruff2003evolution}. 
Nowadays, LIN is a complement to {\can} and widely used in subsystems of {\ivn}s, which do not have the high communication speed requirement.

FlexRay is a new communication bus, which is released in 2009, and it is developed to support faster and more stable communication than {\can}. Compared with {\can}, the main advantages of FlexRay are the higher maximum data rate (10 Mbps) and deterministic time-triggered standard (i.e. time division multiple access (TDMA))~\cite{makowitz2006flexray}.

MOST is developed mainly for the transmission of multimedia data, and its maximum data rate is 150 Mbps.
Hence it is much more suitable to the multimedia data than {\can}.

As a mature and reliable standard communication bus, {\can} has been widely used in various vehicles for over 30 years. Usually, it controls the core part of the {\ivn}. Whereas, LIN, FlexRay, and MOST are generally used as a supplement or auxiliary to {\can} in the vehicle.
However, {\can} has various security limitations~\cite{checkoway2011comprehensive,miller2014survey,koscher2010experimental}, and therefore most of the {\ivn} intrusion detection systems are proposed for {\can}.

\newadd{
In addition to traditional vehicle network technologies, we introduce two emerging technologies: CAN with Flexible Data-Rate (CAN-FD) and Automotive Ethernet (AE).
Due to the increase in real-time data produced by control modules and sensors, CAN needs to meet stringent latency limits, thereby increasing its burden. 
Despite several alternatives being proposed, substantial efforts continue to focus on enhancing CAN, which has been upgraded to a CAN with Flexible Data-Rate (CAN-FD). 
This protocol was developed in 2011 and released by Bosch (in collaboration with industry experts) in 2012. Today, CAN-FD is used in modern high-performance vehicles~\cite{douss2023state}. CAN-FD, compatible with existing CAN networks, allows the new protocol to operate on the same network as traditional CAN. It can dynamically switch to different data rates and handle larger or smaller message sizes. The main differences between traditional CAN and CAN-FD are: 1) Increased length: Traditional CAN offers 8 data bytes, while CAN-FD provides flexible data rates ranging from 0-64 bytes per frame without needing to change the CAN physical layer, reducing protocol overhead and increasing efficiency. 2) Increased speed: Standard CAN networks are limited to 1 Mb/s. CAN-FD boosts the effective data rate to 8 Mb/s, which is eight times faster than traditional CAN. 3) CAN-FD can increase communication efficiency among multiple ECUs by up to 30 times, with faster speeds. 4) Higher reliability: One way to ensure reliability is through the use of Cyclic Redundancy Check (CRC). CAN-XL ({\can} eXtended Length) is an advancement over CAN-FD, designed to further increase data transmission rates and flexibility. It supports larger data frames and higher transmission rates, providing enhanced scalability and performance potential for future automotive applications. On March 22, 2024, the ISO released the 11898-2:2024 standard, which elevates the maximum speed of the CAN bus from the industry-recognized 8 Mbps of CAN FD to up to 20 Mbps, with data payloads of up to 2048 bytes.}

\newadd{
Automotive Ethernet is another protocol that could become mainstream in the future. In recent years, significant changes in the automotive industry, including the provision of various vehicle functions and the introduction of autonomous vehicles, have generated massive amounts of data. Automotive Ethernet has been proposed as the communication standard for IVNs because existing traditional protocol-based IVNs cannot cope with the increased bandwidth. Currently, various Ethernet protocols have been or are being used in Ethernet-based IVNs, such as BroadR-Reach, MOST150, IEEE 802.3bw (100BASE-T1), IEEE 802.3bq-2016, and IEEE 802.3ch-2020~\cite{kim2023vehicle}. In fact, with the increase in vehicle intelligence, many automakers, such as Tesla and BMW, have already started using Automotive Ethernet in commercial vehicles~\cite{veapplication}. Moreover, automakers have unified their views on the use of Ethernet, and many diagnostic software applications are compatible with Automotive Ethernet~\cite{vectorae}. As a new automotive network technology, the widespread adoption of Automotive Ethernet in vehicles will not be instantaneous. Automotive Ethernet will not replace existing vehicle network technologies in the short term. After entering the automotive field, Automotive Ethernet technology will initially integrate gradually from specific subsystems and ultimately advance the evolution of automotive network architecture. Automotive Ethernet holds significant potential.
Given that CAN remains the mainstream transport protocol in current commercial vehicle networks, our study primarily focuses on traditional CAN.
}
\section{THREAT MODEL}
\label{sec:threat_model}
\begin{figure*}[t]
	\centering
	\includegraphics[width=.8\textwidth]{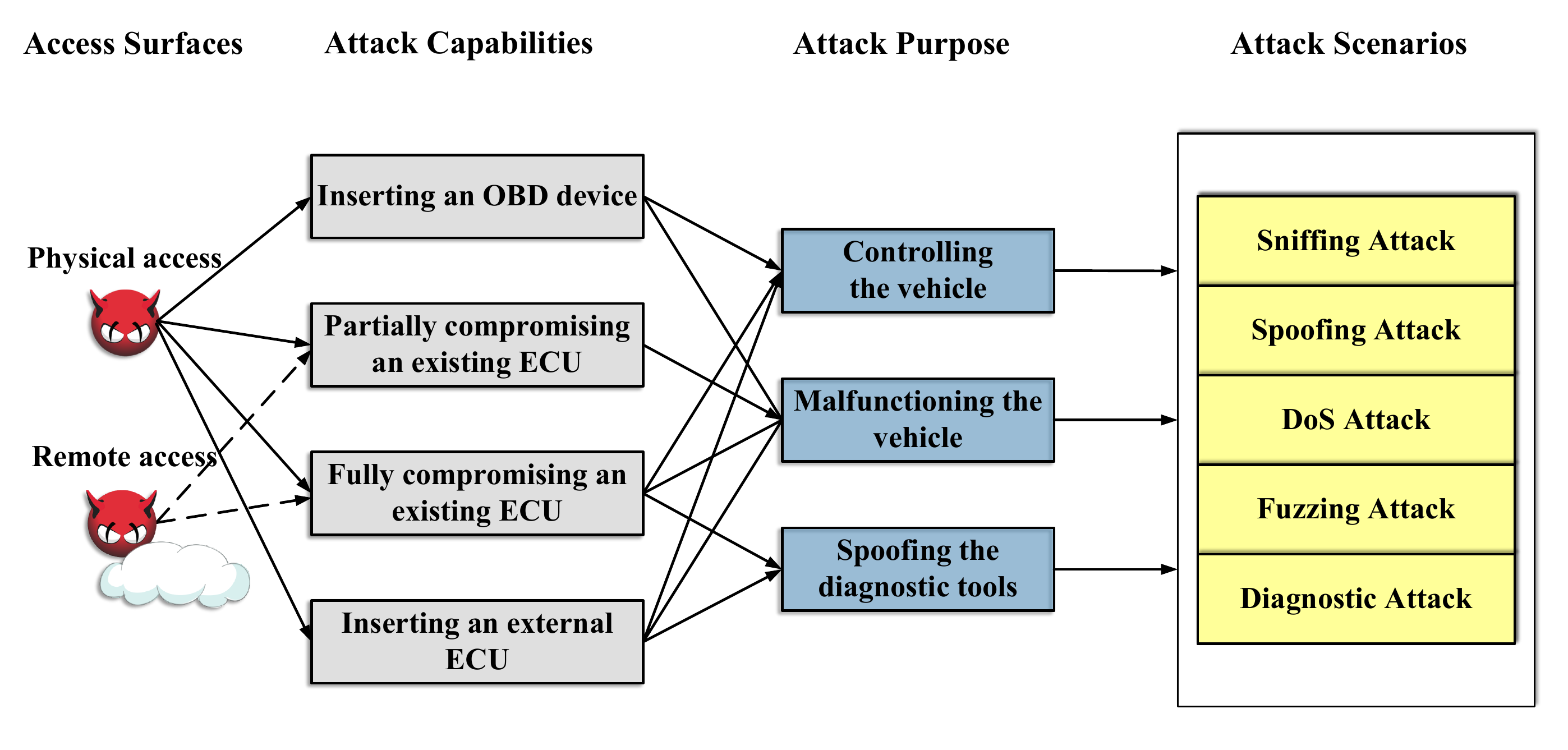}
	\vspace{-2ex}
	\caption{The threat model of the attack.}
	\vspace{-2ex}
	\label{fig:threat_model}
\end{figure*}

\rnewadd{
In this section, our primary focus centers on an in-depth examination of the threat models associated with all VIDS under consideration.
In various variants of VIDS, the authors postulate diverse adversary profiles encompassing varying attack capabilities, thereby influencing the spectrum of detectable attack modalities by the system.
Subsequently, a comprehensive consolidation of the aforementioned assumptions inherent to these systems is presented, establishing a profound linkage with the associated attack scenarios.
Next, a thorough exposition of the threat models is articulated, encompassing three fundamental dimensions: attack surfaces, attack capabilities, and attack purposes. Furthermore, the interplay and synergy between these tripartite facets are visually depicted in Figure~\ref{fig:threat_model}.
First, adversaries access the {\ivn}s through various attack surfaces.
Second, adversaries can own various capabilities to invade the {\ivn}s after they have access to the {\ivn}s.
Finally, adversaries can achieve different attack purposes when they have different capabilities to attack the {\ivn}s.}

\subsection{Access Surfaces}
Initially, the adversaries must establish connectivity with the IVNs through distinct access surfaces.
In accordance with pertinent research conducted by Koscher et al. (2010)~\cite{koscher2010experimental}, it has been ascertained that adversaries possess the capability to gain access to the IVNs via two distinct categories of surfaces, namely physical surfaces and network surfaces.
Physical surfaces commonly pertain to hardware components that establish a direct link with the vehicle, such as diagnostic ports. On the other hand, network surfaces encompass the network connections that bridge the vehicle to the cyberspace, encompassing technologies such as Wi-Fi and Bluetooth. Subsequently, an elaborate exposition elucidating the intricate particulars of these two attack surfaces shall ensue.

\noindent \textbf{Physical surface: }
The adversaries that utilize physical surfaces have to get close to the target vehicle. They can take three methods to attack the vehicle.
Firstly, adversaries can keep malicious devices on the OBD-II port (i.e., a physical diagnostic port which is usually above the accelerator pedal and connected to the {\ivn}~\cite{wen2020plug}.) and attack the {\ivn} directly.
For example, researchers from the Argus Research Team find a way to hack into the Bosch Drivelog ODB-II dongle that is plugged into the OBD-II port, and inject different malicious messages into the {\can} bus. 
They stop the engine of a moving vehicle by connecting to the dongle via Bluetooth~\cite{obd_attack}.
Adversaries can also insert the device briefly and launch an attack by injecting malicious code into the {\ecu} in the vehicle~\cite{christensen2019ethical,wen2020automated}.
Another method is that the adversaries change the firmware of {\ecu}s or install an additional {\ecu} while the vehicle is being repaired~\cite{miller2015remote}.

\noindent \textbf{Network surface: }
There are already many studies conducted on the radio interfaces, which enable the vehicles to accept external inputs and may cause the relevant on-board {\ecu} to be controlled~\cite{koscher2010experimental,costantino2018candy}.
Among the attacks exploiting such attack surfaces, most of the attacks are only effective at short distances due the features of the target communication types~\cite{oman2015relay,kapoor2018detecting}.

\subsection{Attack Capabilities}
The attack capacities of the adversaries are different, and we classify them into the following four major categories, including inserting an OBD device, partially compromising an {\ecu}, fully compromising an {\ecu}, and inserting an external {\ecu}, according to the attack methods.
Both of inserting an OBD device and inserting an external {\ecu} add a new node to the {\ivn}. 
The OBD device connects to the network directly via the OBD-II port while the external {\ecu} connects to the {\ivn} by changing the internal architecture of the vehicle.
During partially or fully compromising an {\ecu}, the attack target is an existing {\ecu} of the {\ivn}s. 
The partially compromised {\ecu} cannot send {\can} messages directly, and the fully compromised {\ecu} is also able to inject forged messages into {\ivn}.

\subsubsection{The malicious OBD device}
The OBD-II port is an important surface for communication between the {\ivn} and external devices.
Since the OBD-II port is exposed to the user, the adversary only needs to plug the attack device into the port without dismantling the vehicle.
However, there are certain limitations when the adversary injects malicious messages through the OBD-II port.
The layout of the {\ivn} can affect the effectiveness of this attack. As the Figure~\ref{fig:IVN_S_1} shows, the gateway in the vehicle can obstruct the broadcast of the normal in-vehicle messages and only allow the diagnostic messages to transmit in some particular vehicle models~\cite{zhou2019btmonitor}.
For example, Zhou et al. \cite{zhou2019btmonitor} find the {\ivn}s of two vehicle models, 2019 Chevrolet Malibu and 2019 Chevrolet Cruze, are not directly connected to the OBD-II port. They can not get the in-vehicle traffic through the port. 
In contrast, the {\ivn}s of another vehicle model, the 2012 Buick Regal, can be monitored directly through the OBD-II port.

\subsubsection{The partially compromised ECU}
Through a partially compromised {\ecu}, it is assumed that the adversary suspends the {\ecu} or puts the {\ecu} in the listen-only mode.
These adversaries who partially compromising an existing ECU can eavesdrop on in-vehicle communications and stop the {\ecu} from sending normal messages to other {\ecu}s, but they can not inject forged messages.
The adversaries can suspend the {\ecu} through the diagnostic commands or hardware vulnerability of the {\ecu}s, and we introduce them in detail.

Nowdehi et al.~\cite{nowdehi2019casad} demonstrate the possibility of such an attack via diagnostic protocols. 
The state of {\ecu} varies in different session modes.
Nowdehi et al. show that when the session mode is changed to programming mode, the {\ecu} can only listen to the bus but not send messages.  
In other words, the adversary can partially compromise an {\ecu} by changing the session mode of this {\ecu} through diagnostic command.

In \cite{cho2016fingerprinting}, Cho et al. propose another method to partially compromise an existing {\ecu}.
They mention that an {\ecu} with Microchip MCP2515~\cite{mcp2515}, which is one of the most common {\can} controllers, can be changed into various operation modes like configuration, normal, and listen-only through Serial Peripheral Interface (SPI). 
Therefore, the adversaries can make the {\ecu} enter different modes such as listen-only mode by utilizing the user-level features for configuring the {\can} controller.

\subsubsection{The fully compromised ECU}
Unlike partially compromised {\ecu}, with a fully compromised {\ecu}, the adversary can control the {\ecu} completely and have access to the memory data.
Apart from listening to the bus and stopping {\ecu} transmission, the adversary can also inject any fabricated messages into the bus while fully controlling the {\ecu}.
Many researchers demonstrate the methods to fully compromise an {\ecu}.

For example, Checkoway et al.~\cite{checkoway2011comprehensive} demonstrate vehicle vulnerabilities in some different external channels. 
In addition, they evaluate the potential for controlling the {\ecu}s via the prototype implementations for TPMS (tire pressure monitoring system), Bluetooth, FM (Frequency modulation), RDS (Radio Data System), and Cellular channels based on the vulnerabilities.
For example, they control the {\ecu} in telematics unit to send {\can} messages by predefined tire pressure sequences.

In~\cite{hoppe2011security}, Hoppe et al. add a few lines of malicious code to the arbitrary {\ecu} to control the vehicle. In the test, once a predefined condition is met, the code replays the {\can} messages containing the flag for opening the driver window.

Koscher et al.~\cite{koscher2010experimental} successfully control the vehicle operations and completely ignore driver input, such as disabling the brakes, stopping the engine, and so on. 
They propose an attack in which malicious code is embedded in the telematics unit, and the attack causes the vehicle to lose control.
Furthermore, they completely erase any evidence of its presence after the attack.

\subsubsection{The malicious additional ECU}
By inserting an external {\ecu}, the adversary can listen to the {\ivn}s and inject self-defined messages into the {\ivn}.
The function of the external {\ecu} is similar to that of the fully compromised {\ecu}, and the adversary can also inject forged messages into the bus through the additional {\ecu}.
However, compared with fully compromising an {\ecu}, inserting an external {\ecu} has to manually install a piece of new hardware equipment into the vehicle. 
The implementation of the attack needs to dismantle the vehicle and requires the adversary to have detailed knowledge of the vehicle architecture.
Additionally, when the adversary injects fabricated messages with different IDs that are supposed to be sent by other {\ecu}s, the risk of detection by fingerprint-based {\vids} \cite{cho2016fingerprinting} dramatically increases.

\subsection{Attack Purposes}
The adversaries launch attacks against the {\ivn} with different purposes, which can be categorized into remote vehicle control, vehicle malfunctioning, and diagnostic data spoofing.
The first type of purposes indicate the adversaries aim to fully control the vehicle remotely.
The adversaries with the seconde type of purpose aim to let the vehicle run out of control, causing driving accidents.
The third type of adversaries target on spoofing the diagnostic devices and concealing the safety issues by injecting fake data into {\ivn}.
Next, we present more details about these attack purposes.

\subsubsection{Controlling the vehicle}
The adversaries try to make the vehicle run as they want and attack it at a specific moment.
They can mislead the vehicle to react as they want by sending the forged in-vehicle messages to the {\ecu}s, and they can also send well-designed diagnostic commands to control the vehicle's actions directly.
For example, researchers from the Argus Research Team stop the engine of a moving vehicle through the diagnostic commands~\cite{obd_attack}, while Miller et al.~\cite{miller2015remote} manage to control the vehicle's turn signals by sending forged in-vehicle messages to the vehicle's {\ecu}.

\subsubsection{Malfunctioning the vehicle}
Instead of taking full control of the vehicle, the adversary can also make the vehicle lose control.
In this attack purpose, adversaries continuously send incorrect messages to the {\ecu}s or prevent the {\ecu}s from sending in-vehicle messages.
For example, Koscher et al.~\cite{koscher2010experimental} find that significant damage to the vehicle can be done by simple fuzzing of packets (i.e., iterative testing of random or partially random packets) because the range of valid {\can} messages is rather small.
Additionally, Cho et al.~\cite{cho2016error} propose a new type of Denial-of-Service (DoS) attack called bus-off attack. 
On two real vehicles, through iterative bus-off attacks, the victim {\ecu} enter the bus-off mode and can not send any messages.
As a result, the two vehicles get out of control.

\subsubsection{Spoofing diagnostic tools}
Another attack purpose is to spoof diagnostic tools and conceal security vulnerabilities in the vehicle.
In this attack purpose, adversaries mask the loss of the safety functionalities which are removed or fail.
This attack can endanger the vehicle's occupants due to the loss of a safety system.
In order to conceal security vulnerabilities, the adversaries manage to emulate the behaviors of a safety system within a diagnostic session by any compromised device with access to the {\can} bus.
For example, Hoppe et al. \cite{hoppe2011security} remove the airbag control system and hide the absence of the system.
They record the reactions to diagnostic queries in the presence of the airbag control module at first, and then they replay these replies when diagnostics software sends the diagnostic queries.
As a result, the diagnostics software reports the airbag control module's presence (including its name, part no., etc.) without any error conditions.

\section{TAXONOMY OF ATTACKS}
\label{sec:attack}
The adversaries usually exploit the vulnerabilities of {\ivn}s to attack the {\ecu}s, and the attacks can prevent normal communication of {\ecu}s on the {\ivn}s or transmit malicious messages to specific {\ecu}s~\cite{miller2015remote,koscher2010experimental}.
There are also attacks that are not related to the {\ivn}s or are not targeted by the {\vids}, and they are not the focus of our attention.
For example, an attack against the remote keyless system may allow the adversary to freely open the door~\cite{francillon2011relay,garcia2016lock}. 
If spoofing the remote keyless system is the ultimate purpose of the adversary, the {\ivn}s are irrelevant, and we ignore the attacks.

\begin{figure}[t]
	\centering
	\includegraphics[width=.45\textwidth]{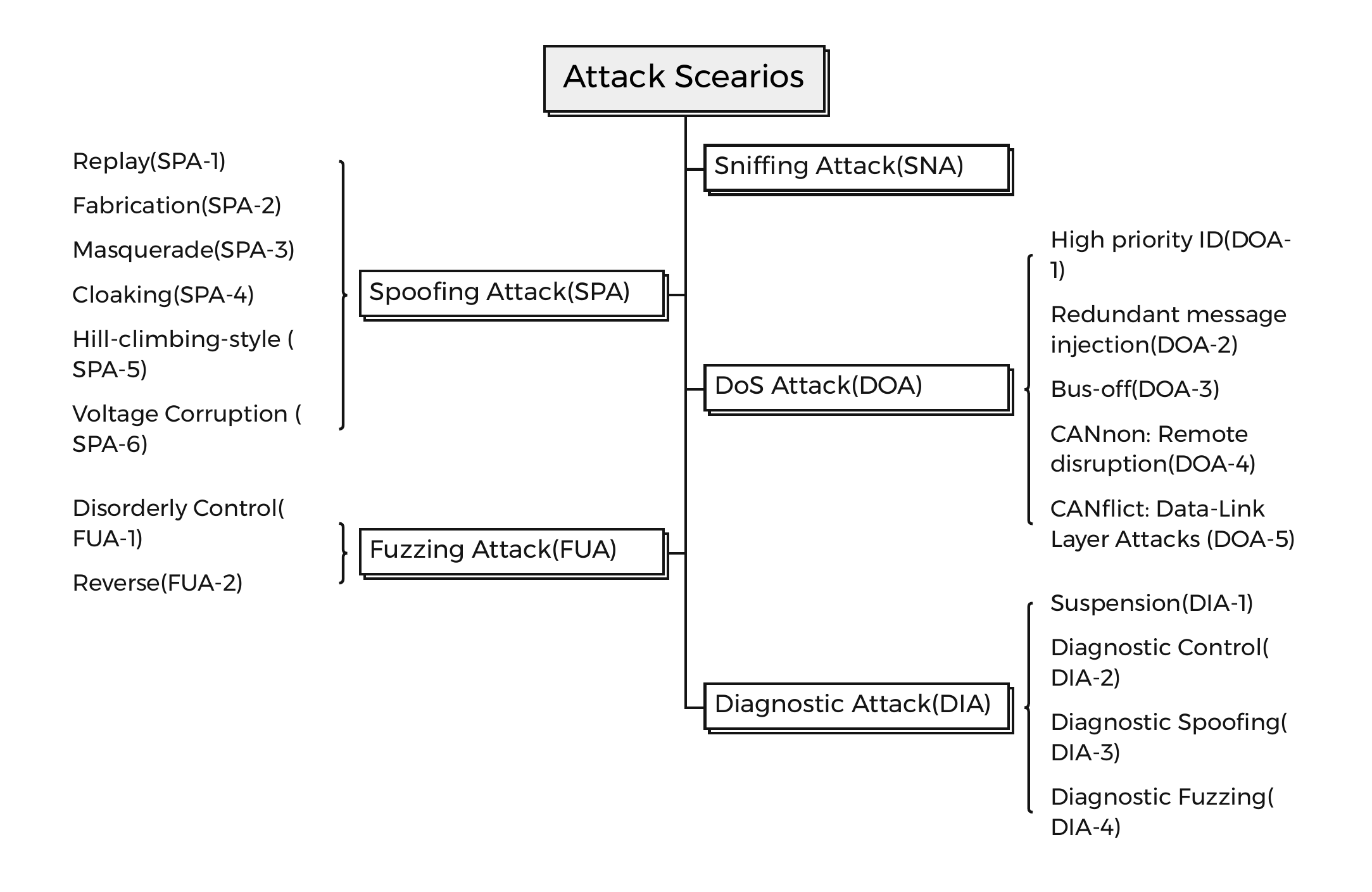}
	\vspace{-1ex}
	\caption{Detailed taxonomy of attack scenarios.}
        \vspace{-1ex}
	\label{fig:attack_taxonomy}
\end{figure}

To provide a taxonomy of {\ivn} attacks, we collect and observe the attacks from the publications we studied.
Afterward, we classify these attacks according to the injected malicious messages, the effects on the {\ecu}s, the attack purpose, and the attack methods. 
Figure~\ref{fig:attack_taxonomy} shows the taxonomy.

Particularly, we classify the attacks according to the attack methods: Sniffing attack (SNA), Spoofing attack (SPA), DoS attack (DOA), Fuzzing Attack (FUA) and Diagnostic attack (DIA).

\subsection{Sniffing Attack}
Since ECUs broadcast all messages in CAN and there is no authentication and encryption in the communication process, any ECU that joins CAN network can listen to all messages~\cite{haas2017automotive}.
The adversary can have access to CAN remotely or physically and listen to the CAN messages transmitted in CAN directly.
Adversaries can directly speculate on the regularity of messages and the semantics of messages.
They can detect specific private information in the vehicle or carry out further attacks based on message semantics.
Existing IDSs are difficult to detect this kind of attack because the attack has little influence on the CAN.

\subsection{Spoofing Attack}
Spoofing attack against the vehicle is launched by forcing the target {\ecu} to accept wrong messages and react in the wrong way. 
An adversary can disrupt the normal operation of the {\ecu} or even take control of the {\ecu} through a spoofing attack.
Additionally, adversaries can take different methods to deceive the target {\ecu}.
Based on the methods to deceive the target {\ecu}, these attacks can be divided into six categories.

\subsubsection{Replay}
The purpose of replay attack is to override the normal messages with the valid messages that have already been transmitted to the {\ivn}.
To mount a replay attack, the adversary needs to fully compromise an {\ecu} or fix an extra {\ecu} in the vehicle. 
Through the attack capability, the adversary can listen to the {\ivn} and replay the target messages. 
What's more, the frequency of forged messages is higher than that of normal messages to occupy the control of the target {\ecu}.
For example, previous research \cite{miller2014survey,song2016intrusion} mentioned that the adversary needs to inject messages 20-100 times faster than the original {\ecu} to make the target {\ecu} listen to the injected messages successfully.

\subsubsection{Fabrication}
The purpose of fabrication attack is to override any periodic messages sent by an uncompromised {\ecu} so that the receiving {\ecu}s are distracted and fail.
Through an in-vehicle {\ecu} fully compromised and an additional {\ecu} added to the vehicle, the adversary can fabricate and inject messages with forged ID, data length, and payload to control the specific {\ecu}. 
A fabrication attack is similar to the replay attack except for sending the messages that have been modified or forged. 
Additionally, the fabrication attack also needs a higher injection frequency, whose reason is the same as that of the replay attack. 
For example, in \cite{taylor2015frequency}, the messages are inserted at 5 times of the transmission rate of normal messages to control the {\ecu}.

\subsubsection{Masquerade}
\rerevised{
Masquerade attacks aim to manipulate unauthorized or compromised ECUs to impersonate legitimate ECUs and affect vehicle operations, utilizing two main approaches. In the first approach, adversaries either connect an unauthorized device to the CAN bus or control an existing ECU, sending forged messages with IDs matching legitimate ECUs while the original legitimate ECUs remain active. The second approach requires compromising two ECUs: one fully compromised and one weakly compromised target ECU (or using an unauthorized device instead of the fully compromised ECU). The fully compromised ECU injects forged messages to replace the weakly compromised target ECU's transmissions ~\cite{cho2016fingerprinting,cho2017viden,studnia2018language}. This is achieved by triggering transmission errors in the weakly compromised target ECU to increase its Transmission Error Counter until reaching bus-off state, forcing it to cease transmission, while the fully compromised ECU simultaneously sends malicious messages that mimic the target ECU's normal traffic pattern, making detection challenging.
}

\revisedtext{
\subsubsection{Cloaking}
Clocking attack is a special attack against specific IDSs.
Cho et al.~\cite{cho2016fingerprinting} proposed a method to identify malicious ECUs by using clock offset as the fingerprint of the ECU. In previous detection systems, they assumed that clock skew could not be imitated.
Sagong et. al. propose the cloaking attack, an intelligent masquerade attack in which an adversary modifies the timing of transmitted messages to match the clock skew of a targeted ECU. 
}

\subsubsection{Hill-climbing-style}
Hill-climbing-style attack is designed to deceive multi-frame based fingerprinting systems~\cite{foruhandeh2019simple}, such as Viden fingerprinting system~\cite{cho2017viden} and Clock-based IDS (CIDS)~\cite{cho2016fingerprinting}.
In a multi-frame based fingerprinting system, a batch of multiple frames has to be collected in order to perform one update of the fingerprinting threshold. Such fingerprinting schemes are vulnerable to the Hill-climbing-style attack, where the adversary is able to control the quantity of attack frames among the batch of frames collected, so that the attacker ECU can both hide its identity and shift the fingerprinting decision threshold gradually.

\subsubsection{Voltage Corruption}
With the specific purpose of evading the existing voltage-based IDSs, the voltage corruption attack is launched through poisoning the training data of such IDSs using two compromised ECUs (i.e., an attacking ECU and an accomplice ECU)~\cite{bhatia2021evading}.
Intuitively, by exploiting the static ID, periodicity, and predictable payload-prefix characteristics of CAN frames from one ECU, the adversary can let the attacking ECU be in the error-passive state and perform simultaneous transmission with a legitimate ECU with the assistance of the accomplice ECU. 
Consequently, the mixed voltages are collected as the training data for fingerprinting and the voltage-based fingerprinting of the victim is corrupted.
Even worse, since the attacking ECU and the victim transmit dominant bits at the same time, such attack cannot be detected by existing IDSs.

\subsection{DoS Attack}
During a DoS attack, the {\ecu} is suspended or unable to receive normal messages. 
Considering the methods of attack against the {\ecu}, we can classify DoS attacks into five major categories.

\subsubsection{High priority message injection}
The goal of the DoS attack with high priority ID messages is to occupy the {\can} bus and block normal messages.
To perform the attack, the adversary must have full control of a normal {\ecu} or an additional {\ecu}. 
Since a lower {\can} ID means higher priority and can get {\can} bus access according to the arbitration mechanism of {\can}~\cite{cho2016error}, the injected attack messages are usually set with low IDs, such as 0x000 that has topmost priority~\cite{muter2011entropy}.
Additionally, the adversary must increase the number of messages to fill the bus.
For example, in~\cite{song2016intrusion}, 6000 topmost priority messages are injected into the bus per second to fill the bus.
In the attack, the valid messages are be blocked, and all {\ecu}s receive none message.
As a result, these {\ecu}s can not work normally, and the vehicle is out of control.

\revisedtext{
\subsubsection{Redundant message injection}
The purpose of redundant message injection attack is to interfere with normal {\ecu}s receiving messages and make the {\ecu}s fail.
In order to mount the attack, the adversary has to fully compromise an {\ecu} or add an extra {\ecu}.
In the attack, the adversary can forge and inject messages massively through the compromised {\ecu}. 
The adversary can inject the traffic to surpass the {\can} bus's maximum capacity, which is only 1 Mbps. 
Additionally, the maximum size of a {\can} message is 128 bits (contains ID, CRC, bit stuffing and all other elements), and there are at least three consecutive recessive bits (i.e.,1) called \textit{`interframe space'} between messages~\cite{canBus}. 
Therefore, the adversary can inject about 8000 messages per second to launch this attack~\cite{song2016intrusion}.
}

\subsubsection{Bus-off}
The purpose of the bus-off attack is to disconnect or shut down an uncompromised {\ecu}.
Through a fully compromised {\ecu} or an additional {\ecu}, the adversary can monitor the transmission of in-vehicle messages and inject malicious messages at a specific moment. 
The bus-off attack utilizes the arbitration and the error handling mechanism in {\can}~\cite{cho2016error}. 
To perform a bus-off attack, the adversary has to transmit forged messages that satisfy the following three conditions.
First, the forged message should have the same ID as the message transmitted by the target {\ecu}.
Second, the forged message has to be transmitted at the same time as the message transmitted by the target {\ecu}.
Third, the forged message has at least one bit position in which its signal is dominant (i.e., 0), whereas victim's signal is recessive (i.e., 1),and all preceding bits of the two messages should be the same.
When an adversary sends forged messages that meet the above conditions to the {\ivn}, the can bus will detect a bit error and the error counter of the target {\ecu} will increase.
Then the adversary will send the forged messages constantly. 
After the error counter accumulates to a certain threshold, the target {\ecu} will turn off itself because of the error handling mechanism. 
As a result, the target {\ecu} can not send or receive {\can} messages until it is reawakened.

\subsubsection{CANnon: Remote disruption}
Kulandaivel et al. introduce a new class of attacks that leverage the peripheral clock gating feature in modern automotive microcontroller units (MCUs)~\cite{kulandaivel2021cannon}. By using this capability, a remote adversary with purely software control can reliably “freeze” the output of a compromised ECU to insert arbitrary bits at any time instance. Utilizing on this insight, they develop the CANnon attack for remote shutdown.

\subsubsection{CANFlict: Data-Link Layer Attacks}
CANFlict is a stealthy attack that can shut down the ECU in a bit-level granular way~\cite{de2022canflict}.
De et al. exploit polyglot frames and pin conflicts to perform data-link layer attacks against CAN, making use of different peripherals already present on the microcontroller.
CANflict enables an attacker to exploit known vulnerabilities of the CAN protocol to remotely implement read and write attacks without any assumption on the periodicity of the transmitted messages.

\subsection{Fuzzing Attack}
Fuzzy attack is a common attack that requires only a small amount of a priority knowledge about the {\ivn}~\cite{song2016intrusion,olufowobi2019anomaly,martinelli2017car}. 
According to the validity of IDs, the fuzzy attack can be divided into two categories according to the {\can} identifiers that the adversary uses, including

\subsubsection{Disorderly Control}
The fuzzy attack with random IDs aims to make all {\ecu}s in the vehicles receive unpredictable messages and get messy. 
To mount an attack by injecting random messages with random IDs, the adversary needs to have a fully compromised {\ecu} or additional {\ecu} added to the vehicle. 
In this attack, the adversary does not need to have a complete understanding of the {\ivn}s or reverse-engineering~\cite{koscher2010experimental}. 
In fact, because the range of valid {\can} messages is rather small, significant damage can be done by simple fuzzing of messages completely (i.e., randomly spoofed identities with arbitrary data)~\cite{lee2017otids,olufowobi2019anomaly}.

\subsubsection{Reverse}
The goal of this attack is to reverse engineer the specific meaning of CAN messages.
During this attack, the attack capabilities that the adversary needs to master are the same as that of the other fuzzy attack. 
In this attack, the adversary has to inject carefully forged messages whose fields are constantly being modified \cite{marchetti2016evaluation}. 
Then, the adversary infers the meaning of each field in the CAN message based on the response of vehicles.

\subsection{Diagnostic Attack}
Diagnostic communication is usually used for vehicle mechanics and developers to test or diagnose the vehicle's state. 
The messages which are used in diagnostic communication (i.e., diagnostic messages), are different from the in-vehicle normal messages used for communication between {\ecu}s, and they are usually injected into the vehicle from the OBD-II port.
We conduct in-depth research on diagnostic communication and come up with some diagnostic attacks.

\subsubsection{Suspension}
Suspension attack in diagnostic communication aims to stop the transmissions of the target {\ecu} and make it be listen-only mode.
To mount a suspension attack in diagnostic communication, the adversary only needs to connect to the OBD-II interface. 
If the adversary can control or add an {\ecu}, he can also carry out such an attack.
The implementation of this attack exploits the diagnostic session in diagnostic communication.
The diagnostic session enables a specific set of diagnostic services and functionality in the {\ecu}s, and an {\ecu} will be in various states for different sessions~\cite{iso14229-1}. 
Nowdehi et al.~\cite{nowdehi2019casad} prove that the {\ecu} can only monitor the bus when the session mode is changed to programming mode.

\subsubsection{Diagnostic Control}
The purpose of the control attack is to control the behaviors at a special moment and do harm to the vehicle and driver.
Control attacks exploit services in diagnostic communications that can control the behaviour of the vehicle (such as stop the engine~\cite{obd_attack}), and the adversary can control the vehicle even when the vehicle is running.
If an adversary sends a dangerous control command at an inopportune moment, this is a huge threat to the safety of drivers.
The adversaries can control the motion state of the vehicle (such as turning off the engine, acceleration, braking and changing vehicle steering) through diagnostic messages directly. 
These control commands are most closely related to the security of the vehicle and difficult to reverse~\cite{miller2015remote}.
In addition, related dynamic parameters (such as speed, RPM, steering wheel angle) are also the focus of researchers and these attacks are easy to detect~\cite{narayanan2016obd-securealert}.

\subsubsection{Diagnostic Spoofing}
The spoofing attack in diagnostic communication aims to deceive the diagnostic devices and conceal the true condition of the vehicle.
To mount this attack, the adversary needs a fully compromised {\ecu} or an additional {\ecu} to send diagnostic messages to the diagnostics devices used by vehicle mechanics.
When the adversary fully compromises an {\ecu}, he can record the reactions to diagnostic queries from diagnostic devices and replay the record messages to spoof the diagnostic devices.
For example, in \cite{hoppe2011security}, Hoppe et al. spoof the diagnostics software and hide the absence of the airbag control module.  
Specifically, they record the reactions to diagnostic queries in the presence of the airbag control module at first. Then, these replies are successfully replayed in the absence of this module. 
As a result, the diagnostics software reports the airbag control module's presence (including its name, part no., etc.) without any error conditions.

\subsubsection{Diagnostic Fuzzing}
In {\ivn}, the diagnostic messages are communicated following specific diagnostic protocols, thus the adversaries can first reverse-engineer the diagnostic protocols and then launch attacks by building the attacks messages following the diagnostic protocols~\cite{fowler2019fuzz}.
Diagnostic fuzzing attack is the first step in reverse engineering, and adversaries can obtain the specific format of the diagnostic protocol through the feedback of the {\ecu}.
In addition, since all such attacks are launched leveraging the messages conforming to the protocols, it is hard to detect these attacks according to the underlying protocols.

\section{DESCRIPTION OF EXISTING INTRUSION DETECTION SYSTEMS}
\label{sec:NIDS}

\begin{figure}[t]
	\centering
	\includegraphics[width=.48\textwidth]{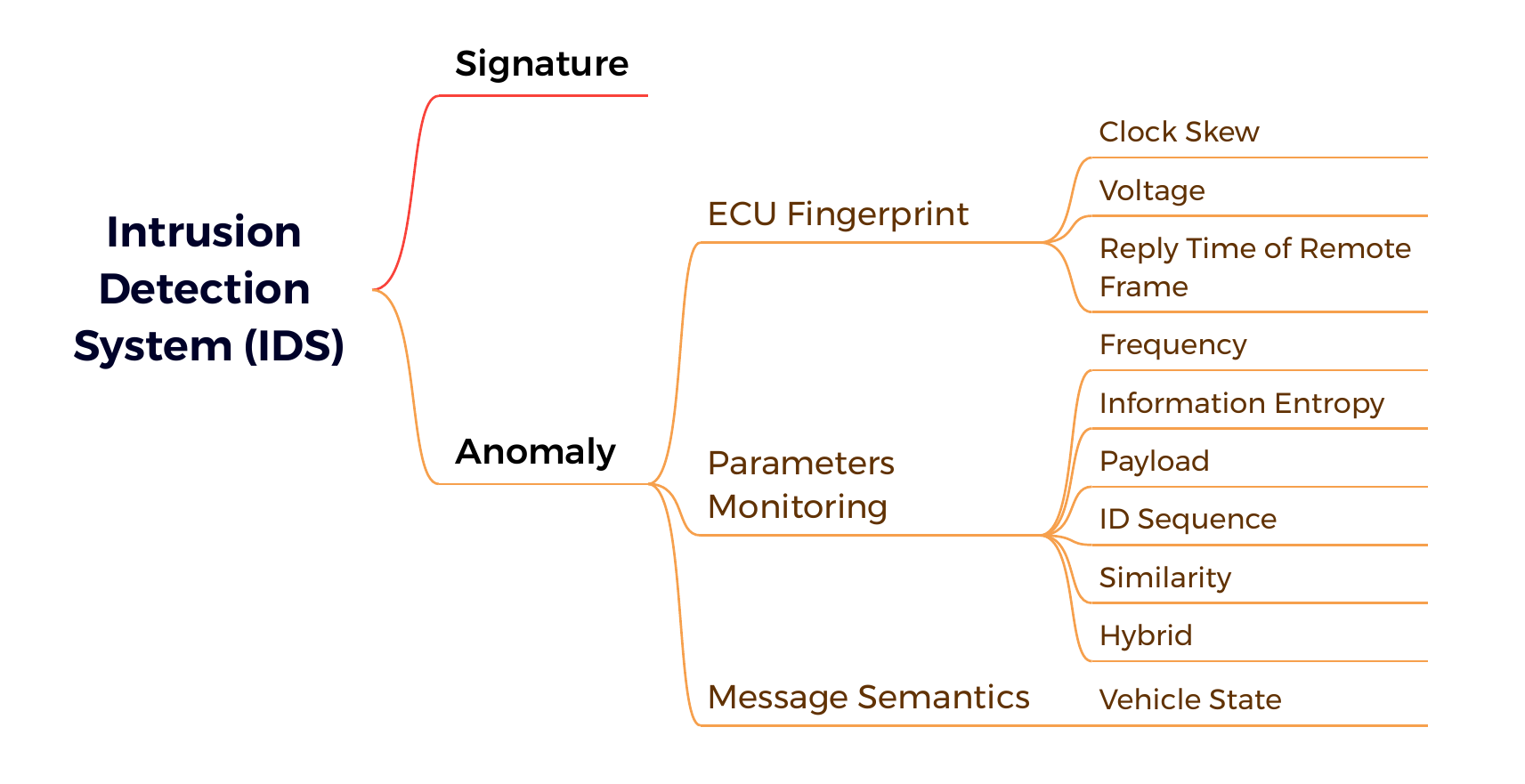}
	\caption{Existing intrusion detection systems.}
	\label{fig:IDS}
\end{figure}

\revisedtext{
Cybersecurity becomes essential for vehicles, and various vehicular {\vids}s have been proposed recently.
We curated a collection of 53 seminal or impactful papers. It is important to highlight that our selection primarily focuses on papers investigating {\vids} for the CAN. We intentionally omitted works related to other network protocols.
In this section, we study the methodologies adopted by the existing {\vids} in detail and Fig.~\ref{fig:IDS} demonstrates our specific approach to classifying these IDSs.
Tab.~\ref{tab-cla-ids} shows the classification results of existing IDSs based on our approach.
}

\subsection{Signature-based VIDS}
The signature-based introduction detection approaches are manly applied to detecting the known attacks.
Intuitively, the traffic features of the known attacks are summarized and set as signatures, and researchers monitor the current network traffic and detect intrusions according to these features.
The messages whose features match these signatures are marked as violations.

The signature-based {\vids} has various advantages.
Since this method does not require {\ecu} to possess powerful computing resources, signature-based {\vids} is easy to deploy in vehicle~\cite{Singh2013ASO}.
Furthermore, this type of detection method can detect known attacks with high accuracy and low error rate, and can determine which type of attack and how many times the {\ecu} is confronting~\cite{jyothsna2011review}.

However, the signature-based {\vids} also has its own limitations.
First, it cannot detect the attack, of which the traffic features are not specified in the predefined knowledge base. 
Many researchers are devoted to proposing the different attack methods on {\ivn} security continually (e.g.,~\cite{miller2013adventures,miller2014survey,miller2015remote,cho2016error}), and these new attacks suggest that it is impractical to consider all attack methods.
It is critical and challenging to keep the signature databases up-to-date with frequent updates.
Despite different problems and difficulties, the researchers also propose some signature-based {\vids}.
We will give details of them below.

In \cite{larson2008approach}, Larson et al. propose a specification-based {\vids} that can be implemented in the specific {\ecu}.
The {\ecu} traffic can be analyzed based on the information derived from the {\ecu}-behavior security specifications and {\can} protocol stacks. 
They evaluate the applicability of the detection and show that most attacks can be detected.
However, the paper is primarily based on rules developed by the \textit{{\can}Open} protocol~\cite{farsi1999introduction}, a typical application layer protocol used on {\can}. 
In reality, the application layer protocols for {\ivn}s are developed by individual OEMs themselves. 
The specifications they propose are not necessarily applicable to all vehicles.

\newtext{
In \cite{studnia2018language}, Studnia et al. propose a language-based VIDS for vehicle embedded networks. They exploit the high predictability of the IVN and extract attack signatures from the behavior models of different ECUs in the vehicle. Using this approach, they can detect malicious sequences of messages transmitted on the IVN.}

\rnewadd{
\smallskip \noindent
\textbf{Brief Discussion:} Signature-Based VIDS offers notable advantages, including high accuracy in detecting known attacks, low false positive rates, and low computational overhead, making it feasible for deployment on resource-constrained ECUs. However, its reliance on predefined attack signatures limits its ability to detect novel (zero-day) attacks, and maintaining an up-to-date signature database requires continuous updates and significant effort. Additionally, these methods often struggle with proprietary vehicle protocols, making them less adaptable across different automotive systems. Due to these limitations, researchers are increasingly focusing on anomaly-based detection methods, which can identify previously unknown threats by learning normal network behavior and detecting deviations, thereby offering greater adaptability and robustness against evolving cyber threats in CAN.
}

\subsection{Anomaly-based {VIDS}}
For anomaly-based intrusion {\vids}, researchers build normal behavior profiles by training the normal model of the system activity at first. 
Then, they utilize the deviations between the profiles and the traffic under test to detect intrusions.
Comparing with signature-based intrusion detection, the anomaly-based intrusion is not based on prior knowledge of the known attacks, and it can detect previously unknown attacks.
However, it's challenging to determine reliable anomaly boundaries because some normal behaviors are not constant, and the adversary can imitate normal behaviors to spoof the detector~\cite{garcia2009anomaly}.

Despite these disadvantages, many researchers pay attention to propose various detection methods to improve detection rates and avoid being evaded.
In the following, we will introduce these methods detailedly. 
These methods can be divided into \textit{{\ecu} fingerprint-based {\vids}}, \textit{parameters monitoring-based {\vids}}, and  \textit{message semantics-based {\vids}}.

\revisedtext{
\subsubsection{ECU Fingerprint-based {VIDS}}
Resulting from the differences in physical properties of {\ecu}s, different {\ecu}s always have different hardware fingerprint.
In the communication among ECUs within the {\ivn}, each ECU possesses one or more unique IDs that only it can use for transmission. It is important to highlight that when developing current IDS, researchers often overlook special messages like diagnostic messages and remote frames.
Therefore, the matching of ID and fingerprint can be exploited to detect the compromised {\ecu}s that send malicious messages with forged ID.
Researchers use various fingerprints of the {\ecu}s to detect the intrusions.
Based on the type of fingerprints, We distinguish between clock skew-based {\vids}s, voltage-based {\vids}s, and reply time of remote frame-based {\vids}s.
}

\paragraph{Clock Skew-based {VIDS}}
The sending time of a {\can} message is affected by the clock frequency of the {\ecu}. 
Due to hardware differences, the clock frequency of different {\ecu}s is slightly different.
In fact, researchers have proposed various schemes for fingerprinting network devices by estimating their clock skews through the timestamps carried in their control packet headers~\cite{jana2009fast}.
Therefore, researchers try to apply the technology to the {\ivn}, and they propose many {\vids}s that use the clock skew to mark off different {\ecu}s and identify the mismatching of {\ecu} and ID.

Cho et al.~\cite{cho2016fingerprinting} propose a clock-based {\vids} (C{\vids}). 
They measure the intervals of periodic {\can} messages at first. 
Then, they extract clock skews from these intervals for fingerprinting specific {\ecu}s and model their clock behaviors using the recursive least squares (RLS) algorithm.
Afterwards, based on the model, C{\vids} detects intrusions via cumulative sum (CUSUM) analysis.

Ying et al. \cite{sagong2018cloaking} also study the effect of clock skew and present a clock skew-based {\vids} based on the widely used network time protocol (NTP). 
Compared with state-of-the-art (SOTA) {\vids}~\cite{cho2016fingerprinting}, this method simplifies updating the average and accumulated skew caused by clock skew.

Furthermore, Ying et al. continue to study this feature. 
They propose the cloaking attack in~\cite{ying2019shape} which is aimed to avoid the detection of clock skew-based {\vids}s and provide formal analyses of the attack for two clocks skew-based {\vids}s, i.e., the SOTA {\vids}~\cite{cho2016fingerprinting} and the NTP-based {\vids}~\cite{sagong2018cloaking}.
The experimental results find that the average prediction error is within 3.0\% for the SOTA {\vids} and 5.7\% for the NTP-based {\vids}.

Zhou et al. \cite{zhou2019btmonitor} directly measure the bit time of the {\can} frames, which is determined by the {\can} controller and transceiver. 
In contrast to previous {\vids}s based on clock skew, the approach does not have to worry that an attacker will use software to simulate the clock skews of victim {\ecu}. 
However, this method requires additional equipment to monitor the {\can}'s electrical signals and a separate detector for each {\can} path.

\paragraph{Voltage-based {VIDS}}
Instead of clock skew, researchers also focus on other fingerprints.
In \cite{murvay2014source}, Murvay et al. use the Mean Squared Error (MSE) of voltage measurements as fingerprints of ECUs, but they use the voltages measured on a low-speed (10Kbps) CAN bus, which is far from contemporary vehicles that usually operate on a 500Kbps CAN bus.
Researchers attempt to apply this technology to {\ivn}s, and they propose many voltage-based {\vids}.

Cho et al.~\cite{cho2017viden} propose a novel scheme that can identify the attacker {\ecu} by measuring and utilizing the subtle difference voltages between the normal {\ecu}s, called Viden.
Via the first phase, Viden exploits the voltage measurements to construct and update the normal transmitter {\ecu}s' voltage profiles as their fingerprints. 
According to the fingerprints, Viden can distinguish the normal {\ecu}s from the attacker {\ecu}s or compromised {\ecu}s. 
However, Viden does not consider the complexity of the vehicle environment (such as variable temperature and power voltage) when the method is implemented. Furthermore, it uses two separate electrical signals, {\can} high and {\can} low~\cite{specification1991robert}, and the electrical signals used separately are less resistant to interference. 
These cases can both reduce its performance of anti-jamming.

Avatefipour et al.~\cite{avatefipour2017physical} also propose a physical-fingerprint model that identifies both channel and ECU.
They extract 40 features based on both time and frequency domain signals, and then employ the features to train a neural network-based classifier. 
They evaluate the {\vids} by using a dataset collected from 16 different channels and four identical {\ecu}s transmitting same messages.
Experimental results indicate that the proposed method achieves correct detection rates of 95.2\% and 98.3\% for channel and {\ecu} classification, respectively.

Choi et al. \cite{choi2018identifying} introduce a {\vids} that can identify a message's origin using an additional fixed 18-bit value in the extended identifier field. 
Their approach increases the total number of bits transmitted per message, and the extended identifier can not be used for the other purpose. 
This method is difficult to apply to existing automobiles because it needs to modify the modern vehicles' existing {\can} protocol.
Subsequently, Choi et al. also present another {\vids}~\cite{choi2018voltageids} named Voltage{\vids} that has improved the previous method in which the additional ID field is no longer needed.
Furthermore, Voltage{\vids} is evaluated in two vehicles and achieve identification rates ranging from 90.01\% to 99.61\%.

Kneib et al. \cite{kneib2018scission} also propose a {\vids} called Scission, which uses fingerprints extracted from {\can} frames to identify the sending {\ecu}s. 
Scission utilizes physical characteristics from analog values of {\can} frames to determine whether a legitimate {\ecu} sends it.
Compared with the previous implementation of {\vids} based on voltages \cite{cho2017viden}, Scission uses the differential signal instead of high and low signals and is more reliable in terms of changing conditions such as battery charge or electromagnetic compatibility.

Foruhandeh et al.~\cite{foruhandeh2019simple} demonstrate the vulnerability of the existing multi-frame-based automotive {\vids}s to a hillclimbing-style attack, which allows a compromised {\ecu} to impersonate another. 
Then, they show SIMPLE, a novel {\vids} that uses physical layer features within a single frame to fingerprint the {\ecu}s and is immune to hillclimbing-style attack. 
Additionally, this method requires a relatively low sampling rate and adapts to various environments.

Kneib et al. \cite{kneib2020easi} continue to study the {\vids} based on the voltage characteristics of {\ecu}. 
They believe that the previous research on {\ecu} voltage requires an oscilloscope that needs a high sampling rate of up to 2.5  GS/s to generate the fingerprints, and the high sampling is a big obstacle to the implementation of these algorithms in real vehicles. 
Therefore, they reduce the resource requirements for sender identification using the characteristics of the rising edge. 
Furthermore, to cope with the complex environment on the vehicle, they also build an adjustable model to change signal characteristics during runtime.

Murvay et al. \cite{murvay2020tidal} introduce a novel {\vids} called TIDAL-{\can}. 
Differential delays of bus signals, which are affected by bus characteristics and sender location, are used by TIDAL-{\can}.
TIDAL-{\can} identifies the specific location of the target {\ecu} by comparing the difference in signal arrival time at the two bus ends, and it can successfully identify the attacks that are implemented by compromised {\ecu}s.
The results of their experimental evaluation show that the method provides high identification rates.
Whereas TIDAL-{\can} also needs the equipment whose sample rates reach 250MS/s, it is not easy to implement in modern vehicles.

\paragraph{Reply Time of Remote Frame-based {VIDS}}
In addition to the physical properties of {\ecu}s, researchers also used differences in {\ecu} reaction times and relative positions as {\ecu} fingerprints.

In \cite{lee2017otids}, Lee et al. propose an intrusion detection method based on the remote frame by measuring the offset ratio and time interval between request and response messages. 
Each {\ecu} will reply to the remote frame with the ID of the {\ecu}~\cite{paret2007multiplexed}. 
Because of the different positions of the {\ecu}s on the bus and different transmitting procedures of different IDs, the average intervals between the request and response messages can be used as the fingerprint of different IDs. 
Therefore, they can use the interval time to determine whether the original {\ecu} sends a specific ID message.

\rnewadd{
\smallskip \noindent
\textbf{Brief Discussion:} Fingerprint-based {\vids}s are one of the most popular methods for {\ivn} intrusion detection. Several papers have been accepted at top conferences in the security field.
These methods take advantage of the physical characteristics of the {\ecu} and can effectively detect most compromised malicious {\ecu}s and added {\ecu}s. 
However, if a compromised {\ecu} still sends the same ID and only changes the frequency or payload, it cannot be detected.}

\newtext{
Additionally, these methods are difficult to implement on existing vehicles because they require complex equipment and operations. 
For example, voltage-based {\vids}s require high-precision oscilloscopes to listen for changes in the voltage of {\can} messages.
Therefore, how to reduce additional operations is the next challenge to be considered for the fingerprint-based approach.
Furthermore, the aging of the hardware and the impact of external environment (e.g., temperature) on the {\ecu}'s physical characteristics must also be considered.}

\subsubsection{Parameters Monitoring-based {VIDS}}
Parameters monitoring-based {\vids}s utilize the change of special in-vehicle parameters (such as frequency of {\can} messages) while detecting attacks.
In these approaches, researchers do not consider the meaning or sender of the messages. They only need to monitor the traffic on the {\ivn} and extract specific traffic parameters.
Based on the used parameters, these {\vids}s can be taken into the following six categories.

\paragraph{Frequency-based {VIDS}}
In a normal vehicle, {\ecu}s send their messages to the {\can} bus periodically, and the transmission frequency is fixed~\cite{song2016intrusion}. 
When adversaries want to attack the {\ivn}s by injecting special messages, the frequency of these messages will change. 
Additionally, since the {\ecu}s will still send normal messages, adversaries need to inject forged messages to the bus at a faster frequency to override the normal messages~\cite{miller2014survey}. 
As a result, the rate of messages on the bus increases significantly and is easily detected.
For example, in \cite{miller2015remote}, Miller and Valasek report that they need to inject at a rate of at least 20 times faster than normal for their attack to be successful.
According to this phenomenon on the {\ivn}, some researchers come up with some {\vids}s based on the frequency of {\can} messages.

Ling et al. \cite{ling2012algorithm} present a method for detecting {\can} malicious messages based on the invariance of {\can} IDs and the constant frequency of each ID.
They aim to detect the injection attack and DoS attack. However, they can not deal with the attack where legitimate {\can} ID messages are injected at a low speed. 
Additionally, their experiment is implemented in {\can}oe, a special bus simulation tool, and they do not apply the method to the real vehicle. 
The method can cause some misjudgments because of the complex situation of the real vehicle.

Taylor et al. \cite{taylor2015frequency} introduce a {\vids} by measuring inter-packet timing over a sliding window and compare the timing to historical averages to yield an anomaly signal. 
In this method, the authors use a one-class support vector machine (OCSVM) to classify normal messages and malicious messages. 
Furthermore, they also show that a similar measure of messages' data contents is not effective for identifying anomalies.

Song et al. \cite{song2016intrusion} propose a lightweight intrusion detection method for the {\ivn} according to the time intervals of {\can} messages. 
They capture {\can} messages from a real car and perform three kinds of message injection attacks. 
They prove that time interval is a meaningful and effective feature to detect injection attacks in the {\can} traffic.

Gmiden et al. \cite{gmiden2016intrusion} introduce a simple {\vids} based on the analysis of {\can} message time intervals. 
The advantage of the method is that it does not require a modification in the hardware layer and can be implemented in each {\ecu}.

Moore et al. \cite{moore2017modeling} also propose an anomaly detection system based on the regularity of normal signals. 
They find that for each {\can} ID, the time of a message only depends on the previous message's time, and the wait time following a fixed distribution. 
Thus, they train models for each {\can} ID based on the interval of two continuous messages. 
Each model will flag unusually short/long intervals as an anomaly while monitoring the traffic and the system produces an alert upon three consecutive anomalies.

Tomlinson et al. \cite{tomlinson2018detection} use a time-defined window to detect message changes in {\can} resulting from injection and reflash attacks. 
They analyze three methods (ARIMA, Z-score, and supervised method) that compare each interval for {\can} messages within the window against the averages for all packets with the same ID within that window. 
This method reduces the calculation cost because it only needs to calculate the average at the end of the moving window. 
However, if attackers understand this detection mechanism, they can inject malicious messages at rates similar to the normal messages within a window to deceive the detection system.

Olufowobi et al. \cite{olufowobi2019anomaly} present an {\vids} based on change-point detection techniques using adaptive CUSUM algorithm to detect statistical changes and intrusions in {\can} bus message stream. 
The method also judges the intrusion by detecting the abnormality of the message sending time. 
The attack can not be detected if the adversary does not change the frequency of the messages.

Young et al. \cite{young2019automotive} demonstrate that the basic assumption that all {\can} messages have consistent timing intervals is not true. 
In normal vehicles, the timing intervals of some special ID can change due to normal driving operations, and the change can make {\vids} based on constant timing intervals inaccurate.
Furthermore, they propose and evaluate a frequency-based {\vids}. 
They prove that this method could solve the problem raised by interval-based approaches.

In \cite{olufowobi2019saiducant}, Olufowobi et al. present an approach for detecting intrusions in {\ivn}s using the pattern of message sending, called SAIDu{\can}T. 
They build a specification based on messages and worst-case response time analysis of the {\can} bus at first and use the specification to detect the abnormal messages.
SAIDu{\can}T considers the jitter and retransmission phenomenon in the {\can} bus to more accurately define the transmit time of the message compared with other frequency-based {\vids}. 
It achieves a better F1 score compared with interval-based and frequency-based approaches with less detection delay. 
However, this method can not solve attacks such as masquerade attacks that imitate the time of the victim {\ecu}.

\paragraph{Information Entropy-based {VIDS}}
The information entropy, often just entropy, is the average amount of information contained in any random variable, which can be interpreted as the intermediate level of "information," "surprise," or "uncertainty" inherent in the variable's possible outcomes~\cite{shannon2001mathematical}.
In the context of network and Internet systems, the concept of entropy-based intrusion detection has been considered in various publications~\cite{gu2005detecting,lee2000information}, but entropy approach has a high rate of false positives because of the randomness of the standard computer networks~\cite{kemmerer2002intrusion}.
Instead, the traffic in {\ivn} is much more stable, and injected malicious messages will significantly change the entropy of traffic.
The researchers can use the change of entropy to detect the injected malicious messages.

In \cite{muter2011entropy}, Michael M\" uter et al. introduce the {\vids} based on information entropy for the first time.
They suggest to measure the entropy of {\ivn} and use it as the specification of the normal operation for the network.
They use the entropy of a set of {\can} IDs and special states to detect the intrusion.
In this paper, they describe three attack scenarios to show the usefulness of their approach. 
Furthermore, they put forward different methods to detect these attacks.
Firstly, they increase the message's frequency with a specific {\can} ID while the engine is running (replay attack). 
To detect this attack, based on the concept of relative entropy~\cite{cover1991entropy}, they calculate the relative distance of the system's normal behavior and the behavior to be detected.

Secondly, they attack the availability of a bus system by performing a flooding attack on the {\can} bus. The implementation is done by sending a mass of messages containing the most dominant identifier 0x000 (Dos attack). In this scenario, they measure entropy during the vehicle's regular operation and compare this value to that during the attack phase.

Thirdly, in this attack scenario, they think the adversary tries to disturb the system by injecting selective, spurious speed signals, e.g., to impact the {\ecu}s that need the signals. 
To defend against the attack, they utilize the conditional self-information theory~\cite{cover1991information} to check the coherence to the speed signal's expected behavior and the previous value.

In \cite{marchetti2016evaluation}, Marchetti et al. propose and evaluate an entropy-based method for detecting anomalies in {\can} messages generated by a real vehicle. 
They find that if only one detector is used to detect anomalies, they can only detect attacks that inject many anomalous messages. 
On the other hand, to detect low-volume attacks, in which the attacker injects only 1 packet per second, they need to set up a detector for each ID.

In \cite{wu2018sliding}, Wu et al. present a novel {\vids} based on information entropy, which uses a fixed number of messages as sliding windows. 
Compared with the above entropy-based {\vids}s, the method uses a simulated annealing method~\cite{bertsimas1993simulated} to get the best parameters (i.e., the best sliding window size, standard deviation, and corresponding sensitivity) at first.
The experimental results demonstrate that the method can effectively improve the accuracy and effectiveness of intrusion detection for DoS and injection attacks on {\ivn}s.

\paragraph{Payload-based {VIDS}}
Many works utilize the data fields of {\can} messages for anomaly detection. On the {\ivn}, the payload syntax and semantics of the same ID are the same~\cite{markovitz2017field}. 
Furthermore, the changes in vehicle status such as speed are continuous and uniform. 
Reflected on the vehicle messages, the changes in data content are regular and stable. 
Therefore, the intrusion can be detected according to the dramatic changes in data fields of {\can} messages.

Stabili et al. \cite{stabili2017detecting} propose a novel method that can identify malicious {\can} messages injected by adversaries in the {\can} bus.
In particular, this detection method studies the payloads of all messages transmitted on the bus. 
It compares the Hamming distance between consecutive payloads of the same ID to build a valid range of the Hamming distance for each ID in the training phase.
Furthermore, since the proposed method has very low computational complexity and small memory footprints, it can be implemented in the real vehicle.

Taylor et al. \cite{taylor2016anomaly} consider the data interdependence between IDs and develop an anomaly detector by learning to predict the next data word originating from each sender on the bus on the base of long-short-term memory (LSTM) recurrent neural network for {\can} bus anomaly detection. 
The message that that differs significantly from the predicted result isflagged as anomaly. 
After implementing this detector, they evaluate it by abnormal data created by modifying the {\can} bus data.

Kang et al. \cite{kang2016intrusion} propose a novel {\vids} by utilizing a deep neural network (DNN) to enhance the security of the {\ivn}.
In this paper, Kang et al. choose the data field that includes 64-bit positions (i.e., 8 bytes) in the {\can} message and calculate the distribution of bit-symbols. 
They use the probabilities of bit-symbols as the features to distinguish normal or malicious messages.

Xiao et al. \cite{xiao2019robust} propose a novel and robust {\vids} by using spatiotemporal information enabled time series prediction.
The proposed IDS analyzes the CAN traffic generated by the {\ivn} in real time and identifies the abnormal state of the vehicle practically.
In this method, the authors use the ConvLSTM model~\cite{shi2015convolutional} to exploit the association between multiple {\can} messages to find more effective features for intrusion detection.
Experiment results show the performance of the model and the effectiveness
against various attacks.

Kukkala et al. \cite{kukkala2020indra} present a novel {\vids} called INDRA that utilizes a Gated Recurrent Unit (GRU) based recurrent autoencoder~\cite{chung2014empirical} to detect anomalies in {\can}.
They use the change of the payload to train the model and detect the anomalies.
Additionally, they evaluate their proposed framework under different attack scenarios.

\begin{table*}[t]
\centering
\begin{tabular}{|c|ccc|}
\hline
\multicolumn{1}{|c|}{Signature} & 
\multicolumn{3}{c|}{\cite{larson2008approach},\cite{studnia2018language}}   \\ \hline
\multirow{10}{*}{Anomaly} & \multicolumn{1}{l|}{\multirow{3}{*}{ECU Fingerprint}}  
& \multicolumn{1}{l|}{Clock Skew}    
& \cite{cho2016fingerprinting},\cite{sagong2018cloaking},\cite{zhou2019btmonitor} \\ \cline{3-4} 
& \multicolumn{1}{l|}{} & \multicolumn{1}{l|}{Voltage}   
& \cite{cho2017viden},\cite{avatefipour2017physical},\cite{kneib2018scission},\cite{choi2018identifying},\cite{foruhandeh2019simple},\cite{kneib2020easi},\cite{murvay2020tidal}\\ \cline{3-4} 
& \multicolumn{1}{l|}{} & \multicolumn{1}{l|}{Reply time} 
& \cite{lee2017otids}  \\ \cline{2-4} 
& \multicolumn{1}{l|}{\multirow{6}{*}{Parameters Monitoring}} 
& \multicolumn{1}{l|}{Frequency}     
& \cite{ling2012algorithm},\cite{taylor2015frequency},\cite{song2016intrusion},\cite{gmiden2016intrusion},\cite{moore2017modeling},\cite{tomlinson2018detection},\cite{olufowobi2019anomaly},\cite{koyama2019anomaly},\cite{young2019automotive},\cite{olufowobi2019saiducant}, \\ \cline{3-4} 
& \multicolumn{1}{l|}{} & \multicolumn{1}{l|}{Information Entropy} 
& \cite{muter2011entropy},\cite{marchetti2016evaluation},\cite{wu2018sliding}  \\ \cline{3-4} 
& \multicolumn{1}{l|}{} & \multicolumn{1}{l|}{Payload}             
& \cite{taylor2016anomaly},\cite{kang2016intrusion},\cite{markovitz2017field},\cite{martinelli2017car},\cite{stabili2017detecting},\cite{studnia2018language},\cite{seo2018gids},\cite{xiao2019robust},\cite{nowdehi2019casad} ,\cite{pawelec2019towards},\cite{kukkala2020indra}, \\ \cline{3-4} 
& \multicolumn{1}{l|}{} & \multicolumn{1}{l|}{ID Sequence}         
& \cite{marchetti2017anomaly},\cite{islam2020graph}  \\ \cline{3-4} 
& \multicolumn{1}{l|}{} & \multicolumn{1}{l|}{Similarity}     
&\cite{ohira2020normal}  \\ \cline{3-4} 
& \multicolumn{1}{l|}{} & \multicolumn{1}{l|}{Hybrid}
&\cite{theissler2014anomaly},\cite{tian2017intrusion},\cite{weber2018embedded},\cite{wang2018distributed},\cite{tomlinson2018using},\cite{koyama2019anomaly},\cite{zhu2019mobile},\cite{hanselmann2020canet} \\ \cline{2-4} 
& \multicolumn{1}{l|}{Message Semantics}    & \multicolumn{1}{l|}{Vehicle State} 
& \cite{narayanan2016obd-securealert},\cite{li2017poster},\cite{ganesan2017exploiting} ,\cite{wasicek2017context},\cite{ben2019detection},\cite{casillo2019embedded},\cite{277216}  \\ \hline
\end{tabular}
\
\caption{The classification of existing IDSs.}
\label{tab-cla-ids}
\end{table*}

\paragraph{ID Sequence-based {VIDS}}
The sequence of messages transmitted on the {\can} bus can also be used for intrusion detection. 
The traffic on the {\can} bus is constant and the messages are sent periodically for each ID.
Hence, the sequence of message IDs observed in the CAN Bus is duplicated~\cite{marchetti2017anomaly}. 
Researchers can use this feature to detect attacks.

Marchetti et al. \cite{marchetti2017anomaly} present an effective method based on the analysis of the sequence of normal {\can} bus traffic. 
This method is implemented by limited memory and low computational complexity and can be applied to current vehicles.

Islam et at.~\cite{islam2020graph} consider the {\vids} in~\cite{marchetti2017anomaly} vulnerable to intelligent attacks and they propose a four-stage intrusion detection system that uses the chi-squared method~\cite{ugoni1995chi} and incorporates graph theory~\cite{bollobas2013modern}.
The proposed methodology exhibits up to 13.73\% better accuracy compared to existing ID sequence-based methods~\cite{marchetti2017anomaly}.

\paragraph{Similarity-based {VIDS}}
Due to the stability of the in-vehicle messages, the distribution of the IDs should be similar across different windows. Some researchers want to use this similarity to detect the malicious messages.

In \cite{ohira2020normal}, Ohira et al. propose a method based on the similarity of sliding windows that can detect every type of DoS attack by using the messages distribution of sliding windows. 
The method uses the Simpson coefficient~\cite{flynn1986faunal} to calculate the similarity of message distribution in the train set and test set.
The method can detect the DoS attack in 100\% of the cases in their experiment, and the detection time is up to 93\% (14 us) shorter than the conventional method.
However, this method still can not solve the masquerade attack, which has little impact on the {\ivn}.

\revisedtext{
In \cite{nguyen2023transformer}, Nguyen et al. propose a novel multi-class IDS using a transformer-based attention network (TAN) for an in-vehicle CAN bus. Their model builds on the self-attention mechanism, removing RNNs and classifying attacks into multiple categories. Furthermore, the proposed models can detect replay attacks by aggregating sequential CAN IDs. }

\paragraph{Hybrid}
As we have shown before, different features in the {\can} network can be used to detect attacks in the vehicle accurately. 
However, when these individual features are used to detect intrusions, they often bring some loopholes. 
For example, frequency-based {\vids} usually cannot detect attacks in which the adversary imitates the victim {\ecu}'s frequency, and it can also produce false positives for event messages and messages with large periodic fluctuations. Furthermore, {\vids} based on the payload cannot achieve a high detection rate when the attacker changes little to the data flow of {\can} network .
Therefore, many researchers try to propose some {\vids}s that contain multiple features at the same time.

Theissler et al. \cite{theissler2014anomaly} propose a {\vids} that uses enhanced one-class Support Vector Machines (SVM) to detect intrusions~\cite{amer2013enhancing}. 
This method directly uses normal multivariate time series from {\ivn}s to learn the normal behavior of vehicles and detect intrusions based on deviations.

Tian et al. \cite{tian2017intrusion} introduce an {\vids} that utilizes a regression Decision Tree with Gradient Boosting (GBDT) technique~\cite{xu2005decision,friedman2001greedy} for {\can} bus.
Additionally, they propose a new feature based on entropy as the feature construction of the GBDT algorithm in which they consider the entropy of {\can} ID and the payload of data.
The experiment results show that the method achieves a high true positive (TP) (97.67\%) and a low false positive (FP) (1.20\%), which means the system has a good performance and can be used to protect the {\can} bus.

Wang et al. \cite{wang2018distributed} propose a distributed {\vids} using hierarchical temporal memory (HTM)~\cite{george2005hierarchical,padilla2013performance}, a machine learning algorithm aimed to capture the structure and algorithmic features of the new cerebral cortex. 
The method uses a standard HTM system and standard parameters to predict {\can} data flow based on the bit sequences from a single ID data domain. 
The experiment results show that this method achieves good performance in AUC score, precision, and recall.

Tomlinson et al. \cite{tomlinson2018using} use a one-class compound classifier that combined euclidean distance and nearest neighbor algorithms~\cite{batchelor1978classification,batchelor2012machine} to detect the {\ivn} attacks. 
They only target a single type of attack test-\textit{ fuzzing test}, where the {\can} messages are filled with random messages. 
However, the experiment results are relatively poor, and the best detection rate is the only 65\%.

Weber et al.~\cite{weber2018embedded} introduce a hybrid anomaly detection system, which combines the advantages of an efficient rule-based system with the advanced detection measures provided by machine learning.
Firstly, they perform a static check based on the format and transmission standard (e.g., transmission frequency and the payload range)~\cite{muter2010structured}.
Secondly, they use an unsupervised anomaly detection algorithm, called Lightweight On-Line Detector of Anomalies (LODA)~\cite{pevny2016loda}, to cooperate with the static check.

Koyama et al. \cite{koyama2019anomaly} present a lightweight {\vids} based on the quantized intervals for periodic {\can} ID and the absolute difference of payloads. 
The results of their experiments show that the system achieves a high detection performance: a true positive rate of 97.55\% and a false positive rate of 0.003\%. However, the attack in which a small number of malicious messages are injected can not be detected.

Zhu et al. \cite{zhu2019mobile} propose a multi-task LSTM {\vids} which utilizes mobile edge computing (MEC)~\cite{chen2018efficient} to assist in the identification of intrusions in the {\ivn}. 
In this system, both the dimension of time and the dimension of data are combined to enhance detection accuracy.
With the assistance of mobile edge computing (MEC), the detection can be finished with 0.61 milliseconds and achieve 90\% of accuracy.
However, the algorithm is still complicated for onboard {\ecu}s, and it is difficult to apply to existing vehicles directly.

Hanselman et al. \cite{hanselmann2020canet} present {\can}et, a novel {\vids} based on a neural network architecture that is trained in an unsupervised manner. The method builds the first deep learning model in the literature that can naturally deal with the data structure of the high dimensional {\can} bus. 
The basic idea is to introduce an independent LSTM input model for each ID that can capture the corresponding signals' temporal dynamics.
Due to the comprehensive features, the true negative rate of {\can}et is fairly high, usually over 0.99.

\rnewadd{
\smallskip \noindent
\textbf{Brief Discussion:} Parameter-based monitoring {\vids} is the most commonly used {\vids}. The biggest advantage of these {\vids}s is that they are easy to implement. Just by listening to the normal data inside the car, these {\vids} do not need some additional equipment.
However, these methods are typically targeted at specific attacks and may be less effective at detecting unconsidered security risks.
Furthermore, due to the complexity of the {\ivn} and the external environment, the monitored parameters will change, affecting the accuracy of detection.}

\subsubsection{Message Semantics-based {VIDS}}
In addition to the fingerprint-based {\vids} and parameters monitoring-based {\vids}, the researchers also propose {\can} message semantics-based {\vids}.
In these {\vids}s, the researchers need to reverse the meaning of the {\can} messages.
Researchers mainly detect whether a vehicle is attacked based on abnormal changes in the vehicle states that are reversed from the {\can} messages, and we call the methods `vehicle state-based {\vids}'.

\paragraph{Vehicle State-based {VIDS}}
Normally, Some {\can} messages on the {\ivn} always contain different vehicle states, such as vehicle speed, acceleration, RPM, the pedal position, and brake pedal position. When the vehicle runs normally, these states have a high correlation.
For example, the rapid growth of vehicle speed means that the acceleration and the pedal position are greater than 0. If the relationship between these states changes, it means the vehicle is attacked.

Wasicek et al. \cite{wasicek2017context} propose a context-aware {\vids} (CAID) framework, which can recognize the control of the physical system through the {\ivn}.
CAID uses sensor information, which can be captured from the on-board diagnostics (OBD-II) interface and parsed according to the OBD protocol~\cite{iso15765-4} to build models of the physical system by using an unsupervised Artificial Neural Network (ANN)~\cite{hassoun1995fundamentals}.
Afterwards, CAID checks the correctness of current sensor data against the reference models.
Thereby, it ensures the safety of the controller’s operations.

Narayanan et al. \cite{narayanan2016obd-securealert} introduce OBD SecureAlert, a system that detects abnormal behavior in vehicles when they are being operated. 
They successfully extract data from various real automobiles by connecting to their OBD-II port. 
With the collected dataset, they generated a Hidden Markov Model~\cite{bar2004analyzing} to predict anomalous states in vehicles.
These techniques can be applied to identify anomalies and unsafe states in vehicles.

Casillo et al. \cite{casillo2019embedded} show an embedded {\vids} for vhicle, which adopts a Bayesian Network~\cite{friedman1997bayesian} approach for the quick identification of malicious messages. 
It uses sensor data collected from the {\ivn}s to detect commands sent by an attacker. Their experiments were carried out using an automotive simulator, CARLA, which can emulate a real vehicle and its interaction with the environment, along with some other matching equipment. They tested the effectiveness of the system against malicious commands on this device.

\rnewadd{
\smallskip \noindent
\textbf{Brief Discussion:} Message semantics-based {\vids} is the most promising detection method. These methods typically reverse diagnostic messages or in-vehicle messages to obtain vehicle states and then detect intrusions through anomalies in vehicle states change.
This method can detect harmful messages to the vehicle and is unaffected by the {\ecu}'s errors and external environment.
However, the greatest challenge for the method is how to accurately reverse in-vehicle or diagnostic messages. 
While some of the protocols are publicly available (OBD protocols~\cite{iso15765-4}), most of them are designed by the OEMs themselves. There is no research work available that provides a way to completely reverse these protocols. 
Therefore, obtaining accurate vehicle states is the focus of message semantics-based {\vids}.
}

\section{EVALUATION OF EXISTING {VIDS}}
\label{sec:evaluation}

\begin{table*}[]
\caption{Evaluation}
\centering
\begin{tabular}{m{1cm}<{\centering}m{2cm}<{\centering}m{4.2cm}<{\centering}cm{2.2cm}<{\centering}cm{2.4cm}<{\centering}}
\toprule
VIDS & Features   & Detection technology  & Attacks covered  & Platform     & Results       \\
\toprule
\rowcolor{LightGray}
\cite{larson2008approach}      & Priori knowledge of {\can} protocols           & Rule-based(specification)                              & DoS,Spoofing,Fuzzing  & Simulation                    & Unobtainable                      \\ 
\cite{muter2011entropy}      & Entropy of ID                     & Rule-based(threshold)                                  & DoS,Spoofing        & Real car                      & Unobtainable                      \\
\rowcolor{LightGray}
\cite{ling2012algorithm}    & Frequency                          & Rule-based(threshold)                                  & DoS,Spoofing  & Simulation                     & Unobtainable                   \\
\cite{theissler2014anomaly}     & Multivariate time series messages & Machine-learning(SVDD)                                 & Spoofing            & Real car                      & Precision=32.3-100\%              \\
\rowcolor{LightGray}
\cite{taylor2015frequency}  & Frequency    	&   Machine-learning(OCSVM)  & 	DoS,Spoofing  & 	Simulation  & 	AUC $\geq$ 96.20\%    \\
\cite{cho2016fingerprinting}       & Clock skew                        & Rule-based(threshold)                                  & Spoofing,DoS & Real car,Prototype, Simulation & FP=0.055\%,TP=100\%               \\
\rowcolor{LightGray}
\cite{taylor2016anomaly}	&Payload	&Machine-learning (RNN and LSTM)	&Spoofing	&Simulation	    &AUC=17.65\%-100\%          \\
\cite{kang2016intrusion}       & ID,Payload    & Machine-learning(DNN)   & Spoofing            & Simulation                    & FP=1.6\%,Accuracy=97.8\%          \\
\rowcolor{LightGray}
\cite{narayanan2016obd-securealert} & State(OBD protocol)               & Machine-learning(HMM)                                  & Spoofing            & Real car                      & Unobtainable                      \\
\cite{marchetti2016evaluation}      & Entropy of ID                     & Rule-based(threshold)                                  & Spoofing,Fuzzing      & real car                      & TN=93.33\%/88.89\%                \\
\rowcolor{LightGray}
\cite{song2016intrusion}     & Interval                          & Rule-based(threshold)                                  & DoS,Spoofing,Fuzzing  & real car                      & Accuracy=100\%                    \\
\cite{gmiden2016intrusion}      & Frequency           & Rule-based(threshold)                                  & DoS,Spoofing   & Simulation           & Unobtainable              \\
\rowcolor{LightGray}
\cite{lee2017otids}      & Interval(remote frame)            & Rule-based(threshold)                                  & DoS,Spoofing,Fuzzing  & real car,Prototype            & Unobtainable                      \\
\cite{marchetti2017anomaly}      & ID sequence                       & Rule-based(outline)                                    & Spoofing,Fuzzing      & Real car,                     & Detection Rate:100\%              \\
\rowcolor{LightGray}
\cite{wasicek2017context}       & State (OBD protocol) & Machine-learning (Bottleneck ANN)    & Spoofing            & Real car                      & Unobtainable                      \\
\cite{moore2017modeling}      & Interval                   & Rule-based(threshold)                                  & Spoofing,DoS        & Real car                      & FP=0.294\%,TP=99.98\%              \\
\rowcolor{LightGray}
\cite{martinelli2017car}      & Payload                           & Machine-learning (Fuzzy logic techniques)             & DoS,Spoofing,Fuzzing  & Simulation                    & FP=0-3.8\%, Precision=96.3\%-100\%      \\
\cite{avatefipour2017physical}  & Voltage profiles (high, low)  &Machine-learning (ANN) 	&Spoofing &Prototype  &Detection Rate:95.2\%/98.3\%
\\
\rowcolor{LightGray}
\cite{markovitz2017field}       & ID,Payload                        & Rule-based(specification)                              & Fuzzing            & Real car,Simulation           & FP=0                            \\
\cite{stabili2017detecting}      & ID,Payload                        & Rule-based(threshold)                                  & Spoofing,Fuzzing      & Real car,Simulation           & Detected Anomalies:100\%          \\
\rowcolor{LightGray}
\cite{cho2017viden}      & Voltage profiles (high, low)       & Rule-based(threshold)                                  & Spoofing                & Real car,Prototype            & FP=0.2\%, Identification=99.8\%    \\
\cite{li2017poster}      & State (Sensor)                     & Machine-learning (Random Forest)      & Spoofing     & Real car,Simulation           & Unobtainable                      \\
\rowcolor{LightGray}
\cite{ganesan2017exploiting}      & State (sensor)                     & Rule-based(threshold)                                  & Spoofing            & Simulation                    & Unobtainable                      \\
\cite{tian2017intrusion} &	entropy of ID and Payload &		Machine-learning (GBDT)&	Spoofing &	Simulation & TP:97.67\%, FP:1.20\%   \\
\rowcolor{LightGray}
\cite{kneib2018scission}      & Voltage profiles (differential)    & Machine-learning (Logistic Regression)                  & Spoofing                & Real car,Prototype            & FP=0, Identification=99.85\%     \\
\cite{tomlinson2018detection} &	Frequency &	Rule-based (threshold) &	DoS,Spoofing &	Simulation &	Accuracy:99.19\%-100\%
 \\
 \rowcolor{LightGray}
\cite{wu2018sliding} &	Entropy of ID &	Rule-based (threshold) &	DoS,Spoofing &	Simulation &	Accuracy:92.3\%/100\%
\\
\cite{choi2018identifying}      & Voltage profiles (differential)    & Machine-learning (SVM,NN,BDT)                          & Spoofing                & Prototype                     & FP=3.52 \%,Identification=96.48\% \\
\rowcolor{LightGray}
\cite{studnia2018language}      & ID,Payload                        & Rule-based(specification)                              & Spoofing            & Simulation                    & Unobtainable                      \\
\cite{seo2018gids}      & Payload                               & Machine-learning(GAN)                                  & DoS,Spoofing,Fuzzing  & real car                      & Accuracy:100\%/98\%               \\
\rowcolor{LightGray}
\cite{sagong2018cloaking} &	Clock skew&	Rule-based(threshold)&	Spoofing,DoS&	Real car,Prototype&	Prediction error\textless 5.7\%   \\
\cite{weber2018embedded} & ID sequence, Payload,Interval &	Rule-based and Machine-learning(LODA) & Spoofing & Simulation & Unobtainable
\\
\rowcolor{LightGray}
\cite{wang2018distributed} & ID,Payload, transmission time &	Machine-learning (HTM) &	Spoofing &	Simulation &	Precision\textgreater 90\% 
\\ 
\cite{tomlinson2018using}	&Frequency and Payload	&Machine-learning (Euclidean distance and nearest neighbor)	&Fuzzing	&Simulation	&Detection Rate: 65\%,/52\%/45\%
\\  
\rowcolor{LightGray}
\cite{olufowobi2019anomaly}       & Frequency                         & Rule-based(threshold)                                  & DoS,Spoofing,Fuzzing  & Simulation                    & Unobtainable                      \\
\cite{young2019automotive}       & Frequency     & Rule-based(threshold)                                  & Spoofing            & Real car                      & FP=1.4\%,Accuracy=100\%           \\
\rowcolor{LightGray}
\cite{pawelec2019towards}      & ID,Payload                        & Machine-learning (RNN and LSTM)                        & Spoofing            & Real car                      & Unobtainable                      \\
\cite{koyama2019anomaly}      & Interval,Payload                  & Rule-based(State transition)                           & Spoofing            & real car                      & FPR=0.003\% TPR:97.57\%           \\
\rowcolor{LightGray}
\cite{nowdehi2019casad}      & Payload                           & Rule-based(threshold)     & Spoofing,DoS & Real car,Prototype, Simulation & Unobtainable                      \\
\cite{zhu2019mobile} &	Interval,Payload &	Machine-learning(LSTM) &	DoS,Spoofing &	Simulation &	Accuracy:90\%
\\
\rowcolor{LightGray}
\cite{ben2019detection}      & State (reverse)                    & Machine-learning (HMM) & Spoofing            & Real car                      & Unobtainable                      \\
\cite{foruhandeh2019simple} & Voltage profiles (differential) &	Rule-based  (Mahalanobis distance) &	Spoofing &	Real car,Prototype &	EER:0/0.8985\%
\\
\rowcolor{LightGray}
\cite{olufowobi2019saiducant} &	Frequency &	Rule-based(specification) &	Spoofing &	Simulation &	Accuracy\textgreater 90\%
 \\ 
\cite{casillo2019embedded}	&State(reverse)	&Machine-learning (Bayesian)	&Spoofing	&Simulation	    &Precision:85\%
\\    
\rowcolor{LightGray}
\cite{zhou2019btmonitor}	&Bit time	&Machine-learning (MLR)	&Spoofing	&Real car,Prototype	&Detection Rate: 99.76\%
\\
\cite{xiao2019robust} & ID,Payload, Timestamp &Machine-learning (ConvLSTM) & DoS,Spoofing,Fuzzing  & Simulation &  F1-score:96\%
\\
\rowcolor{LightGray}
\cite{kneib2020easi} & Voltage profiles (differential) &	Machine-learning (LR,Naive Bayes,SVM) &	Spoofing, &	Real car,Prototype &	Identification rate:99.94\%   \\
\cite{murvay2020tidal} &	Differential Timing & Rule-based(threshold) &	Spoofing,&	Prototype &	Identification rate:100\%
 \\
\rowcolor{LightGray}
\cite{hanselmann2020canet} & ID,Payload, Frequency & Machine-learning (LSTM)  &	DoS,Spoofing, &	Simulation &	True negative $\geq$  99\%
\\ 
\cite{ohira2020normal} & Sliding Windows Similarity &	Rule-based(threshold) &	DoS,Spoofing,Fuzzing &	Simulation  &Accuracy:100\%
\\
\rowcolor{LightGray}
\cite{islam2020graph} & ID sequence & Rule-based(graph) & DoS,Spoofing, & Simulation 
& Accuracy: 94.74\%/100\%/95.24\%
\\
\cite{kukkala2020indra} & Payload &Machine-learning(GRU) & DoS,Spoofing,Fuzzing & Simulation & False positive rate:2.5\%
\\
\rowcolor{LightGray}
\cite{song2020vehicle} & ID & Machine-learning(DCNN) & DoS,Spoofing,Fuzzing & Simulation & FNR:0.05-0.35\%,ER:0.03\%
\\
\cite{277216} & State(reverse) & Rule-based(threshold) & DoS,Spoofing,Fuzzing & Real car,Simulation & Accuracy:100\%
\\
\bottomrule 
\end{tabular}
\label{tab:evaluation-of-vids}
\end{table*}

Researchers propose {\vids}s based on various detection principles and use different validation strategies to evaluate detection effectiveness of their methods.
In this section, we complement the survey by introducing a taxonomy of the {\can} {\vids} in Table~\ref{tab:evaluation-of-vids}.
Next, we compare these {\vids}s from different perspectives: the used features, the detection technology, the attack covered, validation strategy, and detection result.

\subsection{Feature}
Feature used in the {\vids} is a fundamental part of an intrusion detection system. 
Different features require different data to be collected.
We count the features used in all the papers, which are the timing interval of the consecutive messages (can be classified as frequency), clock skew, voltage profiles, the data field, the entropy, the sequence of ID, the state of the vehicle, specification. 
Researchers use these features based on different principles.
We give more details of these principles.

Researchers take advantage of the periodicity and stability of {\can} message transmissions, and they use the stable time interval between {\can} message transmissions, the information entropy of the data stream, and relatively fixed order of the {\can} IDs to detect intrusions~\cite{song2016intrusion,muter2011entropy,marchetti2017anomaly}.
The {\vids}s based on these features only need to collect the data stream from the {\ivn}s. 
These systems calculate the pattern of normal data based on these data. 
During the actual testing, the {\vids}s issue a warning if the test data violates this pattern.

Researchers also use the continuity of data fields in {\can} messages to detect malicious messages. 
The data in a {\can} message usually contains some practical meaning, such as sensors and counters~\cite{markovitz2017field}.
Therefore, they use the difference between neighbouring data or the prediction of the next data to determine whether the message is malicious~\cite{taylor2016anomaly,stabili2017detecting}.
The method also only requires the collection of data streams from the in-vehicle network.

Furthermore, researchers can utilise the semantic information in {\can} messages to detect attacks on vehicles.
They reverse the {\can} messages to obtain the vehicle status and determine whether the vehicle is under attack based on the change of the vehicle status~\cite{wasicek2017context,casillo2019embedded}.
While using semantic information to detect intrusions, the researcher not only needs to get messages from {\ivn} but also needs to know how to reverse these messages.

Finally, researchers propose many {\vids}s based on the fingerprint of {\ecu}s~\cite{cho2016fingerprinting,cho2017viden,choi2018identifying}.
They use the unique features of each {\ecu}, such as clock skew and voltage, to determine whether a message source from the correct {\ecu}.
When using clock skew as the feature, researchers only need to collect data streams from the {\ivn}s. 
However, researchers need sophisticated equipment such as oscilloscopes to capture the voltage values of {\can} messages when voltage is used to detect intrusions.
The complex data collection affects the deployment of this {\vids} in real vehicles.

\subsection{Detection Technology}
Different intrusion systems take different technologies to model and build the system.
Overall, these {\vids}s take two technologies: rule-based technology and machine-learning technology.

\noindent \textbf{Rule-Based Detection:} 
Usually, they can be distinguished into two types.
There are two popular types of rule-based detection technologies.
For the first type, the researchers draw up some specifications according to the prior knowledge of attacks or standard protocols and detect malicious messages with these specifications~\cite{larson2008approach}.
For the other type, the researchers can build the outlines or thresholds of normal behaviors through features mentioned before by the mathematical formula or statistical experiment (e.g.,~\cite{moore2017modeling}).
The thresholds or outlines can determine whether the target messages are malicious.
These {\vids}s, which use rule-based techniques, require only comparison operation during detection, and so that they consume fewer resources and delays.
However, as the threshold is fixed and the changes of the {\ivn}s are complex, this method can cause false alarms when there are large changes in the condition of the vehicle.

\noindent \textbf{Machine-Learning-Based Detection:}
The machine-learning algorithms are already widely used in {\vids}s, and they are quite suitable for solving the classification and modeling issues in {\vids}s.
Furthermore, three types of machine learning models are widely adopted by the {\vids}, including traditional machine learning~\cite{martinelli2017car}, recurrent neural network (RNN)~\cite{taylor2016anomaly}, deep neural networks (DNN)~\cite{song2016intrusion}.
Compared to rule-based algorithms, machine learning algorithms consume more time and computational resources~\cite{taylor2015frequency}. 
Therefore, machine learning-based {\vids}s require powerful computing power from the {\ecu}s.
However, the robustness of these systems is greatly enhanced because machine-learning algorithms can train large amounts of data containing a variety of scenarios~\cite{taylor2015frequency}.

\subsection{Attack Covered}
When the researchers designed an {\vids}, they must consider the targeted attack scenarios before design.
Since none of these algorithms explicitly mention sniffing attack and diagnostic attack, we only list attacks in in-vehicle communications.
The attack scenarios are presented in Section \ref{sec:attack}: \textit{DoS attack},  \textit{Spoofing attack},  \textit{Fuzzing attack}.
The various {\vids}s target different attacks according to their own detection principles.
For example, Kang et al.~\cite{kang2016intrusion} propose a {\vids} based on the continuity of data fields under the same ID.
They target attacks that change the data field for a specific ID.
However, they can not detect attacks that insert malicious messagse with undetected IDs (such as DoS attack and Fuzzing attack).

\revisedtext{
\subsection{Test Platform}

In this section, we assess the reliability of the evaluation method by comparing the utilization of test platforms in various works.
Researchers evaluate the effectiveness of their {\vids} in different ways.
We can divide them into three categories based on the platforms used in the experiment: real vehicle, prototype, and simulation.
When a vehicle is in operation, the changes of vehicle states are complex and varied. 
For the best validation, some researchers deploy their {\vids}s directly on real cars~\cite{cho2016fingerprinting,muter2011entropy,moore2017modeling}.
However, this method requires a real vehicle that can be modified.
Furthermore, the attack test can cause damage to the vehicle and even threaten the safety of the driver.
Therefore, some researchers create a prototype with some microcontrollers and {\can} modules to simulate the {\ecu}s and {\can}~\cite{choi2018identifying,murvay2020tidal}.
\revised{
This method also validates the detection effect of an {\vids} in the {\can}, but a microcontroller such as a Raspberry Pi cannot completely replace an {\ecu}.}
The reliability of this method is poorer than that of real cars.
Finally, some researchers only use a {\can} bus simulator, such as {\can}oe, or use some online datasets~\cite{kang2016intrusion,olufowobi2019anomaly,martinelli2017car}.
Researchers modify the data to simulate the generation of attacks.
This verification strategy is simple but not very reliable.
}

\subsection{Detection Result}
The results are an essential indicator for evaluating the effectiveness of an intrusion detection system. Since different papers describe their results from different perspectives (such as false positive or precision), we compared the relevant results for each paper with their own perspective during the assessment phase. Whereas some items do not perform detailed experiments or do not list the actual results, we use \textit{`Unobtainable'} to represent them.

As the detection results of these systems are based on different data sets, it is not possible to directly compare the results. 
Specifically, The length of the data set, the proportion of positive and negative samples, and the vehicle states when the data is collected can affect the effectiveness of the detection.
Furthermore, some {\vids}s only use a single evaluation index to show the effectiveness of the test, but such results are unreliable.
For example, Song et al.~\cite{song2016intrusion} only calculate the accuracy of the test. 
However, if the test data set is asymmetric, the accuracy rate can not represent the true detection effect.

\section{EXPERIMENT}
\label{sec:experiment}
In this section, we select multiple typical {\vids}s and laterally compare their detection performance by designing specific experiments and using a uniform testing dataset. 
First, we introduce the {\vids}s and the dataset that we select.
Second, we describe the challenges encountered in the process of running these {\vids}. 
Third, we present the detection results of these {\vids} with the same attack dataset.
Finally, we discuss the potential attacks that can evade these {\vids}s.

\subsection{{VIDS} Selection And Implementation}
To further explore the performance of {\vids}, we choose the typical {\vids}s from the papers listed in Table~\ref{tab:evaluation-of-vids} to imitate their algorithm and implement them following two criteria.

We select the methods that can be tested with a unified dataset. 
Parameters monitoring-based {\vids}s usually only need to use a CAN module to collect all CAN frames, and these methods can directly use the same data set for horizontal comparison.
However, ECU fingerprint-based {\vids}s require special datasets for defense.
For example, Kneib et al. \cite{kneib2018scission} need an advanced oscilloscope to collect the voltage of CAN signals to construct a fingerprint of a normal ECU.
These methods typically collect voltages at different fields in the CAN frame at different sampling rates, which are introduced in \S \ref{sec:NIDS}. 
Data source limitations prevent us from evaluating them using a common dataset.

Taking the above reasons into account, we select these papers (\cite{gmiden2016intrusion,song2016intrusion,moore2017modeling,young2019automotive,olufowobi2019anomaly,cho2016fingerprinting,tomlinson2018detection,muter2011entropy,marchetti2016evaluation,wu2018sliding,ling2012algorithm,marchetti2017anomaly,ohira2020normal,taylor2015frequency,taylor2016anomaly,kang2016intrusion,stabili2017detecting})  from Table~\ref{tab:evaluation-of-vids}.
It is particularly worth noting that some papers contain multiple methods.
In \cite{muter2011entropy}, muter et al. proposed two different methods to detect intrusion. 
One uses the change in relative entropy (\textit{\cite{muter2011entropy}(1)}), and the other uses the change in overall entropy (\textit{\cite{muter2011entropy}(2)}).
Tomlinson et al. \cite{tomlinson2018detection} introduce two new unsupervised detection methods. One uses Z-score (\textit{ \cite{tomlinson2018detection}(1)}) and the other uses ARIMA (\textit{ \cite{tomlinson2018detection}(2)}).


\begin{figure}[t]
	\centering
	\includegraphics[width=.45\textwidth]{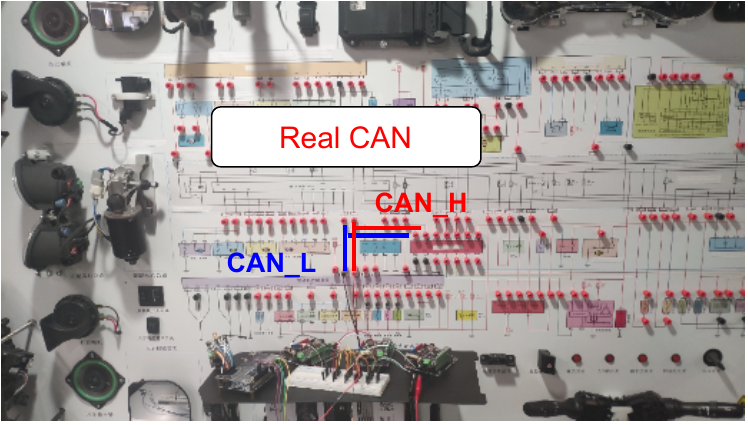}
	\caption{The topology of the real CAN.}
	\label{fig:testbed}
\end{figure}

\subsection{Dataset}
\newtext{
We need to collect the {\can} bus data in an attack-free state and under various attacks to implement these {\vids}.
We collect the data from a real CAN bus, which is shown in Fig.~\ref{fig:testbed}.
The testbed consists of electronics from a 2014 Toyota Corolla. 
Both the hardware and the network are actually used by this car.
We apply it to evaluation instead of actual vehicles due to security and safety considerations.
During evaluation, we connect the CAN analyzer to the \texttt{CAN-H} and \texttt{CAN-L} of the CAN bus.
Various attacks, such as DoS attack and Fuzzy attack, are launched on the real CAN.
Also, we obtained the CAN information from its OEM (Original Equipment Manufacturer). This real CAN contains three internal ECUs. 
Meanwhile, 23 types of frames are transmitted (i.e., 23 various CAN IDs).
We select the ID representing the speed as the target of the spoofing attack.}

Then we describe the methods that we create the datasets.
We do not test these defenses with all attacks.
First, sniffing attack does not have any impact on the data of the {\ivn}. Therefore we do not consider such attacks.
Second, some attacks target specific defense methods and are not suitable as a unified test data set.
For example, voltage corruption attack (SPA-6) is mainly aimed at voltage-based intrusion detection systems.
The adversaries try to pollute the training set of the voltage fingerprint model, so it is outside the detection range of our chosen system.
Based on this consideration, we selected suitable attacks for testing all selected methods.

\smallskip \noindent
\textbf{Replay attack (\textit{SPA-1}))}: We randomly intercept a fixed-length data segment from the normal data at first. Then, we repeatedly insert this data segment into the real CAN. Each attack lasts 10 seconds and keeps the time interval of attack data unchanged.

\smallskip \noindent
\textbf{Fabrication attack (\textit{SPA-2})}: In this attack, we choose a message which represents the speed of the vehicle. Then, we modify and insert the selected message into the CAN every 1ms for 10 seconds. We repeat this operation continuously afterward.

\smallskip \noindent
\textbf{Masquerade attack (\textit{SPA-3})}: This attack requires us to pause the specific ECU for a long time, which is a very big challenge. So we make modifications in the existing normal dataset without attack. 
We select the messages with a specific ID and change their payload. Similarly, we also modify a piece of normal data every 10 seconds.

\smallskip \noindent
\textbf{Disorderly Control attack (\textit{SPA-2})}:
In this attack, we inject messages of totally random CAN ID and payload every 0.5 milliseconds. 
Each intrusion performed for 10 seconds.

\smallskip \noindent
\textbf{Reverse attack (\textit{FUA-2})}: In this attack, we inject malicious messages that are composed of normal IDs and random payload. These messages are inserted into the CAN bus at 1-millisecond intervals and last for 10 seconds.

\smallskip \noindent
\textbf{DoS attack with high priority ID (\textit{DOA-1})}:
In this attack, we inject messages of ‘0x000’ CAN ID every 0.3 milliseconds. 
0x000 is the highest priority ID and most of the works inject messages of this ID to attack the vehicle.

\smallskip \noindent
\textbf{Redundant message injection (\textit{DOA-2})}: The purpose of this attack is to fill the {\can} bus with messages which have normal IDs. We inject malicious messages with normal IDs and random payloads into the CAN bus until the maximum load of the bus is reached.

\smallskip \noindent
\textbf{Control attack in diagnostic communication (\textit{DIA-2})}:
First, we get the threatening control commands from the diagnostic devices. 
In our experiment, we use the diagnostic equipment (i.e., Launch X431~\cite{launchx431}) to control the vehicle and reverse the messages exchanged by the equipment and vehicle.
Afterwards, we inject a control command every 20 milliseconds into CAN bus as the malicious message.

\smallskip \noindent
\textbf{Spoof attack in diagnostic communication (\textit{DIA-3})}:
First, we use diagnostic equipment~\cite{launchx431} to query the speed of the vehicle and record the response messages from the {\ecu}.
Then, we continuously inject these response messages into the CAN bus to constitute the spoof attack in diagnostic communication.

\smallskip \noindent
\textbf{Fuzzy attack in diagnostic communication (\textit{DIA-4})}:
We inject messages of totally random diagnostic {\can} ID and payloads every 20 milliseconds to the real CAN bus.

\smallskip \noindent
\textbf{Dataset with normal diagnostic messages (\textit{DIA-Normal})}:
In order to determine that these {\vids}s can distinguish between normal diagnostic messages and malicious diagnostic messages, we insert normal diagnostic messages in the data set for comparison.
We choose the speed query command in the open diagnostic protocol to ensure that the injected messages can not harm the vehicle~\cite{iso15765-4}.
We inject the selected commands every 20 milliseconds to the CAN bus.

Through these operations, we obtain 11 datasets for testing. The number of normal data and malicious data in these datasets is displayed in Table~\ref{tab:attack_dataset}. Because the amount of malicious data in these attacks is less than normal data, we use various indicators to measure the algorithm's effectiveness, such as accuracy, precision, recall, and f1-score.

\begin{table}[t]
\centering
\setlength\tabcolsep{3mm}
\caption{The composition of the attack dataset.}
\begin{tabular}{c c c}
\toprule
File Name  & Normal Messages & Malicious Messages \\
\toprule
\rowcolor{LightGray} 
SPA-1    & 549410         & 85320                \\
SPA-2    & 532510          & 159752             \\
\rowcolor{LightGray} 
SPA-3    & 456790          & 9331               \\
DOA-1        & 487650         & 93074              \\
\rowcolor{LightGray} 
DOA-2   & 523480          & 314083              \\
FUA-1      & 513680         & 75486              \\
\rowcolor{LightGray} 
FUA-2     & 497450          & 149236              \\
DIA-2 & 524640 & 6826     \\
\rowcolor{LightGray} 
DIA-3 & 532480   &6929    \\
DIA-4 & 473980 & 6166    \\
\rowcolor{LightGray} 
DIA-Normal &507506  &0        \\
\bottomrule 
\end{tabular}
\label{tab:attack_dataset}
\end{table}

\subsection{Challenges and Solutions in Reimplementation.}
After choosing the appropriate dataset, we implement all the selected {\vids}s based on the built dataset. 
When we implement these {\vids}s, we find out various challenges, and we try to solve them using the following solutions.

\subsubsection{Parameters Definition}
Some {\vids}s have uncertain or unmentioned parameters. When we try to reproduce the algorithm in the paper completely, we have to decide some parameters according to the dataset by ourselves.
For example, in \cite{marchetti2016evaluation}, the threshold of the entropy is decided by three parameters: the average entropy value $\mu_{e}$, standard deviation of entropy $\sigma_{e}$, and a model parameter $k$. 
Among them, $\mu_{e}$ and $\sigma_{e}$ are calculated with the training dataset and $k$ is a customized parameter which is used to adjust the threshold.
To achieve the best effectiveness of these {\vids}s, we adjust these uncertain parameters for each dataset separately. 

\subsubsection{Inconsitence Defeination of Attack Models}
Each {\vids} has its own attack model and attack scenario. 
For example, for {\vids} proposed in \cite{song2016intrusion}, the target DoS attack is launched by transmitting abundant traffic to surpass the maximum capacity of {\can} bus and has no requirement on the ID and data field. 
Whereas, in other papers, such as {\vids} \cite{martinelli2017car}, DoS attack is performed by injecting messages whose ID are 0x000 at high speed. 
Additionally, based on the  \ref{tab:evaluation-of-vids}, we can see that many {\vids}s only target part of the attacks scenarios contained in the selected dataset.
For example, {\vids} \cite{young2019automotive} mainly aims to the replay attack and does not mention DoS attack and fuzzy attack. 
To compare the differences between the methods, we select the same dataset. And for a more comprehensive evaluation of the methods, we have expanded the attack dataset to include the attacks mentioned above.

\subsection{Reproduction}
\newtext{
After implementing the above selected {\vids}s and also choosing the appropriate dataset, we conduct the following experiments to check the effectiveness of these methods.
Particularly, we first apply the data collected during normal vehicle running to obtaining the desired threshold or model for each algorithm, and then we evaluate the accuracy of these methods with various attack datasets.
Afterwards, we present the used evaluation metrics as well as the evaluation results.}

\begin{figure*}
	\centering
	\subfigure[]{
		\begin{minipage}[b]{0.23\textwidth}
			\includegraphics[width=1\textwidth]{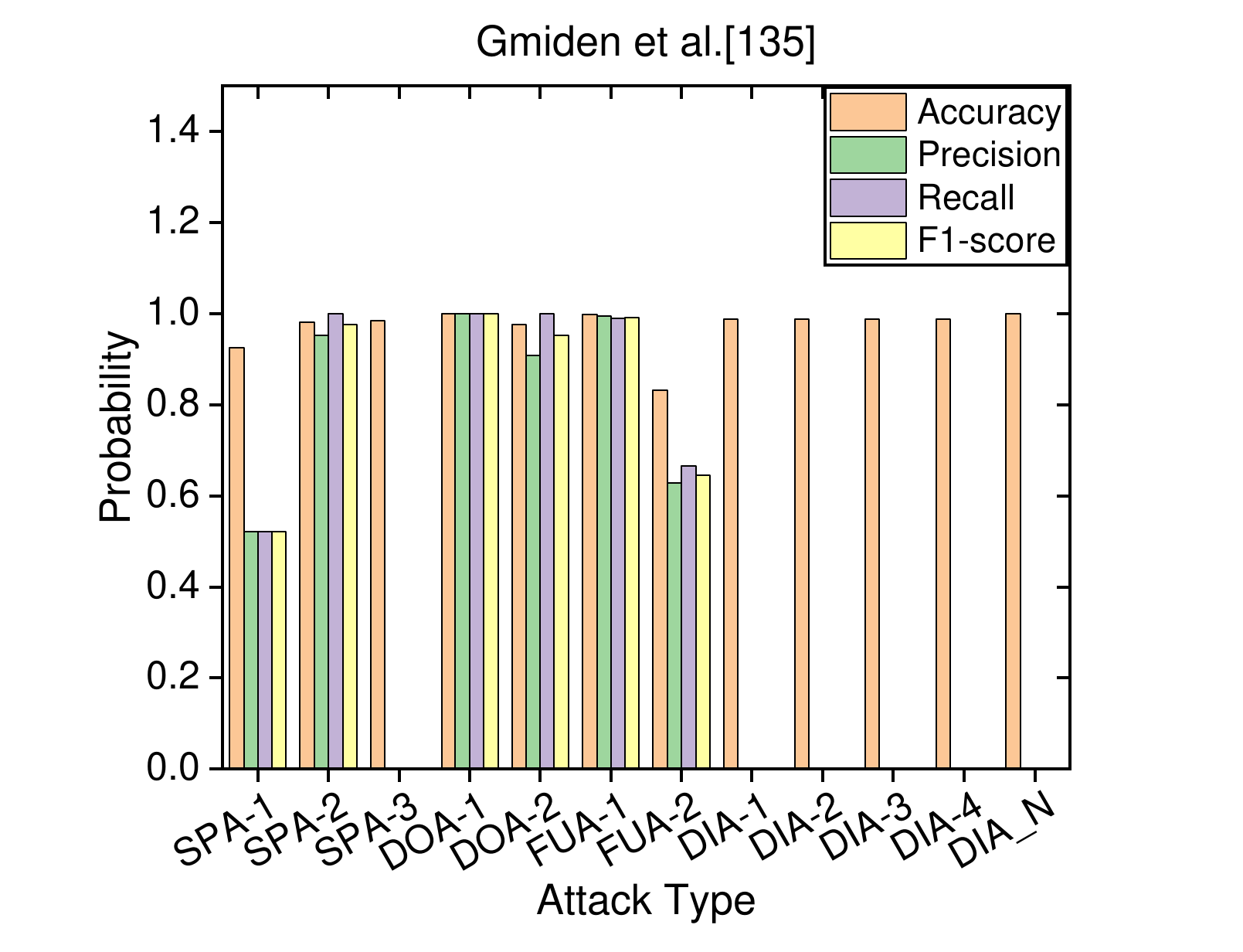} 
		\end{minipage}
		\label{fig:grid_4figs_1cap_4subcap_1}
	}
    	\subfigure[]{
    		\begin{minipage}[b]{0.23\textwidth}
   		 	\includegraphics[width=1\textwidth]{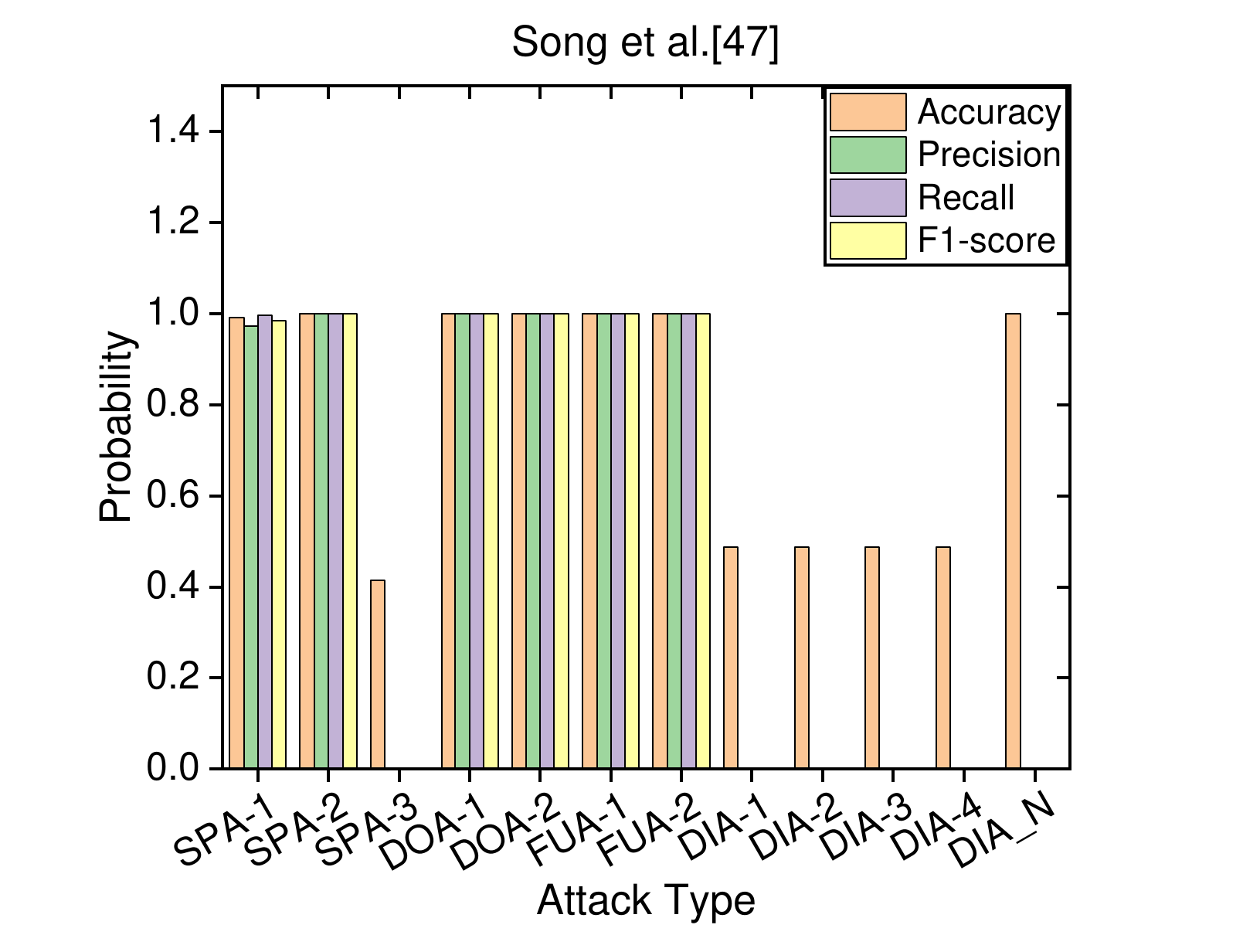}
    		\end{minipage}
		\label{fig:grid_4figs_1cap_4subcap_2}
    	}
	\subfigure[]{
		\begin{minipage}[b]{0.23\textwidth}
			\includegraphics[width=1\textwidth]{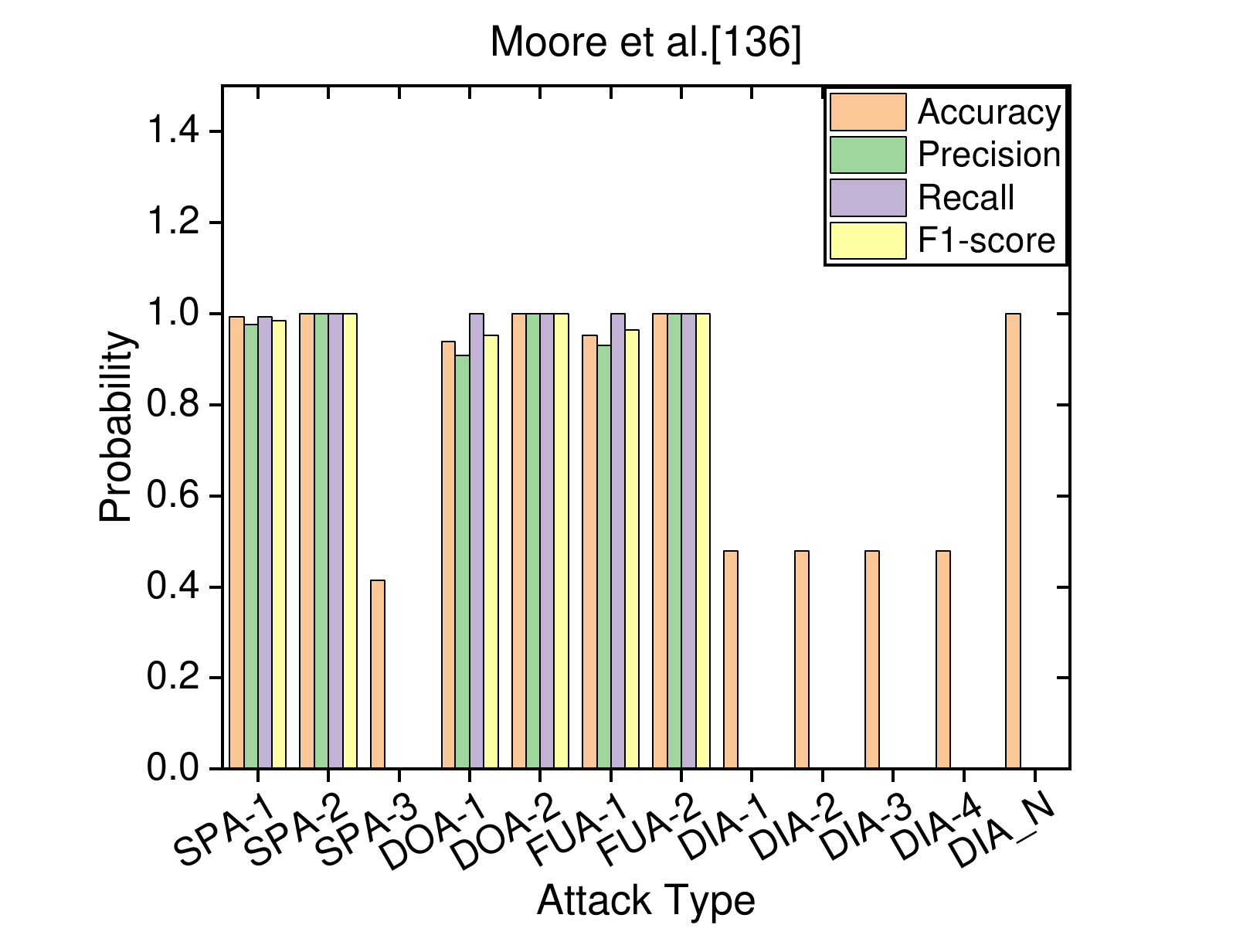} 
		\end{minipage}
		\label{fig:grid_4figs_1cap_4subcap_3}
	}
    	\subfigure[]{
    		\begin{minipage}[b]{0.23\textwidth}
		 	\includegraphics[width=1\textwidth]{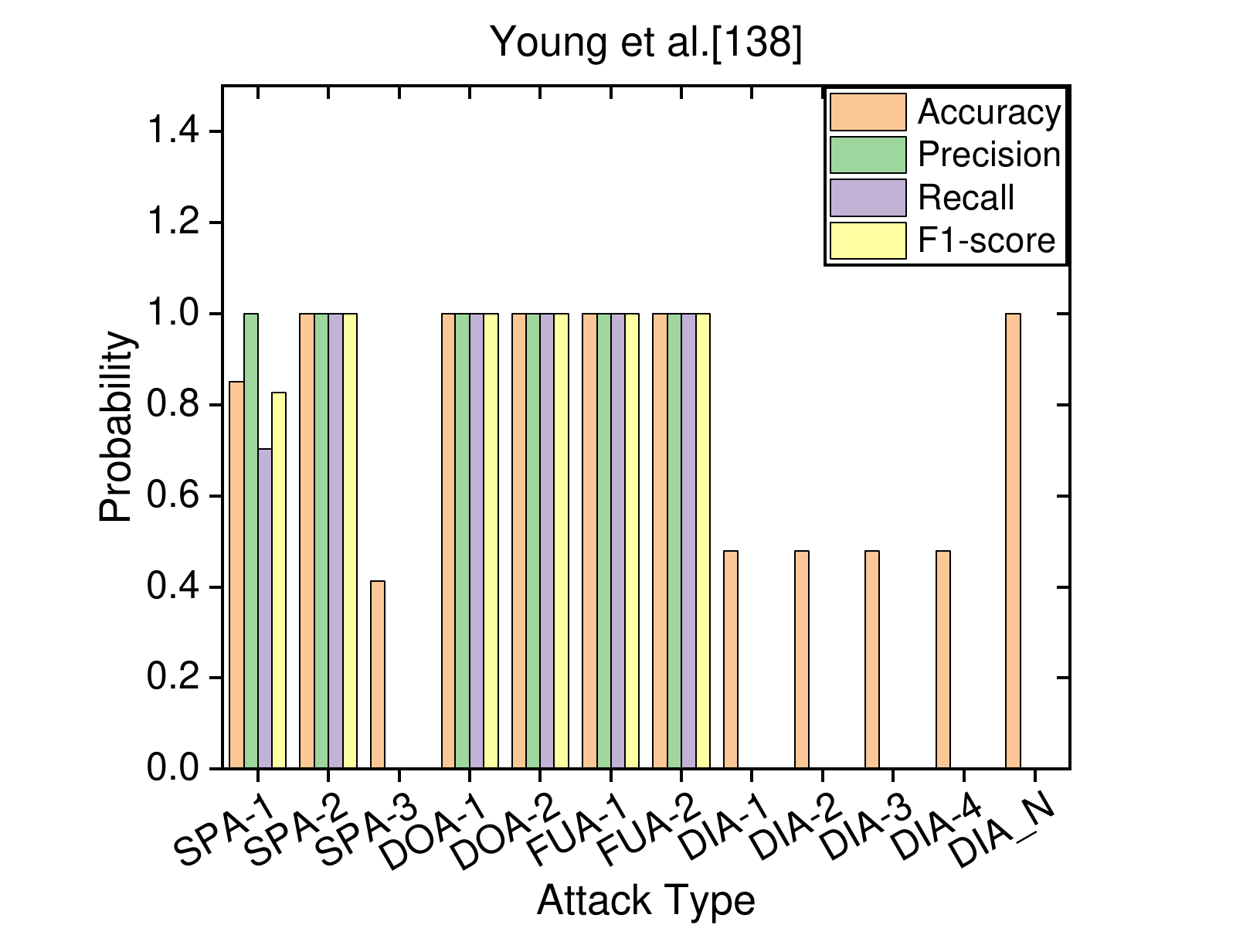}
    		\end{minipage}
		\label{fig:grid_4figs_1cap_4subcap_4}
    	}
    	\subfigure[]{
		\begin{minipage}[b]{0.23\textwidth}
			\includegraphics[width=1\textwidth]{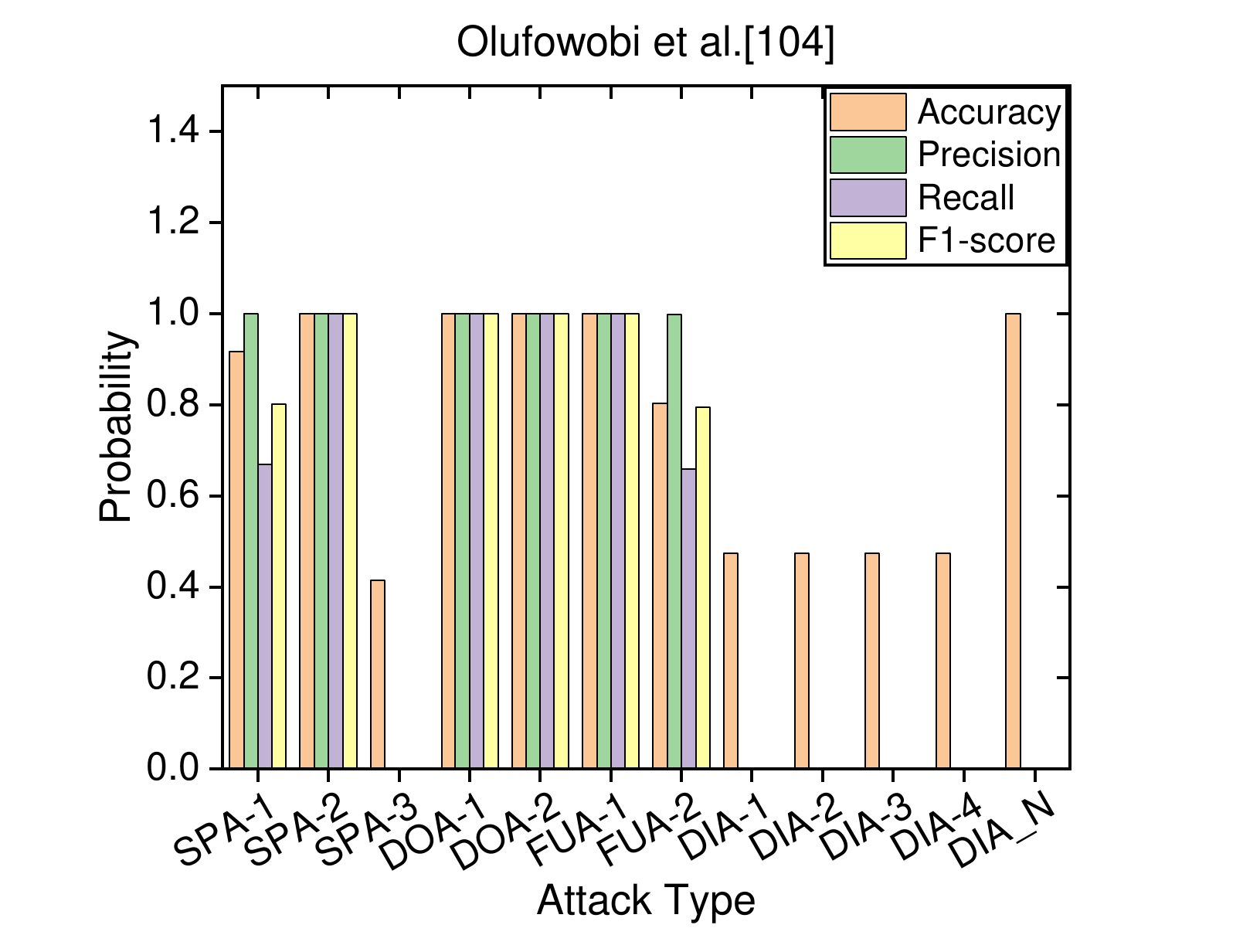} 
		\end{minipage}
		\label{fig:grid_4figs_1cap_4subcap_5}
	}
	\subfigure[]{
		\begin{minipage}[b]{0.23\textwidth}
			\includegraphics[width=1\textwidth]{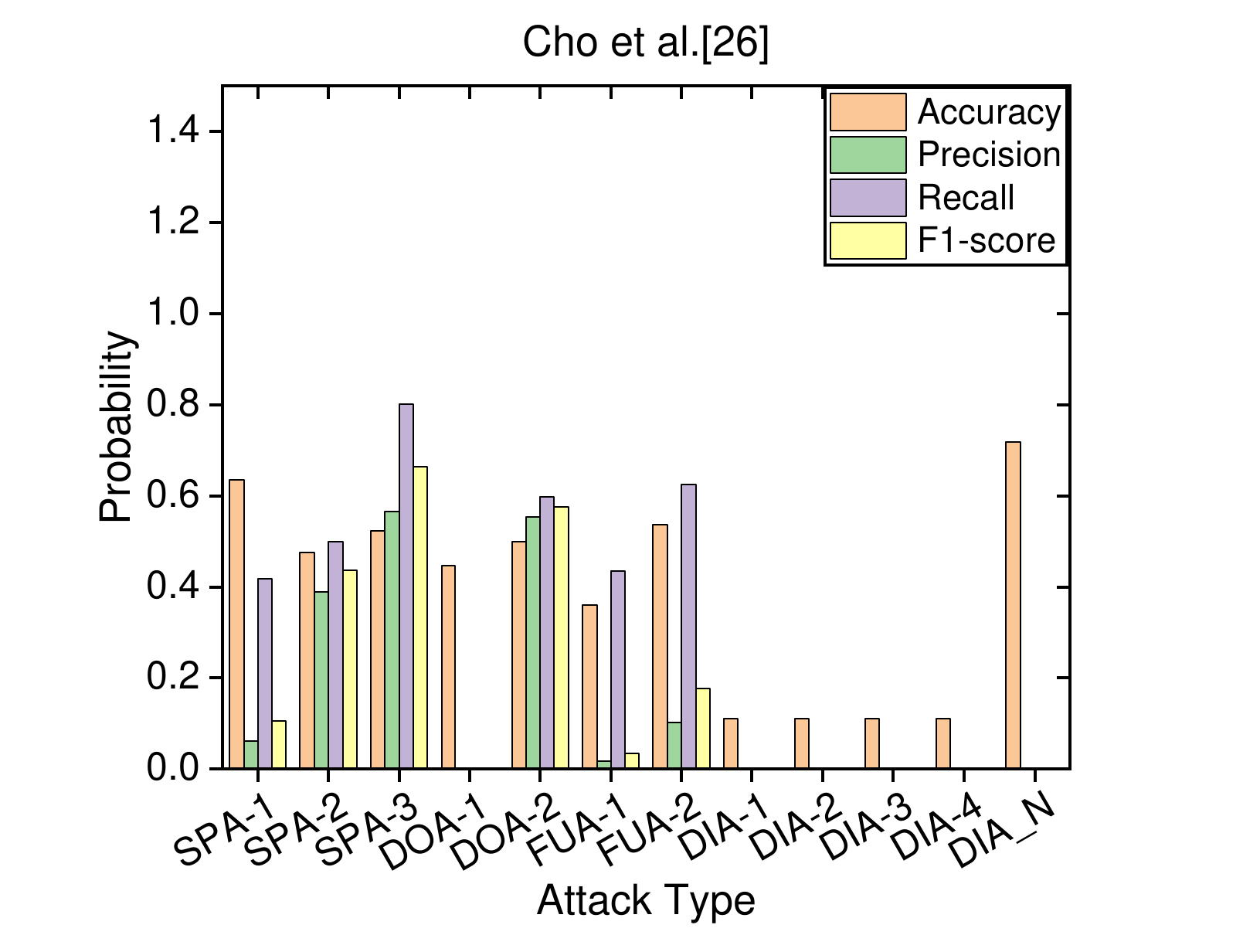} 
		\end{minipage}
		\label{fig:grid_4figs_1cap_4subcap_6}
	}
	\subfigure[]{
		\begin{minipage}[b]{0.23\textwidth}
			\includegraphics[width=1\textwidth]{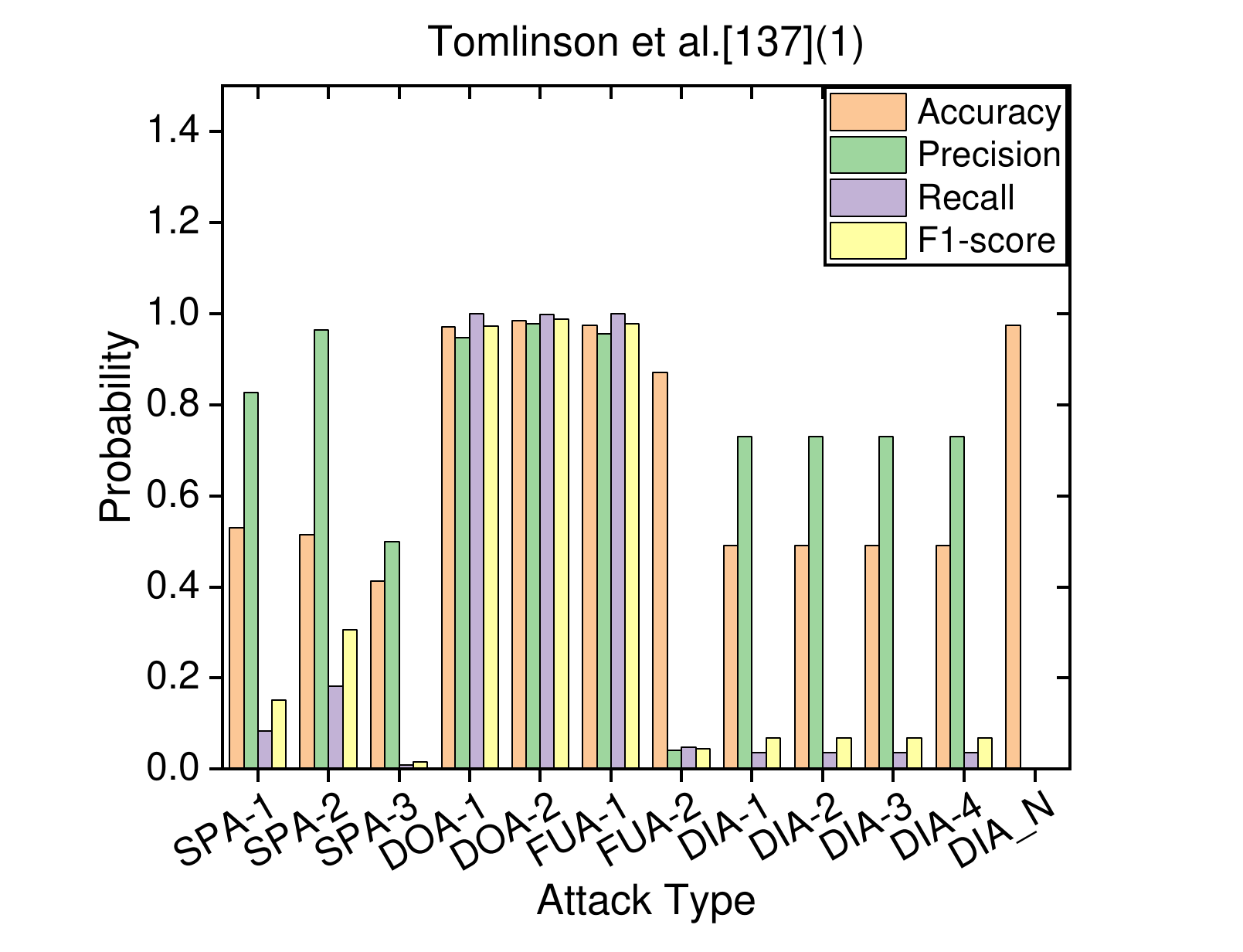} 
		\end{minipage}
		\label{fig:grid_4figs_1cap_4subcap_7}
	}
    	\subfigure[]{
    		\begin{minipage}[b]{0.23\textwidth}
   		 	\includegraphics[width=1\textwidth]{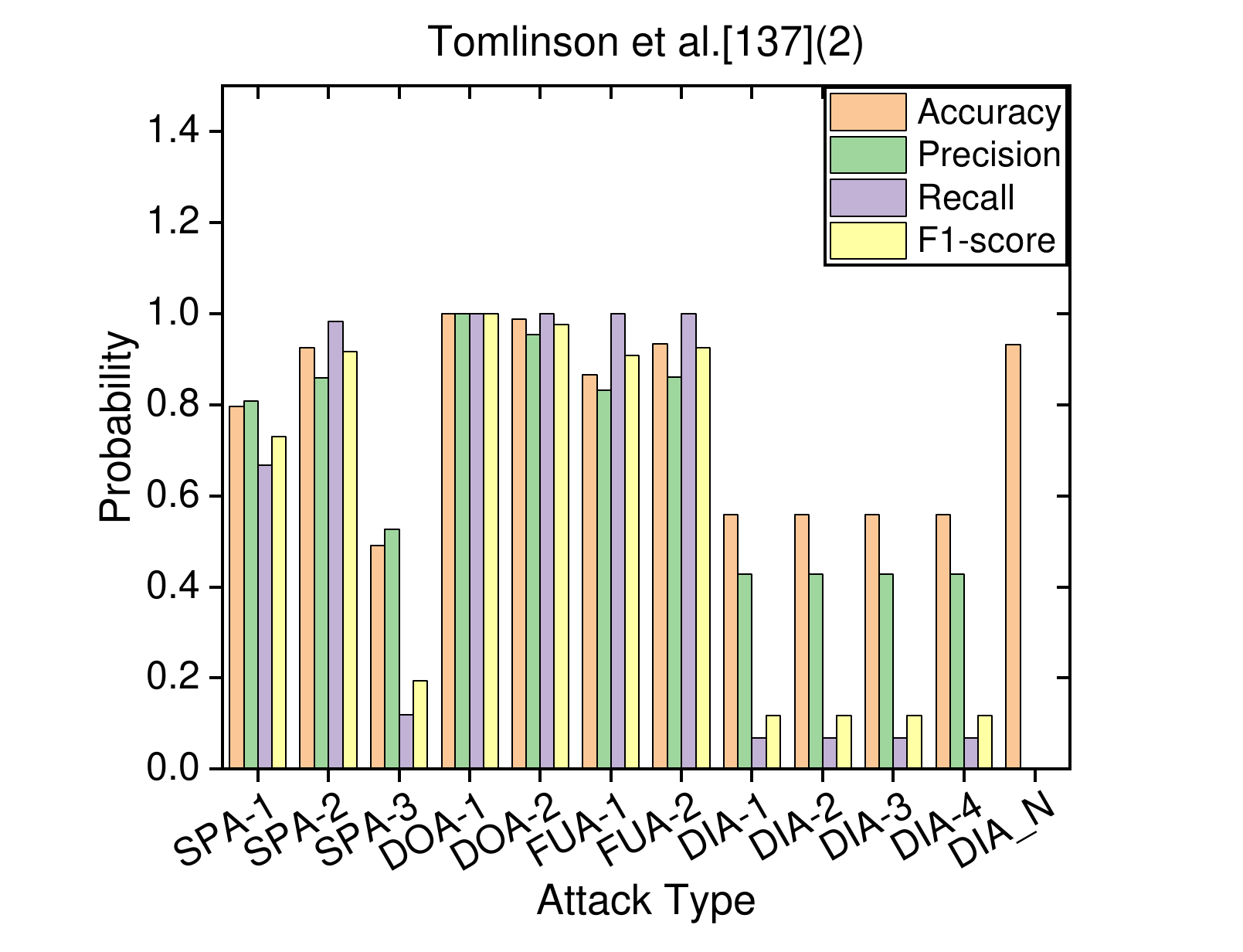}
    		\end{minipage}
		\label{fig:grid_4figs_1cap_4subcap_8}
    	}

	\subfigure[]{
		\begin{minipage}[b]{0.23\textwidth}
			\includegraphics[width=1\textwidth]{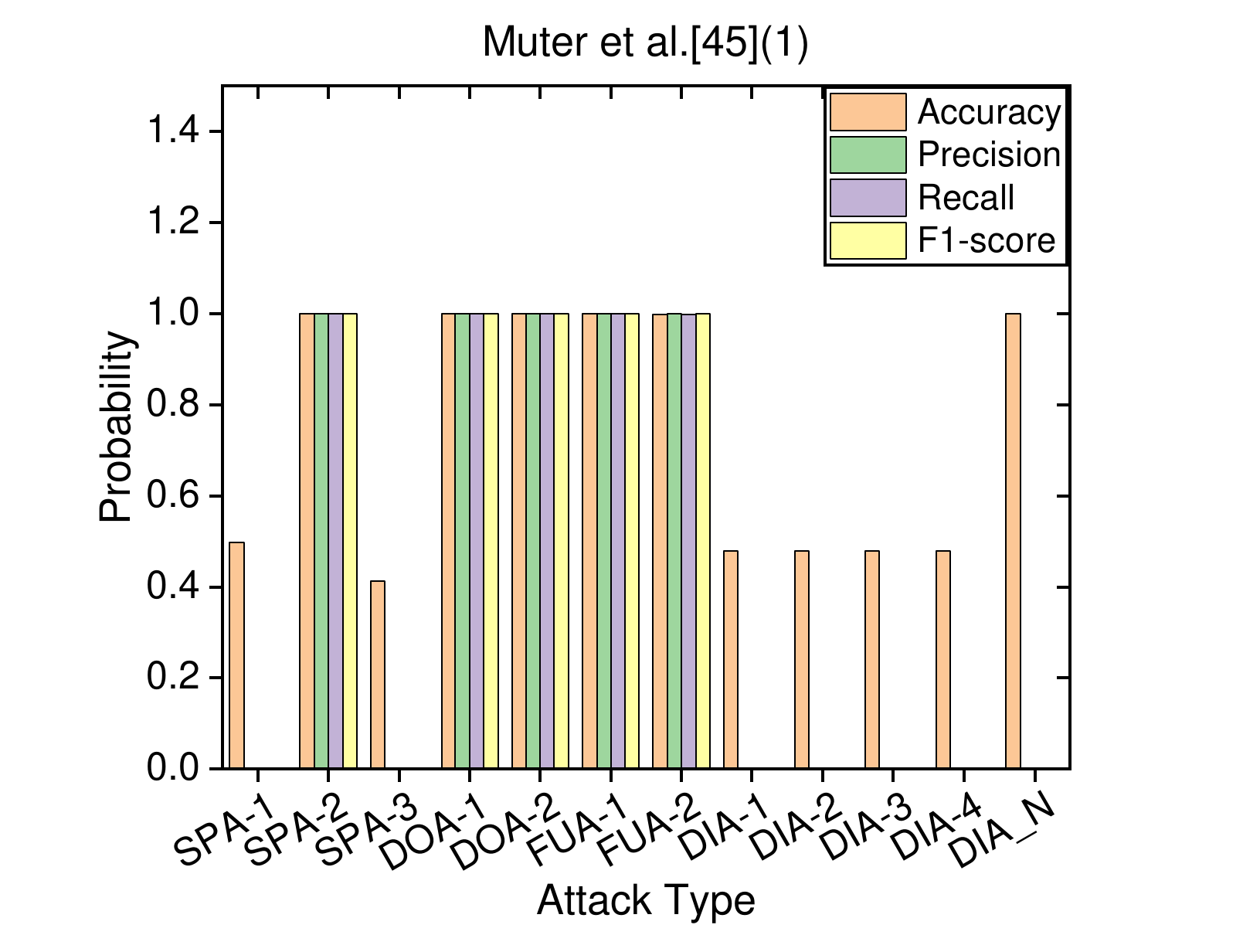} 
		\end{minipage}
		\label{fig:grid_4figs_1cap_4subcap_9}
	}
    	\subfigure[]{
    		\begin{minipage}[b]{0.23\textwidth}
		 	\includegraphics[width=1\textwidth]{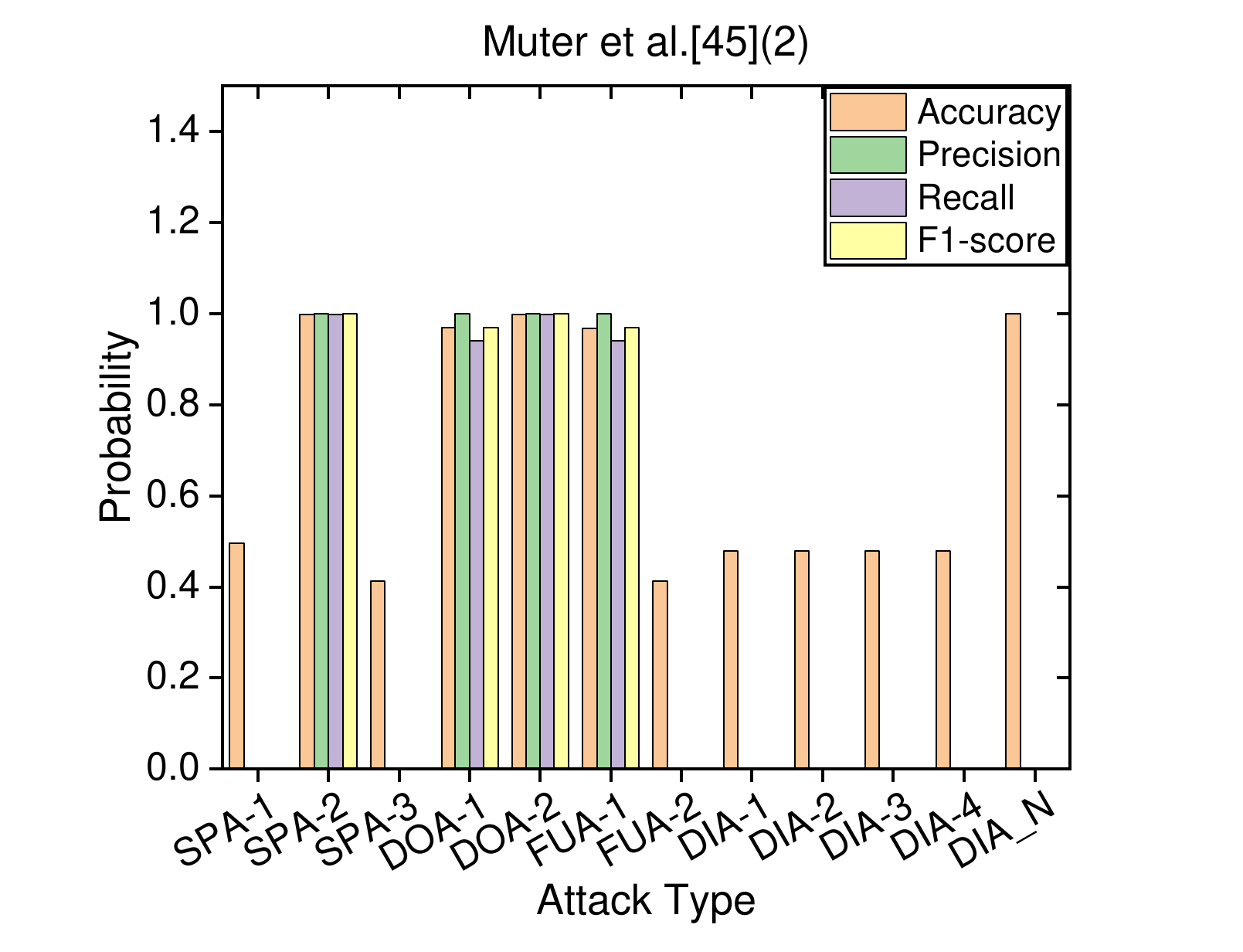}
    		\end{minipage}
		\label{fig:grid_4figs_1cap_4subcap_10}
    	}
    	\subfigure[]{
		\begin{minipage}[b]{0.23\textwidth}
			\includegraphics[width=1\textwidth]{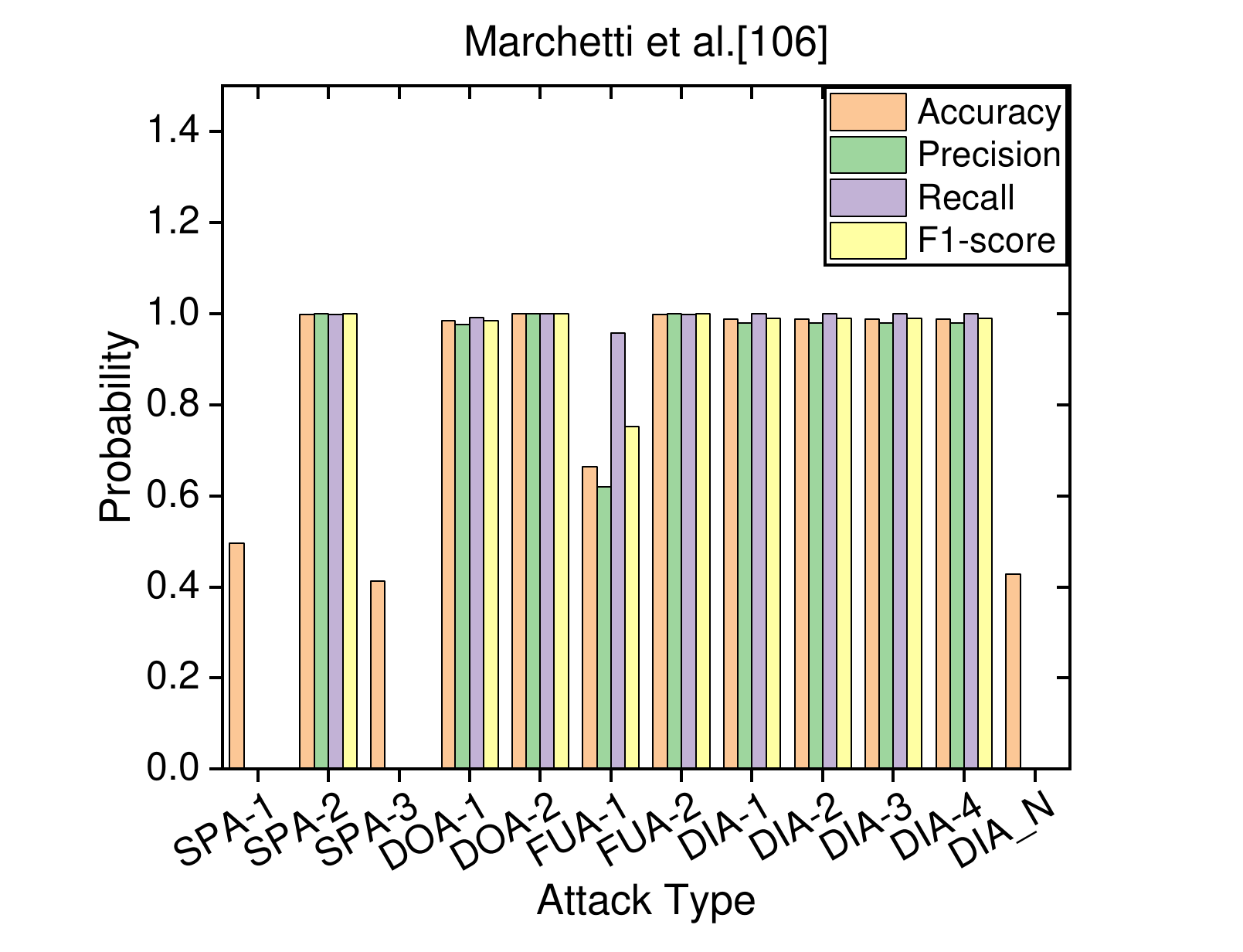} 
		\end{minipage}
		\label{fig:grid_4figs_1cap_4subcap_11}
	}
	\subfigure[]{
		\begin{minipage}[b]{0.23\textwidth}
			\includegraphics[width=1\textwidth]{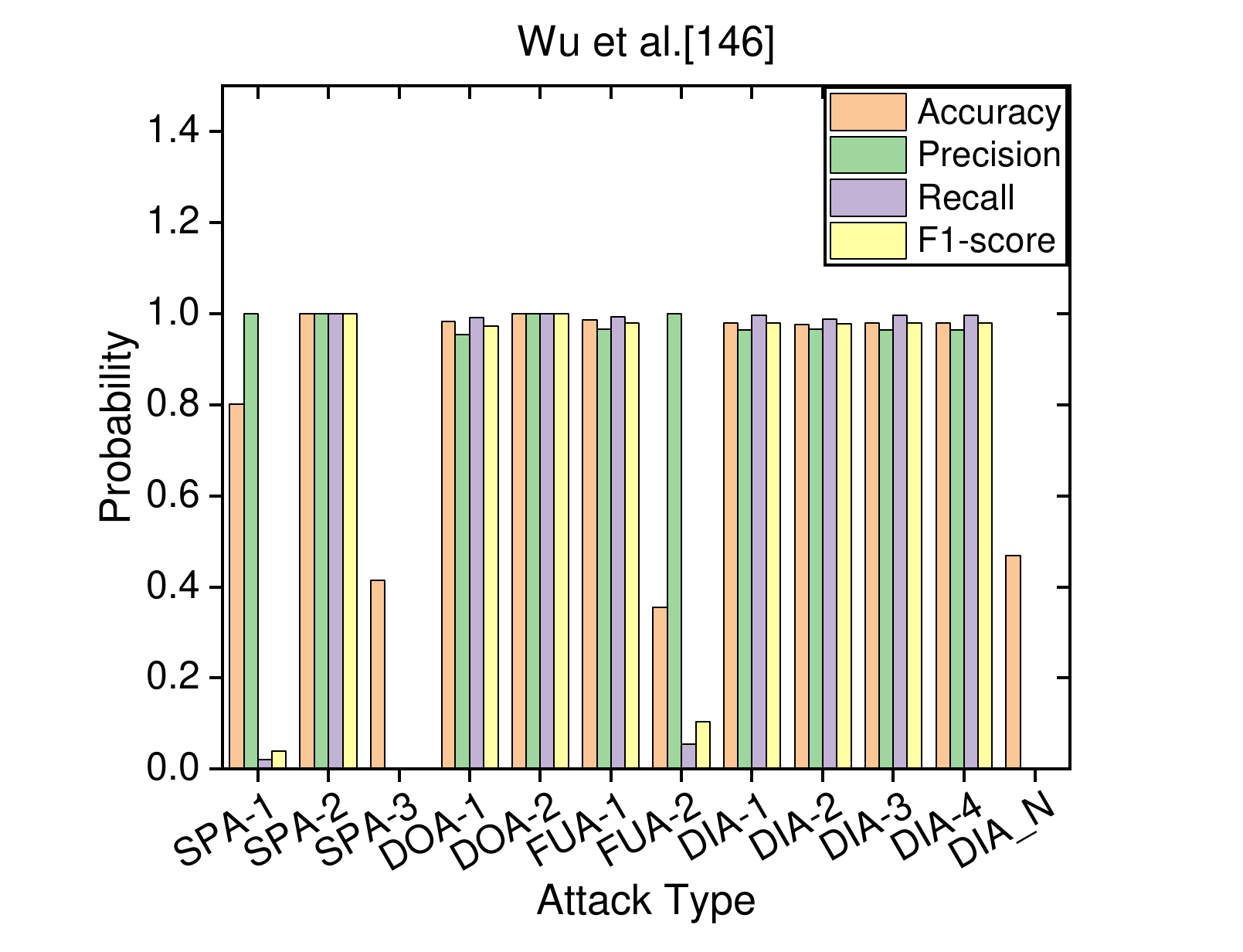} 
		\end{minipage}
		\label{fig:grid_4figs_1cap_4subcap_12}
	}
		\subfigure[]{
		\begin{minipage}[b]{0.23\textwidth}
			\includegraphics[width=1\textwidth]{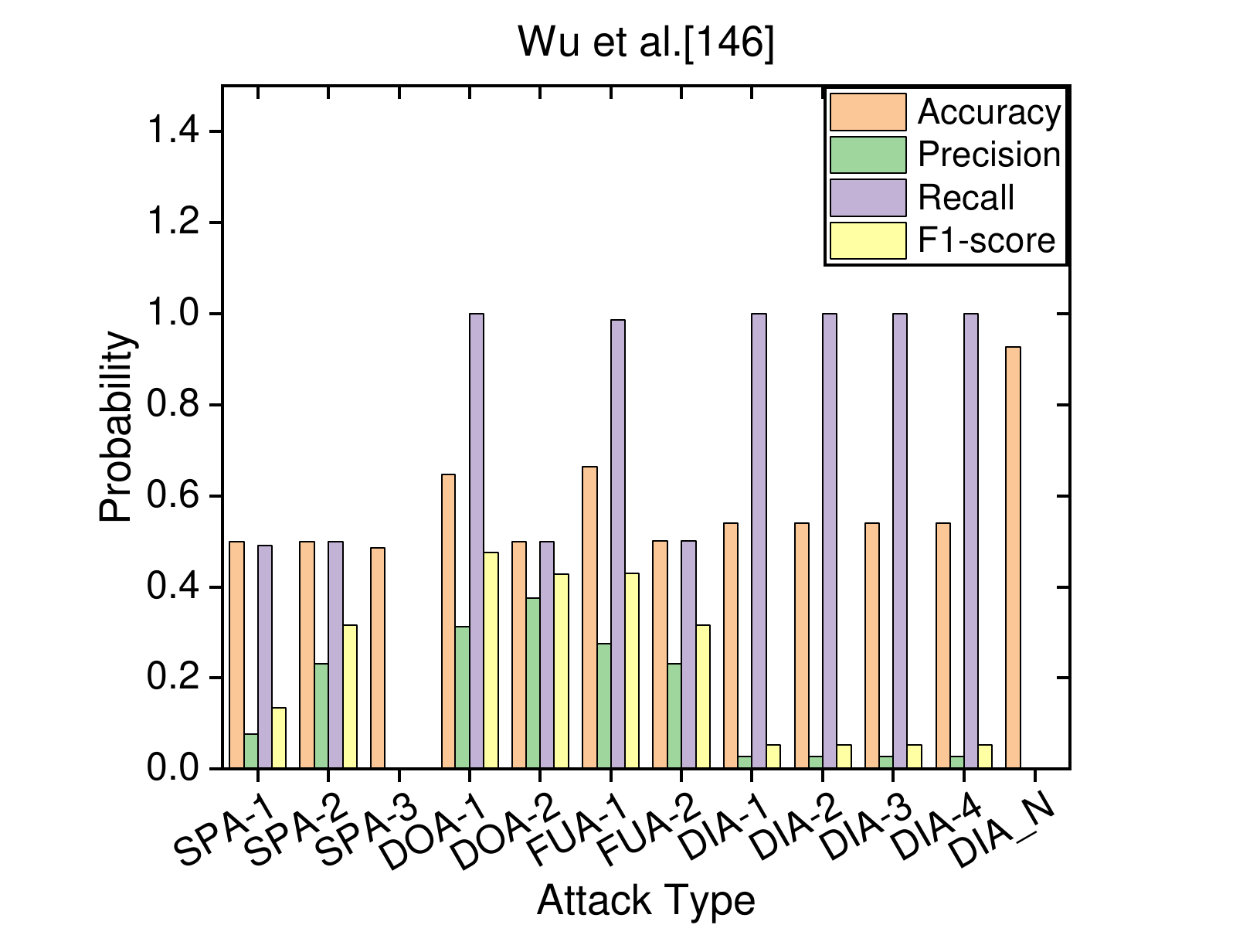} 
		\end{minipage}
		\label{fig:grid_4figs_1cap_4subcap_13}
	}
    	\subfigure[]{
    		\begin{minipage}[b]{0.23\textwidth}
		 	\includegraphics[width=1\textwidth]{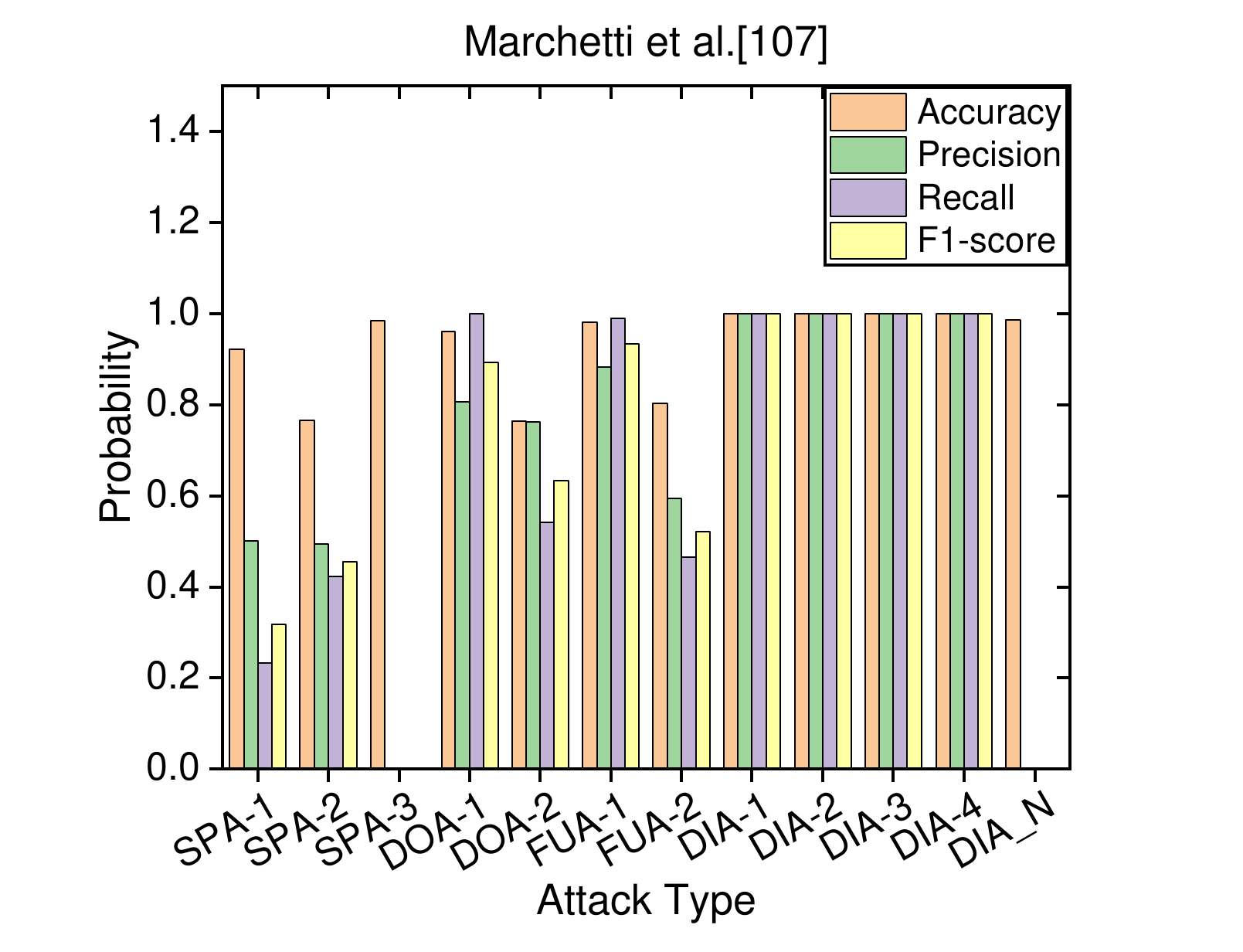}
    		\end{minipage}
		\label{fig:grid_4figs_1cap_4subcap_14}
    	}
    	\subfigure[]{
		\begin{minipage}[b]{0.23\textwidth}
			\includegraphics[width=1\textwidth]{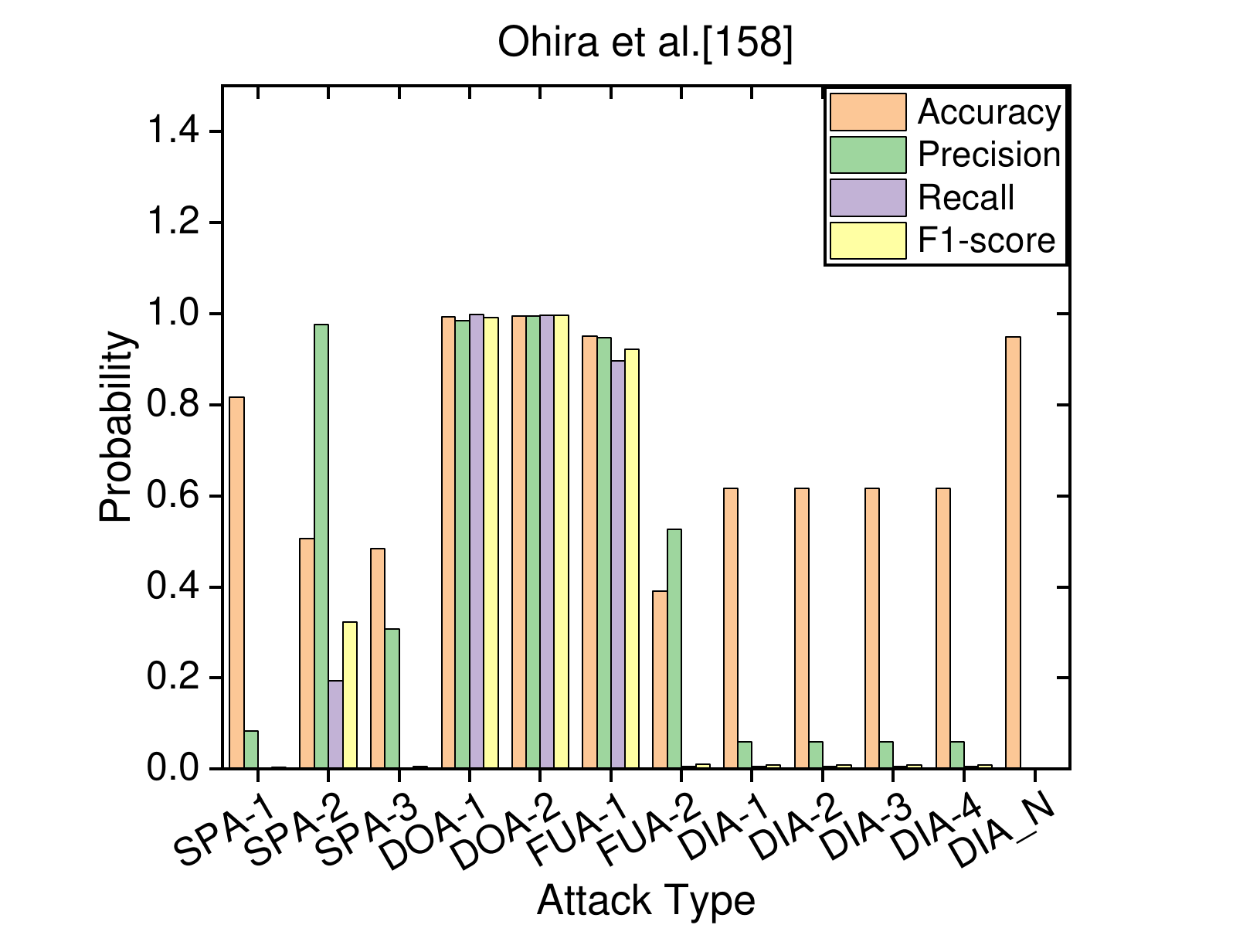} 
		\end{minipage}
		\label{fig:grid_4figs_1cap_4subcap_15}
	}
	\subfigure[]{
		\begin{minipage}[b]{0.23\textwidth}
			\includegraphics[width=1\textwidth]{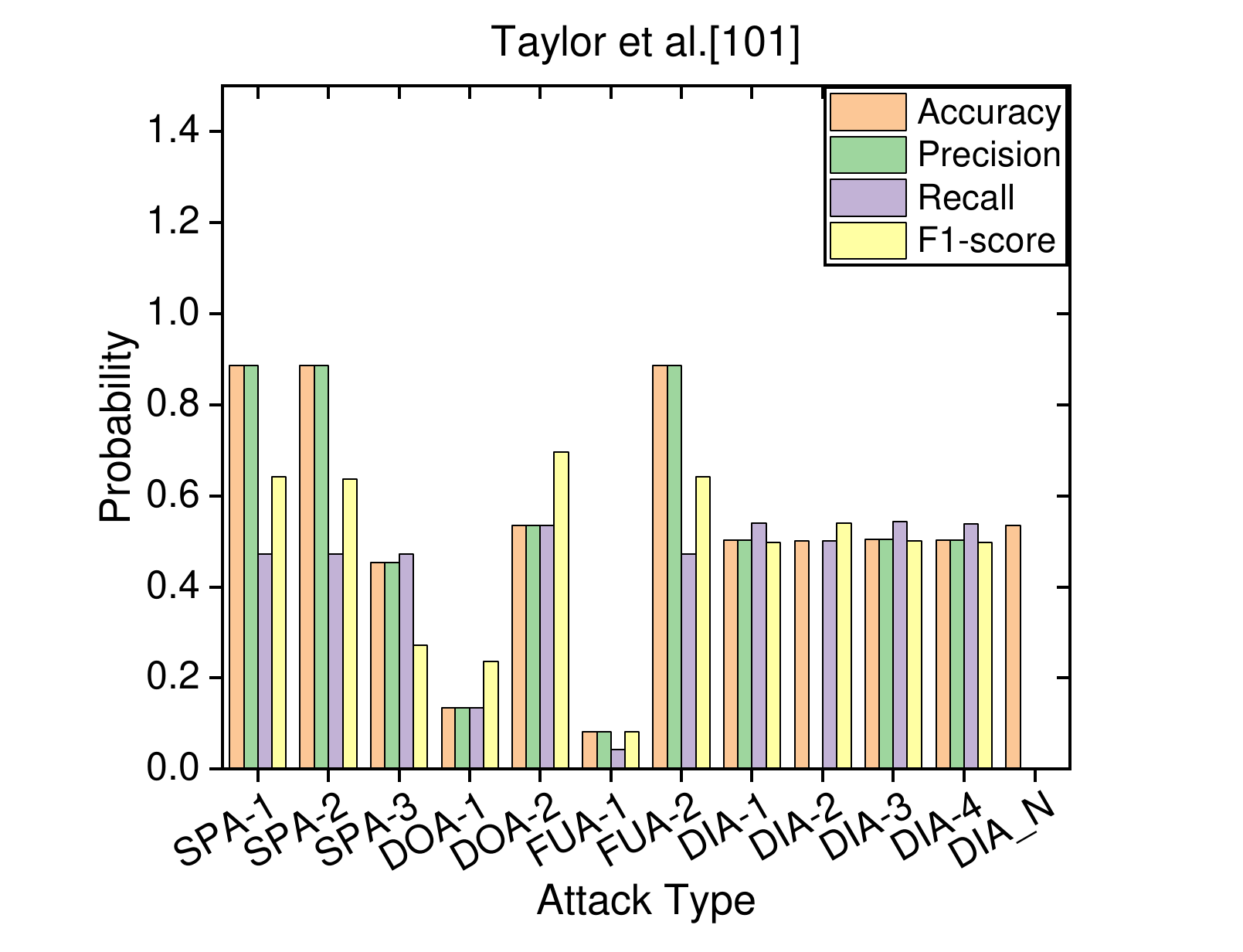} 
		\end{minipage}
		\label{fig:grid_4figs_1cap_4subcap_16}
	}
	\subfigure[]{
    		\begin{minipage}[b]{0.23\textwidth}
		 	\includegraphics[width=1\textwidth]{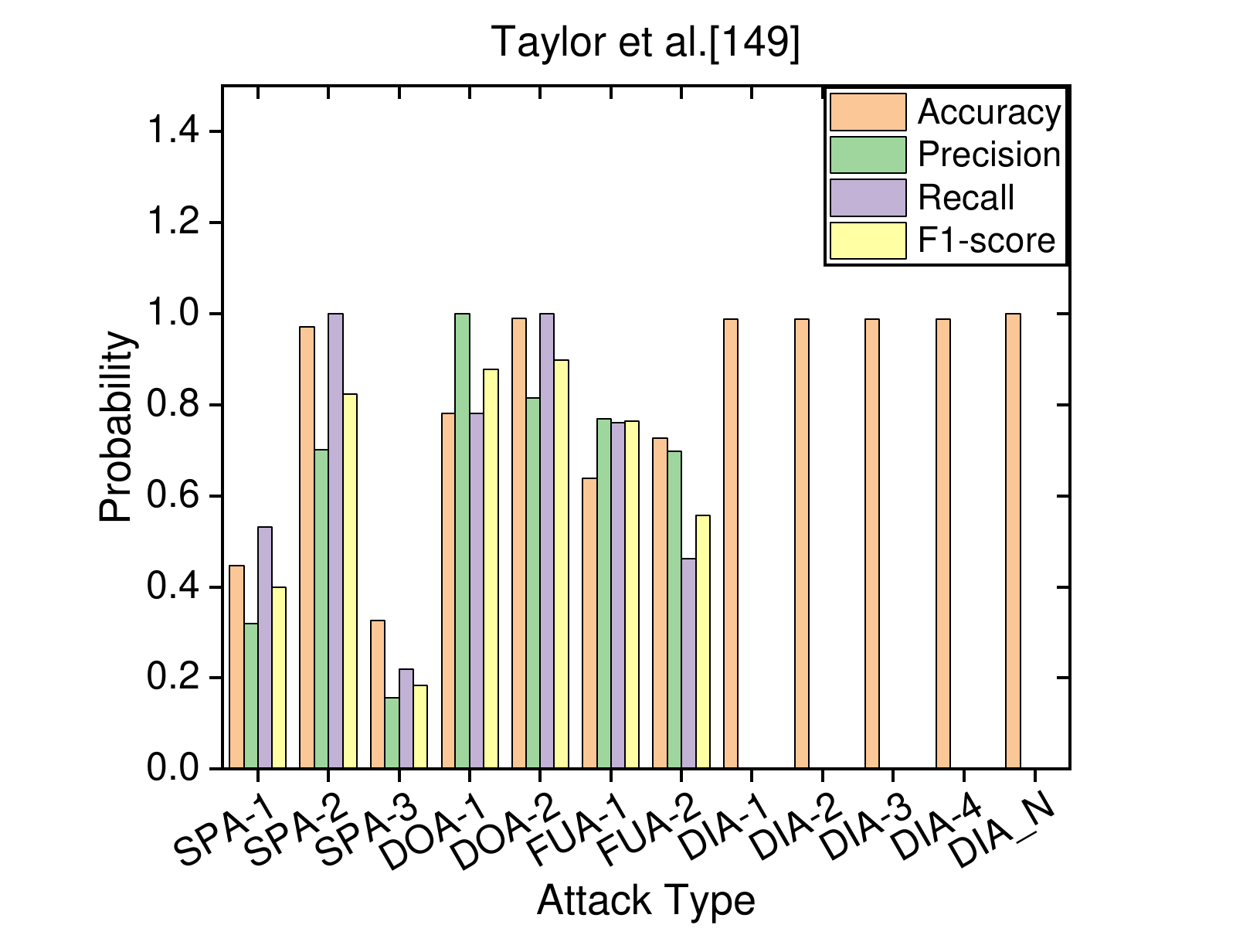}
    		\end{minipage}
		\label{fig:grid_4figs_1cap_4subcap_17}
    	}
    	\subfigure[]{
		\begin{minipage}[b]{0.23\textwidth}
			\includegraphics[width=1\textwidth]{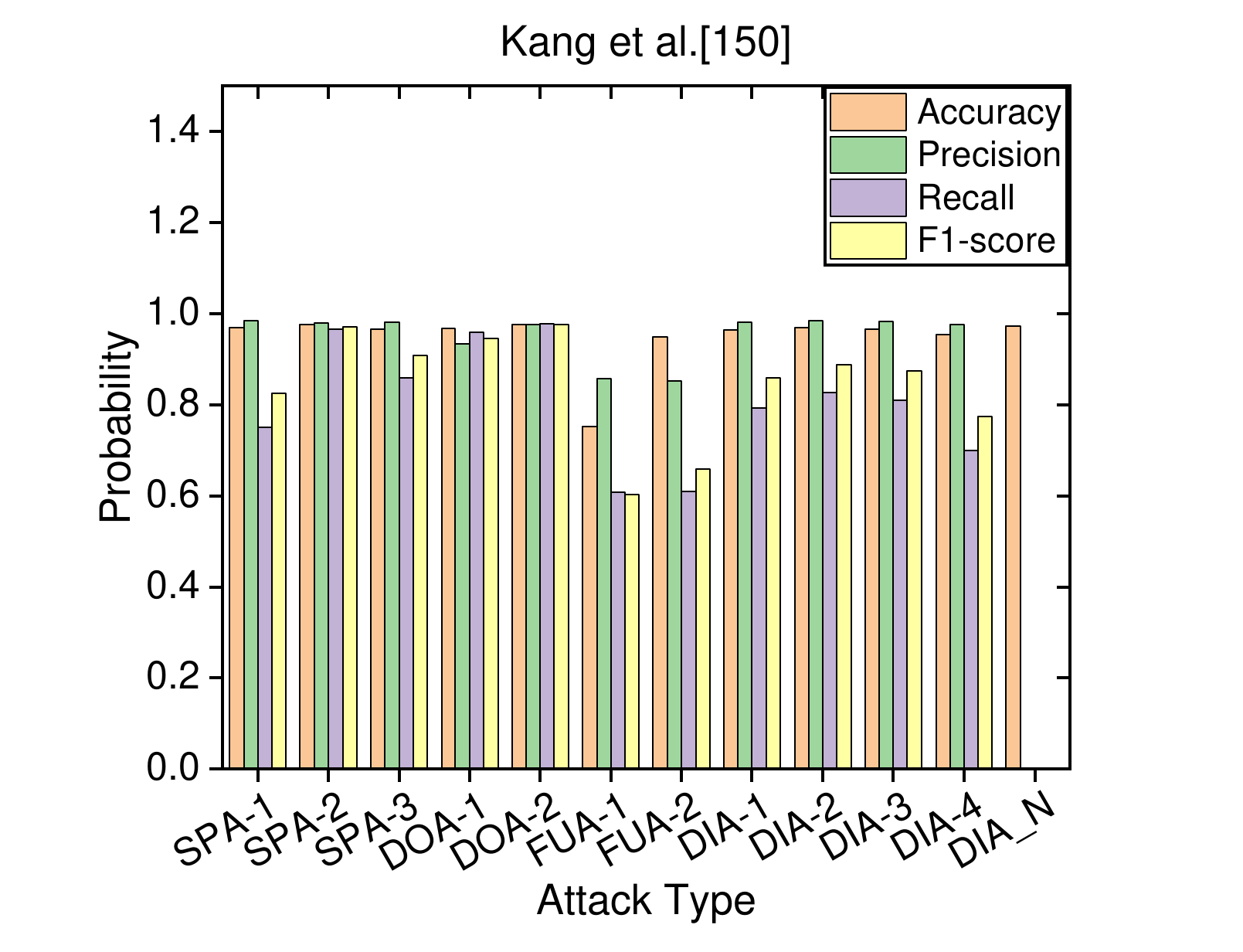} 
		\end{minipage}
		\label{fig:grid_4figs_1cap_4subcap_18}
	}
	\subfigure[]{
		\begin{minipage}[b]{0.23\textwidth}
			\includegraphics[width=1\textwidth]{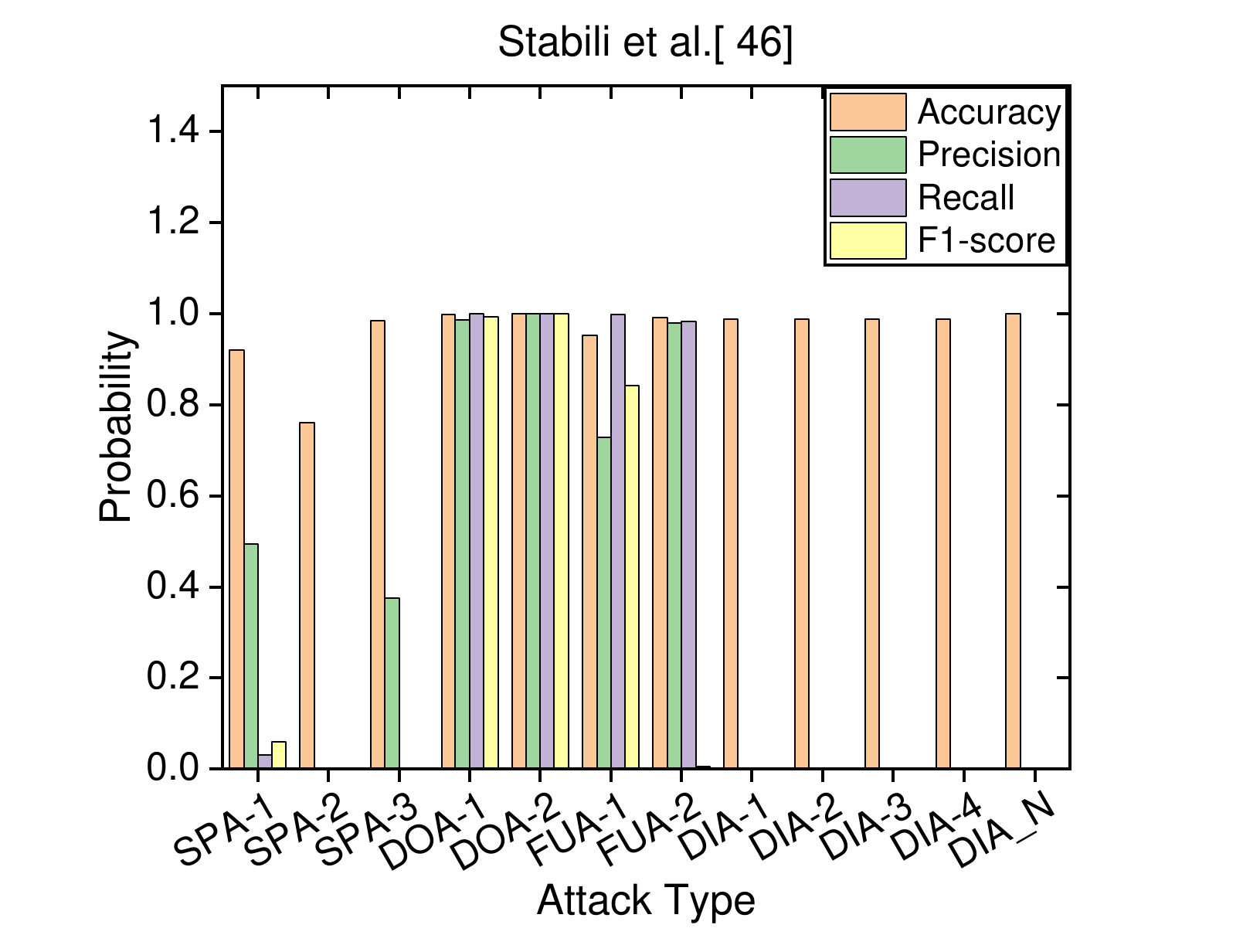} 
		\end{minipage}
		\label{fig:grid_4figs_1cap_4subcap_19}
	}
	\caption{The results of the algorithm evaluation.}
	\label{fig:grid_4figs_1cap_4subcap}
\end{figure*}

\subsubsection{Evaluation Metrics} We experimentally compare the {\vids} based on the following metrics.

    \underline{\textit{Accuracy}} is the most straightforward performance indicator and is simply the ratio of correctly predicted observations to the total number of observations. Accuracy is a good measure, only if we have symmetrical datasets where the values of false positives and false negatives are almost identical. Therefore, we must look at other parameters to evaluate the performance of the model.
    
    \underline{\textit{Precision}} is the ratio of correctly predicted positive observations to the total number of predicted positive observations. Precision demonstrates the system's ability to distinguish between normal messages.

    \underline{\textit{Recall}} is the ratio of correctly predicted positive observations to all observations in the actual class. Recall reflects the system's ability to recognize malicious messages.
     
    \underline{\textit{F1 score}} is a weighted average of ``accuracy'' and ``recall rate''. Therefore, the score takes into account both false negatives and false positives. Intuitively, it is not as easy to understand as accuracy, but the F1 score is usually more useful than accuracy, especially if the class distribution is uneven.

\subsubsection{Detection Result}
Fig.~\ref{fig:grid_4figs_1cap_4subcap} shows the detection results of these {\vids}s for different attacks.
We analyze detection results of the {\vids}s and find out the advantages and disadvantages of these methods.

Let us consider the work by Gmidene et al.~\cite{gmiden2016intrusion} as an illustrative example. The method showcased a relatively high accuracy in countering diverse attacks, owing to the notable true negative (TN) outcomes and the predominance of normal messages over abnormal ones. However, the detection rates exhibited a relatively low efficacy against the \textit{FUA-2} and \textit{SPA-1} attacks. 
This outcome indicates that the method's ability to detect multiple ID attacks is suboptimal. The underlying reason lies in the fact that malicious messages are identified by the frequency changes within a single ID, while the \textit{FUA-2} and \textit{SPA-1} attacks do not significantly alter the frequency of a single ID, thus making them challenging for the method to detect. Consequently, the frequency-based {\vids} struggle to effectively identify these malicious messages.

Furthermore, the precision, recall, and f1-score of the method against the \textit{SPA-3} attack were observed to be zero. This finding indicates the algorithm's incapacity to detect the \textit{SPA-3} attack, where malicious messages replace normal ones while maintaining an unaltered time interval from the preceding message. The frequency-based {\vids}s are unable to detect such attacks due to their reliance on changes in message frequency.

Additionally, the precision, recall, and f1-score of the method against the \textit{DIA-1}, \textit{DIA-2}, \textit{DIA-3}, and \textit{DIA-4} attacks were all found to be zero, with an accuracy of 1. This outcome indicates that the method struggles to differentiate between normal diagnostic messages and malicious ones. This challenge arises from the method's utilization of the stability of in-vehicle messages, while diagnostic messages lack a stable transmission pattern, rendering them indistinguishable.

Another illustrative example can be found in the work by Muter et al. \cite{muter2011entropy}. The study presents two detection methods \cite{muter2011entropy}. The first method leverages the concept of relative entropy among different IDs to identify intrusion instances. However, it is observed that this method fails to detect the \textit{SPA-3} and \textit{SPA-1} attacks. In the \textit{SPA-3} attack, the IDs remain unchanged within the attack dataset, resulting in an unaltered entropy value for the attack data. On the other hand, the \textit{SPA-1} attack introduces new malicious messages with a distribution similar to that of normal messages, resulting in a relatively minor change in the entropy of the attack data. Consequently, the anomalies in entropy go undetected by this method. Moreover, due to the unstable transmission patterns of diagnostic messages within the IVN and the method's inability to obtain a stable relative entropy for these messages, the malicious diagnostic messages remain undetected by the employed {\vids}.

The second method examines the overall entropy changes within the data. However, similar to the first algorithm, it fails to detect the \textit{FUA-2}, \textit{SPA-3}, and \textit{SPA-1} attacks. The inability to detect the \textit{SPA-3} and \textit{SPA-1} attacks is attributed to the same reasons as the first method. In the \textit{FUA-2} attack, the adversary inserts malicious messages with random yet valid IDs. Although these malicious messages increase the entropy within a fixed time window, the overall change in entropy for the entire window is not distinct. Notably, for individual IDs within the window, the change in entropy remains notable. Consequently, while the first algorithm can detect the \textit{FUA-2} attack, the second method fails to do so. Moreover, the detection rates for the \textit{DIA-1}, \textit{DIA-4}, \textit{DIA-2}, and \textit{DIA-3} attacks are poor. Our analysis reveals that the frequency of diagnostic messages is low, thereby limiting their impact on the overall data entropy. Additionally, the method employs a lenient threshold, resulting in the failure to detect malicious diagnostic messages.

Through a comprehensive analysis of various VIDSs, we have reached a comprehensive and conclusive summary finding.

\noindent
\textbf{Brief Discussion:} 
Primarily, it is imperative to emphasize that the direct evaluation of the merits and demerits of these methods based solely on experimental results is not viable, owing to the inherent divergence in their respective threat models. The primary objective of our experiments is to rigorously assess the detection capabilities of these methods in combating prevalent attacks, thereby elucidating their inherent limitations.
Conclusions can be drawn from the empirical findings as follows.

The frequency-based, information entropy-based, ID sequence-based, and similarity-based methods are inherently reliant on the periodicity exhibited by CAN messages, specifically the stability of the ID attribute. 
In scenarios where adversaries transmit supplementary messages to manipulate vehicle control, these methods exhibit a notably high detection rate for identifying malicious messages. However, it is important to note that in instances where adversaries intentionally simulate a normal message cycle, such as in masquerade attacks, these methods may not be able to adequately detect these malicious messages as expected.

The payload-based method, which hinges upon detecting malicious messages through the change in the normal payload, provides a defense mechanism against masquerade attacks. However, owing to the intricate nature of the IVN, establishing a robust and consistent model for the payload of normal CAN messages proves to be challenging. Consequently, discerning the distinction between malicious and normal data becomes arduous, thereby leading to a relatively low detection rate for this method.

\section{DISCUSSION}
\label{sec:discussion} 
Through an extensive review of various VIDS, we identified significant limitations that hinder their practical application. Additionally, we examine the future development of VIDS in the context of emerging automotive technologies.

While our primary focus is on CAN-based VIDS, it is important to consider broader vehicular cybersecurity solutions. This section also explores VIDS in intelligent transportation systems, which utilize large-scale data from connected vehicles and infrastructure, VIDS on J1939 heavy-duty vehicle CAN buses, which face unique challenges due to their distinct network structures, and VIDS for the Internet of Vehicles (IoV), where intrusion detection must adapt to highly connected and dynamic environments using cloud computing, edge processing, and V2X communication. These discussions provide a comprehensive view of vehicular intrusion detection and potential directions for future research.

\newtext{
\subsection{Current Issues}
Initially, an exhaustive compilation of the limitations inherent to all defense methods is presented.
}

\rnewadd{
\subsubsection{Practicality}
In the majority of prevailing vehicle models, conventional ECUs continue to be utilized, employing communication via the CAN bus. In order to safeguard this type of automobile, we think that a plug-and-play incremental protection approach or lightweight protection approach aligns more aptly with the requirements of contemporary OEMs.
To commence, it should be noted that the ECUs found in the majority of vehicle models exhibit constrained computational capabilities and the transmission capacity of the CAN is also subject to limitations. The incorporation of intricate encryption or authentication algorithms within the existing {\ivn} poses a substantial burden, given the aforementioned constraints.}

\rnewadd{
Secondly, the implementation of extensive security updates for legacy automobile models poses considerable challenges for automakers. Integrating over-the-air (OTA) capabilities to existing vehicle models is a rare occurrence, thereby presenting a formidable obstacle in terms of incorporating modified communication protocols and intricate defense mechanisms into the original ECUs.}

\rnewadd{
Another paramount consideration revolves around cost implications. Heightened computing power and accelerated communication technologies entail elevated expenses. For OEMs, undertaking hardware and software updates incurs substantial financial investment. Additionally, accommodating the requirements of previous vehicle models necessitates additional expenditures.}

\subsubsection{Targeted attack}
According to our survey, it is commonly observed that when selecting target attacks, the prevailing tendency of VIDS is to opt for conventional attack types, such as Spoofing attacks, Fuzzing attacks, and Denial-of-Service (DoS) attacks. It is noteworthy that these attack categories were originally proposed in works dating back a decade~\cite{koscher2010experimental}.
Despite the significant detrimental impact caused by these aforementioned attacks, we contend that VIDS should be geared towards addressing more pragmatic or sophisticated attack scenarios.

Primarily, it is observed that numerous papers make references to real-world instances of car attacks.
However, a distinct shortcoming within the existing research is the dearth of focused investigations pertaining to defenses specifically tailored to counter these real-world attacks.
A noteworthy instance is the research conducted by Miller and Valasek, wherein they successfully employed the vehicle's diagnostic protocol to exert remote control over the car. Regrettably, this particular form of attack has received limited attention from researchers thus far.

Secondly, a considerable number of contemporary studies put forth more sophisticated attack methodologies~\cite{kulandaivel2021cannon,bhatia2021evading}. Within these attacks, adversaries possess the capability to obfuscate the attack traces, thereby evading detection by both users and defense systems.
Moreover, these attacks also present a substantial threat to the overall safety of the automobile and
they warrant significant attention and scrutiny from the research community.

\newtext{
\subsubsection{Detection method based on machine learning}
The utilization of machine learning techniques in VIDS holds great promise.
Nonetheless, the prevailing detection defenses employed within VIDS continue to exhibit a relative simplicity in harnessing the potential of machine learning technology.}

\newtext{
Primarily, these methods typically rely on direct utilization of machine learning algorithms for scrutinizing the abnormality in CAN message frequency or payload. In essence, this approach capitalizes on the inherent periodicity and predictability of CAN messages. However, noteworthy advancements beyond prior rule-based methods have not been significantly achieved.}

\newtext{
It is noteworthy that this methodology remains susceptible to manipulation through carefully crafted falsified messages, thereby impeding its robustness and reliability. Furthermore, in comparison to the rule-based approach, the machine learning-based methodology generally necessitates enhanced computational power and entails greater time consumption for the ECU.}

\subsubsection{Abruptness of CAN messages}
The accuracy of detection can be affected by the abruptness of {\can} messages transmission.
Many research works design their systems according to the stability and sustainability of the {\ivn}.
However, some research works show that these systems falsely detect benign event messages because the event messages deviate from the periodicity~\cite{koyama2019anomaly}. 
Besides, some special situations (e.g., retransmission after competing, bit errors) can also destroy the periodicity of in-vehicle messages. 
Therefore, how to distinguish malicious attacks and event messages or special situations is a great challenge.
The anomaly-based {\vids}s, which use the periodicity of in-vehicle messages, are difficult to solve the problems resulting from the abruptness of {\can} messages transmission.

\subsubsection{Hardware Limitations}
The practical application of certain VIDS is severely impeded by the hardware limitations of ECUs. Conventional low-end ECUs typically comprise microcontrollers with modest computational cores \cite{stabili2017detecting}, operating at frequencies in the range of several hundred megahertz and equipped with a few hundred kilobytes of RAM. 
However, certain approaches (e.g., \cite{kang2016intrusion, martinelli2017car}) demand substantial computing resources, rendering their deployment in present-day automobiles challenging. 
As a consequence, VIDS implementations need to be tailored to accommodate the memory and computational constraints of current ECUs.

Furthermore, a few papers even propose VIDS solutions that necessitate the addition of supplementary equipment, such as oscilloscopes, for monitoring the {\ivn} \cite{cho2017viden, kneib2018scission}. While these methods offer exceptional detection efficacy, their associated costs are deemed unacceptable. Given the reluctance of Original Equipment Manufacturers (OEMs) to modify the existing IVN architecture of contemporary automobiles, the feasibility of implementing such approaches remains unlikely.
Exploring alternative avenues that enable researchers to attain comparable detection capabilities in a more convenient and cost-effective manner, such as employing method EASI~\cite{kneib2020easi}, represents a highly promising trajectory worth considering.

\subsubsection{Private Communication Protocols}
The adoption of proprietary protocols by various OEMs presents a significant obstacle to the development of semantic information-based VIDS. 
While certain papers propose VIDS solutions based on the collection of vehicle status data from normal CAN messages or diagnostic messages, the existing ECU systems and transport protocols for CAN messages are provided independently and secretly by different OEMs. 
Consequently, the parsing of diverse transport protocols on the bus and the acquisition of vehicle status pose substantial challenges.

Although the OBD diagnostic protocol allows for the retrieval of limited vehicle status information, additional diagnostic messages must be injected into the vehicle, thereby impeding normal ECU communication. This limitation necessitates careful consideration as it impacts the practicality of the approach and its potential effects on vehicle safety.

\subsection{Trends}
Despite the existing vulnerabilities and loopholes within the current vehicle network, it is important to recognize the rapid advancements taking place in automotive-related technologies. Ongoing efforts are being made to address hardware limitations and software vulnerabilities within modern vehicles, indicating a gradual resolution of these issues. Consequently, there is a strong possibility of significant breakthroughs in VIDS.
Subsequently, we will outline several proposals for VIDS tailored specifically for current vehicle models. Additionally, we will present a forward-looking perspective on future defense methodologies that integrate seamlessly with intelligent automotive systems.

\subsubsection{Integrating multiple methods}
By integrating multiple methods, the effectiveness of intrusion detection can be enhanced. Each detection method possesses its own limitations, but through their combination, a broader range of attack scenarios and types can be accurately identified. For instance, frequency-based VIDS demonstrates advantages such as resource efficiency, high detection rates, and ease of implementation. However, it may be susceptible to evasion by adversaries employing carefully crafted messages, thus exhibiting a certain degree of unreliability. In such cases, VIDS based on voltage signatures can effectively identify these crafted messages. Consequently, when aiming to detect covert attacks~\cite{bhatia2021evading}, the utilization of voltage-based VIDS in conjunction with frequency-based VIDS can provide valuable support and enhance overall detection capabilities.
The integration of methods in VIDS presents a promising and straightforward avenue for development, offering both simplicity and effectiveness.

\rnewadd{
\subsubsection{Advancing Machine Learning in VIDS: Opportunities and Limitations}
Machine learning has shown great potential in enhancing \vids by leveraging anomaly detection techniques based on CAN message frequency and payload analysis. However, despite these advancements, ML-based \vids still face several limitations that hinder their practical deployment.
}

\rnewadd{
One major challenge is the computational overhead associated with deep learning models, which often require significant processing power and memory, making them difficult to deploy on resource-constrained ECUs. To address this, lightweight ML techniques such as model quantization, pruning, and knowledge distillation can be explored. These techniques reduce the size and complexity of ML models while maintaining detection accuracy. Additionally, TinyML—a framework designed for running ML models on low-power embedded devices—can be investigated to enhance the feasibility of ML-driven VIDS in real-world automotive environments.}

\rnewadd{
Another limitation of ML-based VIDS is their static nature, as most models are trained on a fixed dataset and lack adaptability to new attack patterns. Future research should focus on incremental learning and online learning techniques, enabling models to continuously update and adapt to emerging threats without requiring complete retraining. Furthermore, federated learning (FL) can be employed to allow multiple vehicles to collaboratively improve their intrusion detection capabilities while preserving data privacy. This decentralized approach reduces the need for centralized data storage and minimizes communication overhead.}

\rnewadd{
\subsubsection{Addressing the Impact of Bursty CAN Message Transmission on Detection Accuracy}
Traditional VIDS designs often assume a stable and periodic CAN message transmission pattern, relying on deviations from expected message intervals as indicators of potential attacks. However, real-world CAN traffic exhibits bursty transmission behavior, where event-triggered messages deviate from periodic patterns, leading to increased false positives in anomaly-based detection systems.}

\rnewadd{
To mitigate this issue, advanced time-series analysis techniques, such as Long Short-Term Memory (LSTM) networks, Transformer models, and Hidden Markov Models (HMMs), can be leveraged to improve anomaly detection in bursty CAN environments. These models can learn temporal dependencies and distinguish between legitimate event-driven message bursts and malicious anomalies.}

\rnewadd{
Additionally, multi-modal data fusion can be explored by integrating CAN message analysis with other vehicle sensor data (e.g., wheel speed, braking pressure, and GPS data) to provide additional context for anomaly detection. By correlating information across multiple data sources, VIDS can reduce false alarms caused by benign event-driven deviations.}

\rnewadd{
Another promising approach is the use of adaptive anomaly detection mechanisms, where detection thresholds dynamically adjust based on contextual information. For instance, reinforcement learning algorithms can be employed to continuously refine the decision boundaries of an ML-based VIDS, ensuring that benign variations in CAN traffic do not trigger unnecessary alerts while still detecting genuine intrusions.}

\rnewadd{
\subsubsection{Overcoming ECU Hardware Limitations for Efficient VIDS Deployment}
The computational constraints of traditional ECUs pose a significant challenge for the deployment of sophisticated VIDS, particularly those utilizing deep learning or complex statistical models.}

\rnewadd{
To address these limitations, hardware acceleration techniques such as Field-Programmable Gate Arrays (FPGAs) and Application-Specific Integrated Circuits (ASICs) can be explored to offload computationally expensive tasks from the ECU. FPGA-based implementations of intrusion detection algorithms can significantly improve processing speed while maintaining energy efficiency.}

\rnewadd{
Another viable solution is the adoption of edge computing for VIDS, where computationally intensive tasks are offloaded to dedicated edge nodes within the vehicle (e.g., a central gateway ECU or an onboard automotive AI processor). This architecture enables real-time analysis while reducing the computational burden on individual ECUs.}

\rnewadd{
Finally, lightweight cryptographic techniques such as elliptic curve cryptography (ECC) and hash-based authentication mechanisms should be investigated to enhance security without overburdening ECU processing capabilities. By integrating these efficient cryptographic methods, VIDS can maintain robust security features while remaining feasible for deployment in modern vehicles.}

\rnewadd{
\subsubsection{Addressing Challenges Posed by Proprietary Communication Protocols}
One of the major barriers to the development and adoption of semantic-based VIDS is the lack of standardization in CAN message semantics. Automotive OEMs often implement proprietary communication protocols, making it difficult for intrusion detection systems to interpret and analyze vehicle-specific CAN messages. .}

\rnewadd{
To overcome this challenge, reverse engineering techniques can be explored to infer the meaning of proprietary CAN messages. Recent advancements in unsupervised learning and natural language processing (NLP) techniques may provide new ways to automatically extract semantic information from CAN traffic without requiring access to OEM-proprietary documentation.}

\rnewadd{
Another promising approach is the use of blockchain technology to create a decentralized and immutable repository of CAN message definitions shared across multiple stakeholders in the automotive industry. By leveraging blockchain for secure and transparent data sharing, researchers and industry practitioners can collaborate to build more standardized and interpretable VIDS solutions.}

\rnewadd{
Furthermore, the adoption of standardized automotive communication protocols, such as AUTOSAR Adaptive Platform and Vehicle-to-Everything (V2X) security frameworks, can help mitigate the challenges posed by proprietary protocols. Encouraging industry-wide adoption of open standards can facilitate the development of more effective and interoperable VIDS solutions.}

\subsubsection{Protection Mechanisms for Smart Driving Cars}
The advancement of vehicle intelligence has brought about a transformation in the defense mechanisms employed in automobiles. As previously discussed, conventional vehicles still rely on low-computing ECUs and CAN for vehicle control. However, with the progress of intelligent and autonomous driving technologies,OEMs are increasingly adopting high-performance ECUs capable of supporting intelligent driver-assistance systems or automated driving systems. Furthermore, the realization of intelligent driving necessitates the integration of data from high-precision sensors like cameras and lidars, thereby demanding the utilization of higher-speed networks. Consequently, a number of automotive manufacturers have initiated the adoption of advanced network technologies as substitutes for CAN in their vehicles. Examples of these technologies include CAN-FD ({\can} with Flexible Data-Rate) and vehicle Ethernet, among others.

Through the utilization of enhanced hardware and higher-speed communication buses, a broader range of methods can be employed to safeguard the {\ivn}. 
One such approach involves car manufacturers implementing authentication techniques to ensure the secure transmission of messages. 
OEMs can implement sophisticated authentication techniques that verify the integrity and authenticity of transmitted data, effectively mitigating the risk of unauthorized access or tampering.
Additionally, the use of encrypted data can effectively safeguard the confidentiality of sensitive information within the vehicle. 
Employing encryption methods can effectively safeguard the confidentiality of private information, ensuring that critical data remains inaccessible to unauthorized entities. By leveraging encryption protocols, car manufacturers can bolster privacy protection and instill confidence in users regarding the security of their personal information.
These methods hold promise in surpassing the efficacy of traditional intrusion detection systems, thereby fortifying the security of the {\ivn}.

In reality, despite the presence of intelligent assisted driving system or automated driving system, the {\ivn} of the smart car continues to be predominantly based on the CAN bus at present. As a result, researchers have shifted their focus away from the {\ivn} of smart cars and towards new attack surfaces, including but not limited to the perception modules of vehicles~\cite{cao2022you,jin2022pla}, autonomous driving algorithms~\cite{wachi2019failure}, and the Vehicle-to-Everything(V2X)~\cite{zhou2020distributed,nair2022ai}. 
In forthcoming endeavors, we shall embark upon an exhaustive investigation of all scholarly undertakings pertaining to the safety of intelligent automobiles.

\rnewadd{
\subsection{Vehicular Intrusion Detection Systems in Intelligent Transportation Systems}
In the current intelligent transportation system, the {\ivn}, a crucial component of the internal network, primarily facilitates communication among ECUs within the vehicle using protocols like CAN (or other low-speed protocols). As technology advances, newer communication protocols are likely to replace existing ones.  
In existing research, intrusion detection for {\ivn} and defenses for other networks often progress independently. These methods are typically based on different communication protocols and system models, with limited integration of various intrusion detection approaches. The following are key challenges and considerations when devising an intrusion detection system within the Intelligent Transportation System (ITS) framework.}

\textbf{Data Alignment and Accuracy.} One significant challenge in developing a comprehensive intrusion detection system is addressing the delays in information transmission between various networks. When designing such a system, decision-making often relies on data collected from different networks, encompassing information like vehicle speed, steering angle, and radar data in the CAN. The challenge arises from the disparate data transmission rates across these networks, resulting in a complex and potentially messy dataset. Effectively aligning and maintaining the accuracy of this diverse data pose challenges that must be carefully considered in the design process.

\textbf{Impact Analysis between Networks and Modules.} When devising anomaly detection algorithms, it is crucial to take into account the interactions between different networks and intelligent modules. For instance, in cases where an intelligent assisted driving system bases acceleration and deceleration decisions on sensors like cameras, the instructions are transmitted to the {\ivn}. In this context, the rationality of these instructions can be assessed by considering the status of the {\ivn}. This assessment helps determine whether the intelligent assisted driving system is under attack~\cite{xue2022said}.
Moreover, in traditional intrusion detection designs, there was a prevailing assumption that messages within the in-car network followed a periodic and stable pattern. However, with the introduction of various intelligent driving modules transmitting diverse data and instructions, these messages may disrupt the continuity and periodicity of the original messages. Consequently, when developing detection algorithms, it becomes essential to comprehensively consider the impact of different networks.

\textbf{Utilization of Advanced Technologies.} There is an opportunity to leverage more machine learning algorithms and artificial intelligence technologies. In prior research, a significant challenge restricting the design of intrusion detection algorithms stemmed from the limited performance of the ECU itself. However, with contemporary car manufacturers incorporating higher-performance ECUs for intelligent driving, there is room to employ newer technologies for anomaly detection, such as deep learning and large language models.

\textbf{Verification and Encryption.} It is essential to contemplate the verification and encryption of data. In the past, due to constraints related to the transmission speed of in-car networks and the performance of the ECU, data within the car was typically not encrypted by manufacturers. Researchers could design intrusion detection systems by directly analyzing changes in in-vehicle data. However, as ECUs evolve and higher-speed networks are employed, there is a likelihood that car manufacturers will implement data encryption. Consequently, researchers must develop more advanced detection algorithms aligned with the communication protocols utilized for data transmission.

\textbf{Scalability and Flexibility.} Algorithm design should prioritize scalability and flexibility. The swift evolution of intelligent transportation systems translates to frequent changes in the quantity and types of automotive sensors and network architecture. Intrusion detection systems for intelligent transportation are often deployed across diverse vehicles. Therefore, researchers must account for variations in hardware and software among different vehicle types when crafting intrusion detection algorithms. Simultaneously, the designed system should remain unaffected by upgrades to the car's internal software or hardware.

In summary, addressing these challenges and considerations is essential for designing effective and robust intrusion detection systems in the dynamic and interconnected environment of intelligent transportation systems.

\rnewadd{
\subsection{Vehicular Intrusion Detection Systems on J1939 Heavy-Duty Vehicle CAN Buses}
\label{ssec:IDS_j1939}
In contrast to regular commercial vehicles, heavy-duty vehicles also employ the CAN protocol for transmitting in-vehicle messages. However, heavy-duty vehicles diverge from the traditional CAN approach by utilizing a specialized CAN protocol based on the SAE J1939 standard. This standard incorporates extended frames and a dedicated transport protocol for multi-packet transmission. Only a limited number of recent studies have delved into addressing safety concerns specifically for heavy-duty vehicles utilizing the SAE J1939 standard~\cite{jichici2023control}. Due to the disparate protocols in focus, direct comparisons with the intrusion detection systems discussed for typical commercial vehicles pose challenges. Consequently, we only presented these intrusion detection systems for CAN based on the SAE J1939 standard.}

H. Shirazi et al. transform the original transmission message into specific parameters representing the vehicle's status. Subsequently, they employ machine learning algorithms to construct a model of the normal vehicle. This model is then utilized to identify DoS and fuzz attacks~\cite{shirazi2020using}.
Mukherjee et al. introduced a priority graph-based method for detecting message injection attacks~\cite{mukherjee2017precedence}.
In a recent development, Jichici et al. proposed a two-stage intrusion detection mechanism for J1939~\cite{jichici2022effective}. The initial phase verifies the legitimacy of the encrypted addresses (source and destination) in the CAN ID. The subsequent phase focuses on detecting single-bit alterations in the data field through appropriate range checks. Given the encryption of CAN frame data fields, the avalanche effect of block ciphers aids in identifying adversarial manipulation.
Rogers et al. presented an alternative approach relying on timing and data analysis to identify spoofing and masquerading attacks in J1939 and NMEA2000 networks~\cite{rogers2022detecting}. This mechanism can detect manipulation attacks by scrutinizing unusual changes in electrical potential during the transition from the dominant to the passive state, i.e., a single bit flip.
Popa et al. investigated whether ECU voltage characteristics can serve as fingerprints for detecting spoofing attacks in J1939~\cite{popa2022ecuprint}.

\rnewadd{
\subsection{Vehicular Intrusion Detection Systems for Internet of Vehicles}
\label{ssec:IDS_iov}
In this section, we provide a brief overview of research related to IoV intrusion detection. It is important to note that the CAN bus is a subset of IoV, and intrusion detection for IoV is not fully covered in our survey. However, considering it as a significant direction in the latest advancements in vehicle technology, we briefly mention related advanced works.}

In essence, the Internet of Vehicles represents the integration of Vehicular Ad Hoc Networks (VANETs) and the Internet of Things (IoT) \cite{dureja2020review}. 
Modern connected vehicles utilize IoT to connect to networks, accessing real-time traffic data, navigation, and other driving conveniences. IoV employs various network technologies to enable communication within vehicles and between different entities on the road, fostering intelligent knowledge sharing. However, the extensive connectivity in the Internet of Vehicles, which involves numerous IoT sensors and processors, poses inherent risks. The continuous communication between road entities and the network makes IoV susceptible to intruders \cite{garg2022security}. Security in the Internet of Vehicles is a critical concern, as incorrect information interfering with vehicle decision-making could have severe consequences, even leading to fatalities. Potential attackers might exploit vulnerabilities in network communications to take control of a vehicle, disseminate misleading information, or conduct other malicious activities that compromise the confidentiality, integrity, availability, and authenticity of vehicle systems. An illustrative example is a group of hackers successfully tricking Tesla's Autopilot software into veering into oncoming traffic \cite{wadhawan2020ignore}. Moreover, the wealth of data generated by autonomous driving raises privacy concerns, as this data can be utilized for artificial intelligence (AI) applications and data mining, exposing users' sensitive information to potential risks.

To bolster the security of the Internet of Vehicles, researchers recognize the need for an IDS capable of efficiently detecting anomalous behaviors in the network and promptly alerting authorities or users to potential threats \cite{seo2018gids}.
Deep learning proves effective in discerning the inherent patterns within sample data. It accommodates higher-dimensional learning and prediction needs by establishing a nonlinear network structure with multiple hidden layers. Certain researchers~\cite{spandonidis2022development,busacca2022smart,wan2022edge} employ deep learning methods and edge computing technologies to analyze the traffic and speed of vehicles in the Internet of Vehicles. This analysis furnishes personalized safety information to drivers, thereby laying the groundwork for intrusion detection in the Internet of Vehicles.
The studies~\cite{loukas2017cloud,gui20206g,alladi2021deep} consistently highlight that the application of deep learning methods significantly enhances intrusion detection performance, making it a widely adopted approach in the field of Internet of Vehicles intrusion detection.
Yang et al. \cite{yang2022federated} introduced an intrusion detection method tailored for {\ivn}. Their approach leverages federated deep learning, capitalizing on the periodicity of network messages. The ConvLSTM model is employed to identify network intrusions, and the intrusion detection model is trained using federated deep learning techniques.
Li et al. \cite{li2021transfer} presented an intrusion detection scheme for the IoV that relies on transfer learning. The proposed method incorporates two modes: cloud-assisted update and local update.
Shone et al. \cite{shone2018deep} introduced an unsupervised deep learning intrusion detection technology utilizing an asymmetric deep autoencoder to construct a classification model. However, this method faces challenges in achieving better classification performance in unbalanced samples. 
Xu et al. \cite{xu2020toward} devised a Log-Cosh variational autoencoder method, incorporating a logarithmic hyperbolic chordal function to design a loss term for generating diverse intrusion data, thereby enhancing detection accuracy. Despite these advancements, deep learning-based solutions still encounter a high false-positive rate, primarily attributed to inadequate extraction of relevant features in the IoV.
Intrusion data within the IoV encompasses numerous spatio-temporal features that can reflect certain attacker characteristics. Consequently, researchers have explored the utilization of deep learning methods, such as CNN or LSTM, to extract and process these spatio-temporal features

Hu et al. [\cite{hu2020novel} developed an intrusion detection technique employing CNN with a split convolution module. This approach aims to enhance the diversity of spatial characteristics and reduce the impact of information redundancy across channels on the model. 
Park et al. \cite{park2021host} transformed network traffic into a grayscale image, established a Siamese CNN based on the small sample learning method, and determined the attack type based on the similarity score of the attack samples.
To capture time-dependent dynamic features in network traffic, Zhou et al. \cite{zhou2023intrusion} proposed an incremental LSTM network intrusion detection method. This method introduces state changes into LSTM, processing network data by acquiring the hidden layer state of LSTM dynamic information.
Ashraf et al. \cite{ashraf2020novel} employed a combination of LSTM and autoencoder to extract timing features from Internet of Vehicles network traffic, enhancing the accuracy of intrusion detection in the Internet of Vehicles. While previous solutions often use only CNN or LSTM to process spatiotemporal features, this approach might suffer from insufficient feature extraction. Consequently, some researchers advocate for hybrid models integrating both CNN and LSTM to address this limitation.
Wang et al. \cite{wang2017hast} introduced a hierarchical intrusion detection system based on spatiotemporal features. Initially, CNN is utilized to learn spatial features in network traffic packets, followed by LSTM to learn temporal features between multiple network traffic packets. This sequential approach results in a more accurate spatiotemporal feature vector. However, these solutions overlook the challenge of variable time intervals between packets in the data stream. To tackle this issue, Han et al. \cite{han2020stidm} proposed a space- and time-aware intrusion detection model. They developed a time and length-sensitive LSTM method to capture broader temporal features from intermittent flows.
Shams et al. \cite{shams2023flow} devised an IDS model capable of collaboratively collecting network data from both vehicles and Roadside Units (RSUs). They implemented a multi-class IDS utilizing a Convolutional Neural Network (CNN) with a novel feature extraction method named Context-Aware Feature Extraction-Based CNN (CAFECNN). Leveraging the collected network flow data, the CAFECNN model effectively identifies both passive and active types of attacks. Results indicate that the proposed model demonstrates superior identification capabilities for hard-to-detect passive attacks in comparison to traditional machine learning methods.

The intrusion detection methods leveraging deep learning have proven effective in detecting network attacks in the Internet of Vehicles. However, these AI-based approaches also introduce risks and challenges, including vulnerabilities to adversarial sample attacks and concerns about the security of the intrusion detection system itself. Researchers are increasingly exploring the use of formal methods \cite{seshia2022toward,krichen2022formal} to enhance the reliability of artificial intelligence solutions. By employing mathematical logic, models, and proofs, these methods aim to verify whether the Internet of Vehicles intrusion detection system aligns with design specifications and identify potential errors, ultimately improving the security and dependability of intrusion detection.
In the current landscape, intrusion detection methods relying on spatiotemporal features often utilize deep learning techniques like CNN and LSTM to establish sequential intrusion detection models. However, these methods can be susceptible to the influence of previous models, and there is a tendency to overlook comprehensive spatiotemporal characteristics. There is a need for more comprehensive extraction of spatiotemporal features to enhance the overall performance of these methods.

\section{SUMMARY}
The advancement of the automotive industry has prominently elevated the significance of ensuring cyberspace security within vehicular systems. A proliferation of attacks has been observed, predominantly focusing on the {\can} utilized {\ivn}. In response to these threats, numerous defense strategies have been devised to mitigate attacks and fortify the security of vehicular systems. Nevertheless, the practical implementation of these solutions encounters certain constraints and hurdles that warrant further attention and exploration.

This paper offers a comprehensive investigation into the current landscape of vehicle attack and defense strategies, with a specific focus on the CAN. The objective of this study is to critically evaluate the limitations of existing approaches and provide valuable insights for the future design of VIDS. 
We provide a comprehensive synthesis of existing VIDS from multiple perspectives and conduct evaluations on a unified dataset to assess the effectiveness of selected methodologies.
Our analysis reveals a predominant emphasis on specific attack categories within the examined VIDS, thereby disregarding the more sophisticated and realistic attack scenarios. To address these shortcomings, we put forth a set of defense recommendations based on our research findings. Furthermore, considering the advancement of automotive intelligence, we propose additional cybersecurity recommendations tailored to the domain of smart car technology.

\section*{Acknowledgments}
The authors sincerely thank the associate editor and the anonymous reviewers for their invaluable contributions, which have significantly enhanced the quality of this article. This work was supported by Hong Kong RGC TRS Project (No. T43-513/23-N) and HKPolyU Grant (1-ZVG0, U-ZGGG).

\bibliographystyle{IEEEtran}
\bibliography{main_final.bbl}    

\section{Biography Section}
\vspace{-40pt}
\begin{IEEEbiography}[{\includegraphics[width=1in,height=1.25in,clip,keepaspectratio]{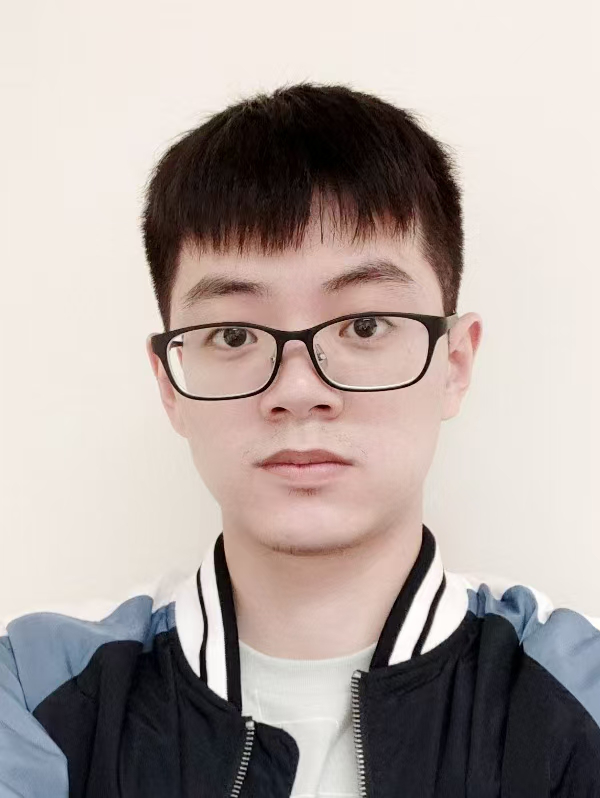}}]{Yangyang Liu} received the bachelor degree in Hunan university, Hunan, China, in 2017. He is currently pursuing the Ph.D. degree at the Hong Kong Polytechnic University since 2021. His research focuses on network measurement and the security of in-vehicle networks with papers published in top-tier venues, such as S\&P, USENIX Security, NDSS, and JSAC.
\end{IEEEbiography}
\vspace{-40pt}

\begin{IEEEbiography}[{\includegraphics[width=1in,height=1.25in]{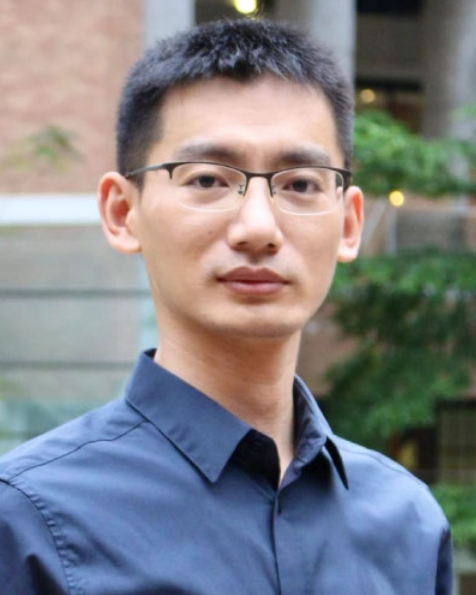}}]{Lei Xue}
 (Member, IEEE) received the Ph.D. degree in computer science from The Hong Kong Polytechnic University. He is an Associated Professor with the School of Cyber Science and Technology, Sun Yat-sen University. He is also a
 member with Guangdong Provincial Key Laboratory of Information Security Technology. He is a Yat-sen Scholar. His current research topics mainly focus on mobile and IoT system security, program analysis, and automotive security.
\end{IEEEbiography}
\vspace{-40pt}

\begin{IEEEbiography}[{\includegraphics[width=1in,height=1.25in]{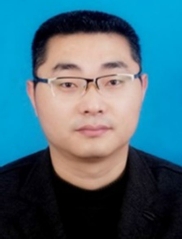}}]{
Sishan Wang} received his Master degree in Wuhan University, China, in 2010. He received his Bachelor degree in Hubei University of Automotive Technology, China, in 2005. He is currently a Lecture at Institute of Automotive Engineers in Hubei University of Automotive Technology. His research interests are Intra-Vehicle Networking, Automotive Ethernet, and Autonomous Vehicle Control.
\end{IEEEbiography}
\vspace{-40pt}

\begin{IEEEbiography}[{\includegraphics[width=1in,height=1.25in]{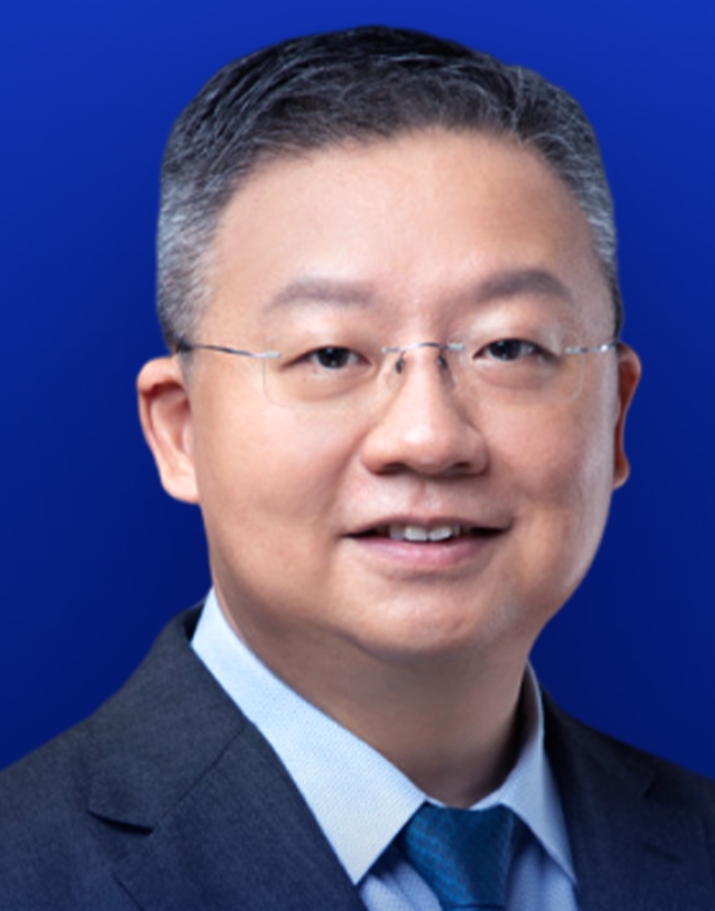}}]{Xiapu Luo}
 (Senior Member, IEEE) is a Professor with the Department of Computing, The Hong Kong Polytechnic University. His research focuses on mobile/IoT security and privacy, blockchain/smart contracts, network/web security and privacy, software engineering, and internet measurement. He has published papers in top security/software engineering/networking conferences and journals. His research has led to more than ten best/distinguished paper awards and several awards from industry.
\end{IEEEbiography}
\vspace{-40pt}

\begin{IEEEbiography}[{\includegraphics[width=1in,height=1.25in]{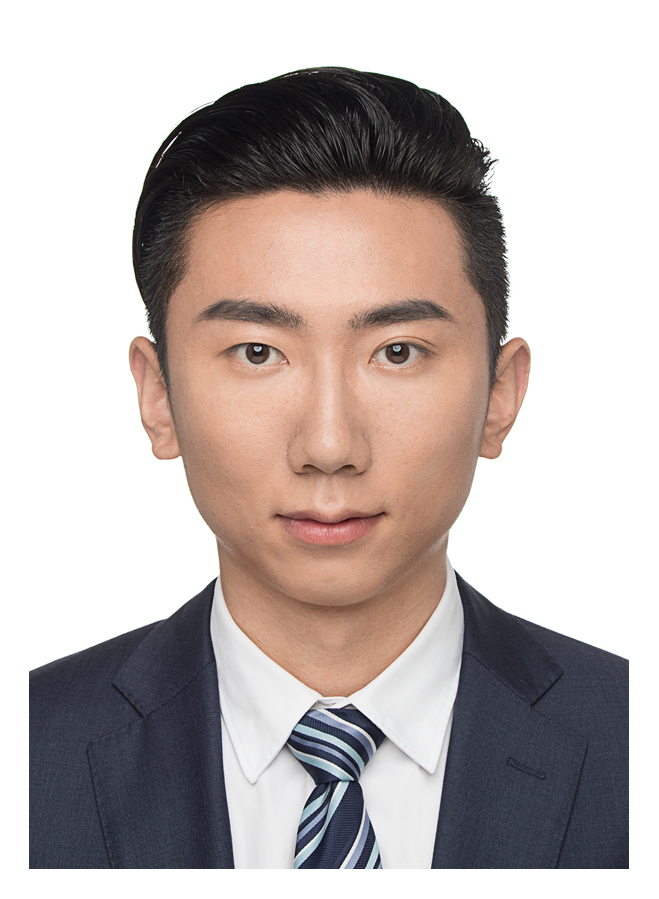}}]{Kaifa Zhao} is a Ph.D. candidate in the Department of Computing at The Hong Kong Polytechnic University, Hong Kong, China. His research interests span LLM4Code, AI4Security, Security4AI, and mobile security and privacy.
\end{IEEEbiography}
\vspace{-40pt}

\begin{IEEEbiography}[{\includegraphics[width=1in,height=1.25in]{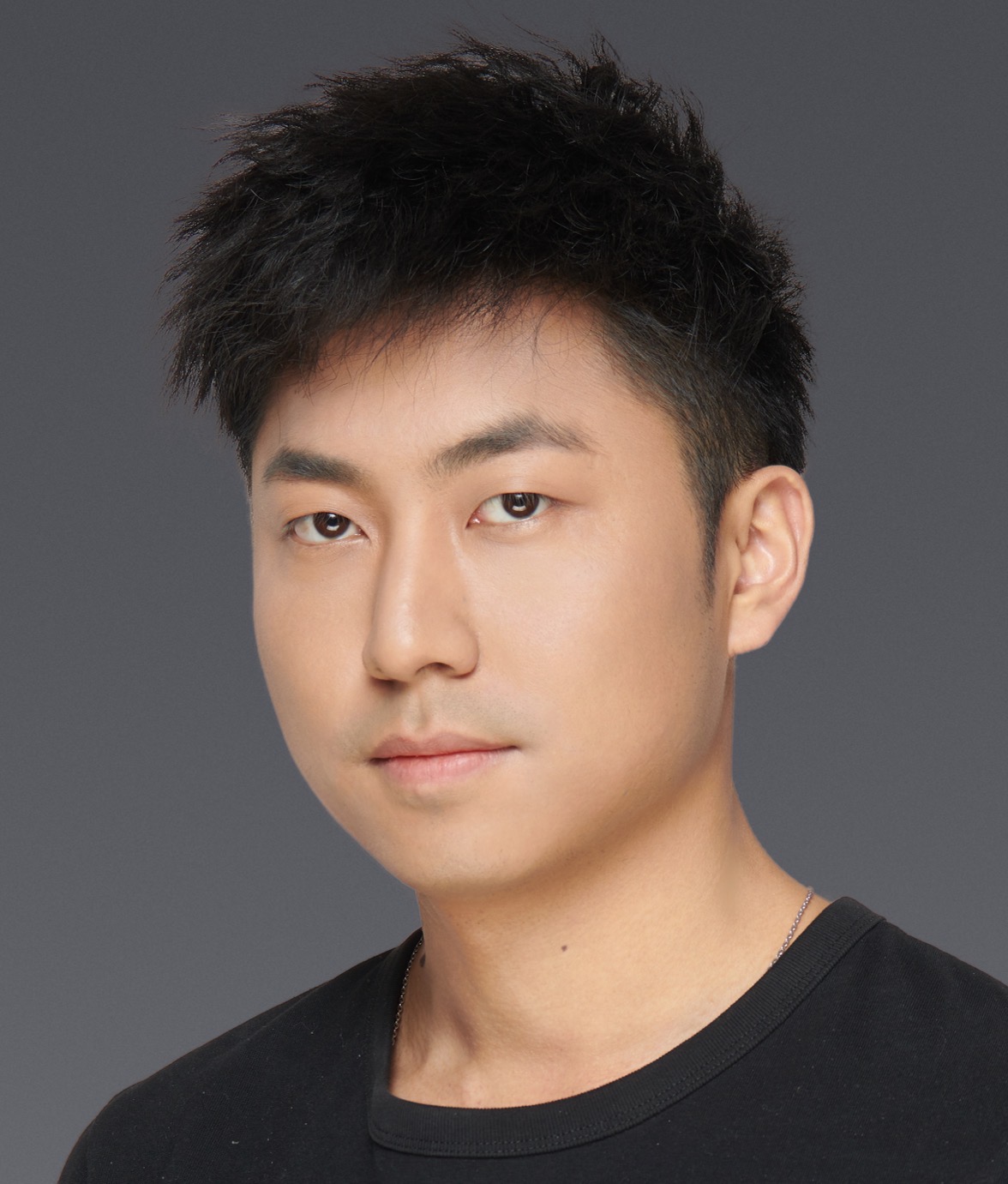}}]{Pengfei Jing} received his Ph.D. degree from the Department of Computing at The Hong Kong Polytechnic University (PolyU) in 2025, under the supervision of Prof. Luo Xiapu. His research interests mainly include the safety and security of modern vehicles, and cutting-edge autonomous driving systems, including end-to-end systems and vision-language-action (VLA) systems.
\end{IEEEbiography}
\vspace{-40pt}

\begin{IEEEbiography}[{\includegraphics[width=1in,height=1.25in]{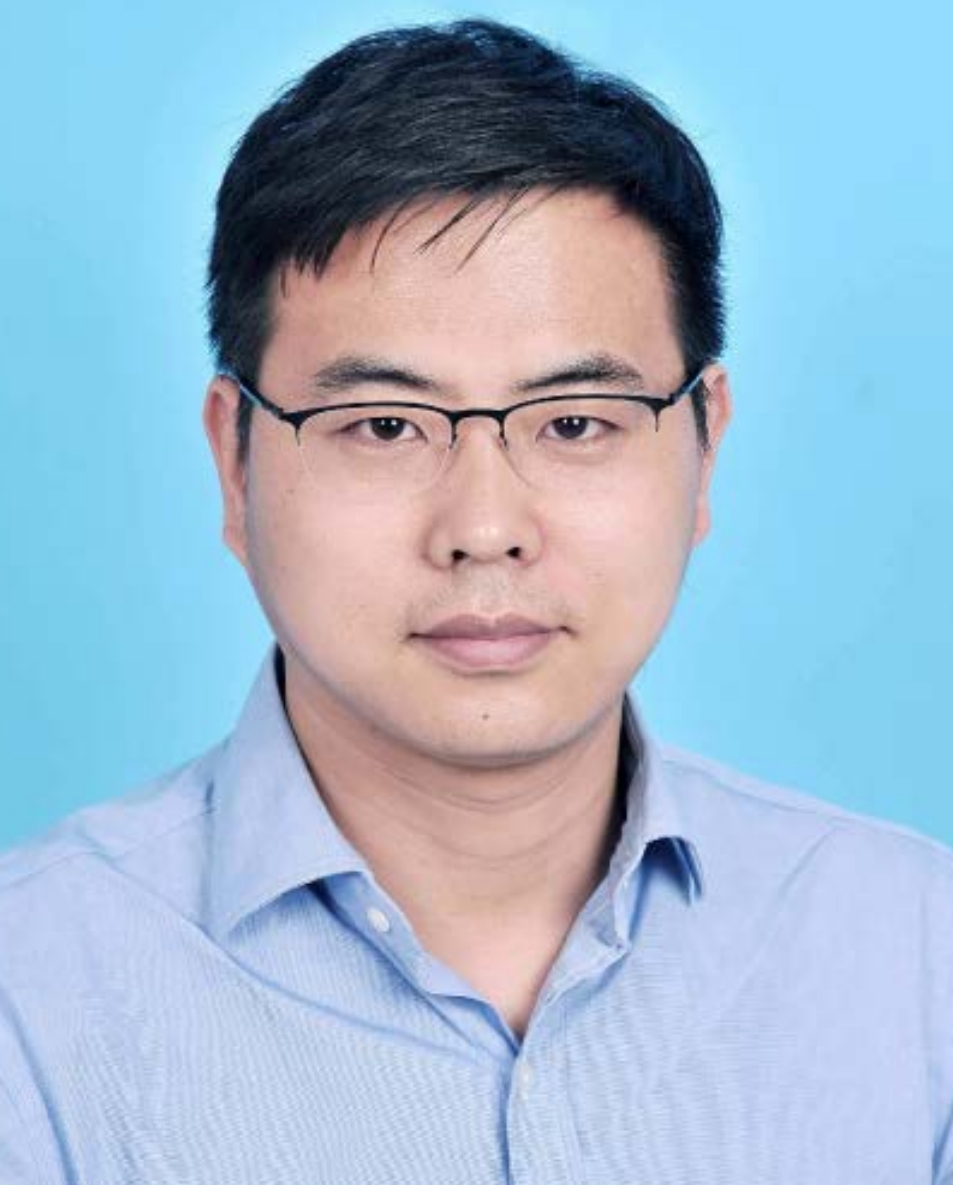}}]{Xiaobo Ma}
(Member, IEEE) received the Ph.D. degree in control science and engineering from Xi’an Jiaotong University, Xi’an, China, in 2014.
He is a Professor with the MOE Key Laboratory for Intelligent Networks and Network Security, Faculty of Electronic and Information Engineering,
Xi’an Jiaotong University. He was a Post-Doctoral Research Fellow with The Hong Kong Polytechnic University in 2015. He is a Tang Scholar. His research interests include internet measurement and cyber security
\end{IEEEbiography}

\vspace{-40pt}
\begin{IEEEbiography}[{\includegraphics[width=1in,height=1.25in]{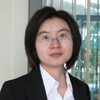}}]{Yajuan Tang} received the Ph.D. degree in radio
physics from Wuhan University in 2006. She is currently an Associate Professor with the Department
of Electronic and Information Engineering, Shantou
University. Her current research focuses on network
security, network privacy, and malicious traffic
analysis in networks using advanced signal processing techniques.
\end{IEEEbiography}

\vspace{-40pt}
\begin{IEEEbiography}[{\includegraphics[width=1in,height=1.25in]{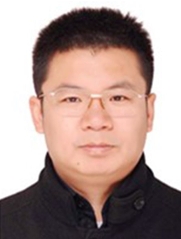}}]{Haiying Zhou} received his B.E., M.E. and Ph.D. degrees in Electronic Engineering from Wuhan University in 1997, 1999 and 2005. He is a Full Professor at Hubei University of Automotive Technology (HUAT, China). His research interests are in the area of new technologies and smart applicants of Internet of Things (IoT), Electric Vehicle and Intelligent Drive.
\end{IEEEbiography}

\end{document}